\documentclass[11pt,a4paper]{article}
\usepackage{jheppub}
\usepackage[english]{babel}
\usepackage[utf8]{inputenc}
\usepackage{graphicx}
\usepackage{amsmath,amssymb,amsfonts,amsthm}
\usepackage{bm,bbm}
\usepackage{tikz,stackrel}
\usepackage{pstricks}
\usepackage{pifont} 
\usepackage{empheq}
\newcommand{\bc}{\begin{center}}
\newcommand{\ec}{\end{center}}
\newcommand{\be}{\begin{equation}}
\newcommand{\ee}{\end{equation}}
\newcommand{\ba}{\begin{eqnarray}}
\newcommand{\ea}{\end{eqnarray}}
\def\bs{\begin{subequations}}
\def\es{\end{subequations}}
\newcommand{\ben}{\begin{equation*}}
\newcommand{\een}{\end{equation*}}
\newcommand{\ban}{\begin{eqnarray*}}
\newcommand{\ean}{\end{eqnarray*}}
\renewcommand{\leq}{\leqslant}

\def\a{\alpha}
\def\b{\beta}
\def\de{\delta}
\def\g{\gamma}
\def\la{\lambda}
\def\k{\kappa}
\def\e{\epsilon}

\def\Om{\Omega}
\def\om{\omega}

\def\G{\Gamma}
  
\def\s{\sigma}

\def\vp{\varphi}
\def\N{\nabla}
\def\cA{\mathcal{A}}
\def\cB{\mathcal{B}}
\def\cC{\mathcal{C}}
\def\cD{\mathcal{D}}
\def\cE{\mathcal{E}}

\def\cG{\mathcal{G}}

\def\cQ{\mathcal{Q}}
\def\cR{\mathcal{R}}

\def\cV{\mathcal{V}}

\def\ds{d_\textsc{s}}
\def\dh{d_\textsc{h}}

\def\p{\partial}

\def\B{\Box}
\newcommand{\Eq}[1]{(\ref{#1})}
\newcommand{\Eqq}[1]{eq.~(\ref{#1})}
\newcommand{\Eqqs}[1]{eqs.~(\ref{#1})}

\def\cob{\color{blue}}

\newcommand{\au}[2]{#1.~#2}
\newcommand{\book}[5]{\emph{#1}, #2, #3, #4 (#5)}

\newcommand{\oarX}[1]{\href{http://arxiv.org/abs/#1}{{\ttfamily\cob arXiv:#1}}}
\newcommand{\arX}[1]{\href{http://arxiv.org/abs/#1}{{\ttfamily\cob arXiv:#1}}}
\newcommand{\doin}[6]{\href{http://dx.doi.org/#1}{{\cob {\it #2} {\bf #3 #4} (#6) #5}}}
\newcommand{\doinn}[5]{\href{http://dx.doi.org/#1}{{\cob {\it #2} {\bf #3} (#5) #4}}}
\newcommand{\doij}[5]{\href{http://dx.doi.org/#1}{{\cob {\it #2} {\bf #3} (#5) #4}}}

\newcommand{\ndoinn}[5]{\href{#1}{{\cob {\it #2} {\bf #3} (#5) #4}}}
\newcommand{\procsinm}[5]{in \emph{#1}, #2 eds., #3, #4 (#5)}

\newcommand{\tia}[1]{\textit{#1},}
\def\rme{e}
\def\rmd{d}
\def\rmi{i}

\newcounter{listcounter}

\allowdisplaybreaks

\begin{document}

\title{Gravitational potential and galaxy rotation curves in multi-fractional spacetimes}

\author[a]{Gianluca Calcagni,}
\emailAdd{g.calcagni@csic.es}
\affiliation[a]{Instituto de Estructura de la Materia, CSIC, Serrano 121, 28006 Madrid, Spain}

\author[b]{Gabriele U.\ Varieschi}
\emailAdd{Gabriele.Varieschi@lmu.edu}
\affiliation[b]{Department of Physics, Loyola Marymount University, 1 LMU Drive, Los Angeles, CA 90045, USA}

\abstract{Multi-fractional theories  with integer-order derivatives are models of gravitational and matter fields living in spacetimes with variable Hausdorff and spectral dimension, originally proposed as descriptions of geometries arising in quantum gravity. We derive the Poisson equation and the Newtonian potential of these theories starting from their covariant modified Einstein's equations. In particular, in the case of the theory $T_v$ with weighted derivatives with small fractional corrections, we find a gravitational potential that grows logarithmically at large radii when the fractional exponent takes the special value $\alpha=4/3$. This behaviour is associated with a restoration law for the Hausdorff dimension of spacetime independently found in the dark-energy sector of the same theory. As an application, we check whether this potential can serve as an alternative to dark matter for the galaxies NGC7814, NGC6503 and NGC3741 in the SPARC catalogue. We show that their rotation curves at medium-to-large radii can indeed be explained by purely geometric effects, although the Tully--Fisher relation is not reproduced well. We discuss how to fix the small-radius behaviour by lifting some approximations and how to test the model with other observables and an enlarged galaxy sample.}

\keywords{Classical Theories of Gravity, Models for Dark Matter, Models of Quantum Gravity}
\preprint{\doij{10.1007/JHEP08(2022)024}{JHEP}{2208}{024}{2022} [\arX{2112.13103}]}

\maketitle



\section{Introduction}

Quantum gravity is a subject of research and debate at multiple levels. Often, the discussion is centered on whether a consistent theory of quantum gravity exists, which is  no longer an issue because it has already been solved in several independent ways \cite{Ori09,Fousp,Modesto:2017sdr}. On the other hand, paradoxically, the big question of whether it can be tested by observations is sometimes dismissed on the grounds that quantum-gravity effects should always be confined to Planck scales, which is true only in the very specific and limited case of perturbative corrections to the action in the form of higher-order curvature operators (e.g., \cite{Calcagni:2020edt} and references therein).

To answer the above question, it may be useful to concentrate on features commonly shared by all or most quantum-gravity scenarios. One such feature is dimensional flow, the change of spacetime dimension with the probed scale. By ``spacetime dimension'' one may indicate different concepts, usually limited to the Hausdorff dimension $\dh$ (the scaling of spacetime volumes with respect to the linear length) and the spectral dimension $\ds$ (the scaling of the dispersion relation of a particle with respect to the momentum). While there is great variation in the details of dimensional flow, such as the numerical values of the dimensions and the scales at which transitions from one value to another take place, the phenomenon \emph{per se} occurs in all known theories of quantum gravity \cite{tH93,Car09,fra1,first,Car17,MiTr}. 

One of the attempts to better understand dimensional flow has taken the straightforward route to implement this feature explicitly at the level of the action. The ensuing class of theories is generically called multi-scale spacetimes \cite{trtls} and a particular realization of multi-scale geometries goes under the name of multi-fractional spacetimes \cite{revmu,Calcagni:2021ipd}. In these scenarios, the measure of the action is factorized in the spacetime coordinates for simplicity and, under very mild assumptions, acquires a universal parametric form \cite{first}. The Hausdorff dimension $\dh$ is determined by the scaling of this measure. The spectral dimension $\ds$, however, depends on the kinetic term of the fields in the action and there is no unique way to fix this. Depending on the symmetries of the Lagrangian, one can realize different kinetic terms with the same anomalous scaling. According to the classification of \cite{revmu,Calcagni:2021ipd}, if the kinetic term is made of ordinary partial derivatives $\p_\mu$, the theory is called multi-fractional theory with ordinary derivatives, labelled as $T_1$ in short. If, instead, one uses ordinary derivatives decorated with measure weights, one realizes the theory $T_v$ with weighted derivatives (if the weights are balanced on the left and on the right of each derivative) or the theory $T_q$ with $q$-derivatives (if the weight is only on the left). Finally, if the kinetic term is made of fractional differential operators, one has several other possibilities, collectively labelled as $T_\g$ and called multi-fractional theories with fractional operators.

Although multi-fractional theories have been explored in many aspects ranging from particle physics to cosmology, there are several important gaps in our knowledge about whether and how dimensional flow can leave an observable imprint. The example we choose in this paper is the gravitational potential, which has not been computed so far. Clearly, a modification of Newton's potential can have many ramifications in the solar system, astrophysics and cosmology. In particular, it can bear consequences for the problem of dark matter. When the gravitational theory changes, it may be possible to get an alternative to particle-like dark matter and explain the rotation curve of galaxies, the matter and galaxy distribution and the content of the universe in terms of a modified dynamics and new degrees of freedom \cite{Mannheim:2005bfa,Li:2019ksm,Modesto:2021yyf}. The cases of $f(R)$ gravity \cite{Capozziello:2012ie} and models of modified Newtonian dynamics (MOND) \cite{Sanders:2002pf,Bekenstein:2004ne,Famaey:2011kh} are well documented. Recently, the idea of invoking anomalous dimensions to realize dark-matter galactic observations has been retaken, implicitly, in Newtonian dynamics with a fractional Laplacian \cite{Giusti:2020rul,Giusti:2020kcv} and, explicitly, in Newtonian fractional-dimension gravity (NFDG) \cite{Varieschi:2020ioh,Varieschi:2020dnd,Varieschi:2020hvp}. A covariant version of NFDG was shown to essentially match with the multi-fractional theory $T_1$ \cite{Varieschi:2021rzk}, so that one may wonder whether multi-fractional theories can directly say anything interesting about the dark-matter problem.

In this work, we start from the covariant equations of motion of the three multi-fractional theories with integer-order derivatives and obtain the Poisson equation relating the gravitational potential and matter. Solving this equation for a point-wise source yields the gravitational potential. Its form turns out to be quite different in different theories. Next, we will calculate the gravitational potential for an extended matter distribution and compare it with astrophysical data from the Spitzer Photometry \& Accurate Rotation Curves (SPARC) database, a catalogue of the rotation curves of 175 galaxies \cite{McGaugh:2016leg,Lelli:2016zqa}. In particular, here we take the three prototypical galaxies presented in \cite{McGaugh:2016leg}: the bulge-dominated spiral NGC7814, the disk-dominated spiral NGC6503 and the gas-dominated dwarf NGC3741, which were used as main examples of the empirical radial acceleration relation in MOND-like models \cite{McGaugh:2016leg} and were also studied in the recent NFDG model \cite{Varieschi:2020hvp}. An explanation of the rotation curves of these and other galaxies would be the first step for a purely geometric alternative to dark matter, the next steps (which we will not take here) being the study of subjects such as galaxy clusters, the Bullet cluster, gravitational lensing, and so on. We will find that, among the multi-fractional theories with integer-order differential operators, only the theory $T_v$ can explain the plateau in the rotation curve of galaxies. This happens for a certain scaling of the measure which, surprisingly, recovers a Hausdorff dimensions of spacetime equal to $\dh=4$ even if the background is static and, ordinarily, would be sensitive to only three dimensions. These findings point towards yet unexplored symmetries of the theory and they are complementary to those of \cite{Calcagni:2020ads} for a homogeneous (i.e., only time dependent, i.e, one-dimensional) background, where the same restoration of $\dh=4$ dimensions was noted.

The plan of the paper is as follows. In section \ref{mearev}, we review the spacetime measure used in multi-fractional theories and discuss its form in spherical and cylindrical coordinates. The Poisson equation and Newton's potential of the theory $T_1$ are derived for the first time in section \ref{secT1}. The Poisson equation, Newton's potential, the theoretical galaxy rotation curves and the observational rotation curves compared with SPARC data for the theories $T_v$ and $T_q$ are discussed in sections \ref{secTv} and \ref{secTq}, respectively. Section \ref{disc} is devoted to conclusions, where we also sketch some preliminary results about the Bullet cluster and ways to fix the small-radius fit of the rotation curves. Appendix \ref{dmrev} is a review of the derivation of the Poisson equation and Newton's potential in Einstein's theory, as well as of SPARC observational results for the galaxies NGC7814, NGC6503 and NGC3741 and of the rotation curves predicted by Einstein's theory, obtained both from a theoretical approximation of the baryonic matter distribution (an exponentially damped energy area density distributed in a thin disk) and from the energy density actually observed. Those familiar with the details can ignore this appendix but we recommend its reading not only because it collects our conventions and notations, but also because it serves as a parallel of the ingredients, calculations and strategies employed in the multi-fractional cases. The Poisson equation in Cartesian coordinates is reported in Appendix \ref{carcor} for all three multi-fractional theories. The circular velocity for the theory $T_v$ is derived in Appendix \ref{app0}. Appendix \ref{appA} contains some technical material about the multipole expansion of inverse-power potentials. Some useful integration formul\ae\ are collected in Appendix \ref{appB}. Appendix \ref{appC} contains the calculations of the thin-disk approximation with exponential matter density for the theory $T_v$, while Appendix \ref{appD} contains the expressions of the thick-disk data-based galactic potential and gravitational field for the theory $T_v$.


\section{Spacetime measure in multi-fractional theories}\label{mearev}

Since the spacetime measure is the same for all the theories we will consider in the following, we recall its main aspects.

Any continuous spacetime with a varying Hausdorff dimension admits an integration measure of a universal parametric form, where the choice of parameter values strongly depends on the specific theory \cite{first,revmu}. Here we consider factorizable measure weights,
\be\label{meagen}
\rmd^D q(x):=\prod_{\mu=0}^d\rmd q_\mu(x^\mu)=\rmd^Dx\,v(x)=\rmd^Dx\prod_{\mu=0}^{d} v_\mu(x^\mu)\,,
\ee
where $v_\mu(x^\mu):=\p_\mu q_\mu(x^\mu)$ is a function of the Cartesian coordinate along the direction $\mu$ and of some fundamental scales of geometry $\ell_l$, $l=1,2,\dots$. This measure breaks Lorentz invariance and recovers it only at certain scales where $q_\mu(x^\mu)\to x^\mu$ and $v(x)\to 1$. In the most general factorizable case, the profiles $q_\mu(x^\mu)$ are an infinite superposition of complex powers, $q_\mu(x^\mu)=\sum_l \g_l|x^\mu/\ell^\mu_l|^{\a_l+\rmi\om_l}$ for all $\mu$, where $\a_l,\om_l\in\mathbb{R}$ and $\g_l\in\mathbb{C}$ are dimensionless constants and $\ell^\mu_l\in\mathbb{R}$ are fundamental length or time scales of the geometry. However, reality of the measure constrains the parameters $\g_l$ and $\om_l$ of this sum to combine into a real-valued expression, a generalized polynomial deformed by logarithmic oscillations.

In most applications ranging from classical mechanics to quantum field theory, factorizable measures are much more manageable \cite{frc2,frc3} than measures preserving, say, only spatial rotation \cite{fra1,fra2}. However, as a first approximation in spherically or axially symmetric configurations, one can rewrite \Eq{meagen} in terms of the radial coordinate $r$ (spherical radius) or $R$ (cylinder radius). Also, in this paper we consider only static configurations and we can ignore the time-dependent part of the measure or, equivalently, set $v_0(x^0)=v_0(t)=1$. Lastly, only one fundamental scale $\ell_1\equiv\ell_*$ is necessary to get dimensional flow, which is the simplest configuration giving rise to non-trivial effects. For a spherical or axially symmetric configuration, we can dub this scale $r_*$ or $R_*$.


\subsection{Measure in radial coordinates}\label{radcor}

On Minkowski background in spherical coordinates, we can symmetry reduce the measure weight to a radial configuration, where only the radial part has an anomalous scaling. In this setting, the universal scaling found in \cite{first,revmu} takes the following form for a one-dimensional measure $q(r)$:
\bs\label{genpo}\ba
q(r) &=& r+\frac{r_*}{\a}\left|\frac{r}{r_*}\right|^\a F(r)\,,\\
F(r) &=& A_0+\sum_{n=1}^{+\infty} \tilde F_{n}(r)\,,\\
\tilde F_{n}(r) &=& A_n\cos\left(n\om\ln\frac{r}{r_*}\right)+B_n\sin\left(n\om\ln\frac{r}{r_*}\right),\label{fnl}
\ea\es
where $r>0$, $\a\in\mathbb{R}$ is a fractional exponent proportional to the Hausdorff dimension of spacetime, $0<A_n,B_n<1$ are constant amplitudes and $\om=2\pi\a/\ln N$ is a frequency taking values on the discrete set $N=2,3,\dots$. The constant $A_0$ can take the values $1$ or $0$, depending on whether one includes the zero mode or not, respectively. We will always set $A_0=1$.

The measure weight $v(r)$ is simply
\be\label{vq}
v(r)=\p_r q(r)\,.
\ee
The log-oscillating part of the measure enjoys discrete scale invariance,
\be
F\left(\rme^{\frac{2\pi}{\om}}r\right) = F(r)\,,\qquad \rme^{\frac{2\pi}{\om}}= N^{\frac1\a}\,,
\ee
a property typical of fractals and multi-fractals. Averaging out the oscillations \cite{frc2,revmu}, $\langle F\rangle =1$ and one finds a simple power-law behaviour capturing the scaling of geometry across an infrared (IR) to ultraviolet (UV) divide set by $r_*$:
\be\label{qavg}
\langle q(r)\rangle = r+\frac{r_*}{\a}\left|\frac{r}{r_*}\right|^\a.
\ee
This averaging is useful in many applications of multi-fractional theories because it removes a modulation of the measure which does not add much to the main geometric effect. We will adopt \Eq{qavg} in most of the paper. However, in certain cases log oscillations do play an important physical role \cite{Calcagni:2017via,Calcagni:2020ads} and we will include them whenever the main effect we want to find (a flattening of the rotation curve of galaxies) is not captured by geometries with the power law \Eq{qavg}.

The exact, full form of the measure in Cartesian coordinates with many fundamental scales as well as the exact form of the amplitudes $A_n$ and $B_n$ can be found in \cite{first,revmu,Calcagni:2021ipd}.

In $d$ spatial dimensions, passing from Cartesian to spherical coordinates requires more work. We have to translate the anomalous scaling of the $q_i(x^i)$ into an anomalous scaling which, again, we take only in the radial direction. To understand the form of the effective coordinate $q(r)$, consider a $d=2$ problem with isotropic coordinates $q_i=q$:
\be\label{qqQ}
q^2(x)+q^2(y) =: \cQ^2(r)\,.
\ee
Ignoring logarithmic oscillations and writing $q(x)=x+ a x^\a$, $q(y)=y+ a y^\a$ and $\cQ(r)=r+ b r^\b$ for constant $a$ and $b$, we have
\be\label{corr1}
q^2(x)+q^2(y) = x^2+y^2+a^2(x^{2\a}+y^{2\a})+O(xy^\a,yx^\a) =: r^2+b^2r^{2\b}+2br^{\b+1}.
\ee
Since, by definition of the theory, the rulers of observers in the fractional frame follow standard geometry, we equate $x^2+y^2= r^2$ to give $r$ the same geometric meaning as usual. Then, from
\ba
r^{2\b}&=&(x^2+y^2)^\b\nonumber\\
&=&x^{2\b}+\sum_{n=1}^{+\infty}\frac{\G(\b+1)}{n!\G(\b-n+1)}x^{2(\b-n)}y^{2n}\nonumber\\
&=&y^{2\b}+\sum_{n=1}^{+\infty}\frac{\G(\b+1)}{n!\G(\b-n+1)}y^{2(\b-n)}x^{2n}\nonumber\\
&=&\frac12\left\{x^{2\b}+y^{2\b}+\sum_{n=1}^{+\infty}\frac{\G(\b+1)}{n!\G(\b-n+1)}\left[x^{2(\b-n)}y^{2n}+y^{2(\b-n)}x^{2n}\right]\right\}\nonumber\\
&=&\frac12\left(x^{2\b}+y^{2\b}\right)+\textrm{(mixed terms)}\,,
\ea
we can realize the correspondence \Eq{corr1} approximately provided $\a=\b$ and $a^2=b^2/2$. Generalizing to $d$ dimensions, $b=\sqrt{d}a$, but we can always reabsorb the factor $\sqrt{d}$ into the scale $r_*$, so that the averaged expression \Eq{qavg} approximates $\cQ(r)$ up to mixed coordinate terms, while the radial derivative of \Eq{defqr} is the measure weight 
\be\label{radialtildev}
\tilde v(r)=\p_r q(r)=1+\left|\frac{r}{r_*}\right|^{\a-1}.
\ee
The same radial composite coordinate \Eq{qavg} can be obtained directly in the fractional frame using an exact procedure \cite{Calcagni:2017ymp}. Whichever route one takes to arrive at \Eqq{qavg}, the scaling of the radial measure is then $\rmd q(r)\,q^{d-1}(r)\simeq\rmd r\,\tilde v(r)\,[r\tilde v(r)]^{d-1}=\rmd r\,r^{d-1} \tilde v^d(r)$, which is the same as the scaling of the radial measure $\rmd r\, r^{d-1} v(r)$ defined by
\be\label{radialv}
v=v(r)=1+\left|\frac{r}{r_*}\right|^{d(\a-1)}.
\ee
Of course, one could define the theory with an exact profile $q(r)$ and derive its anomalous scaling in Cartesian coordinates, i.e., take \Eq{qqQ} as a definition of the left-hand side ($:=$) instead of the right-hand side ($=:$). However, Cartesian coordinates are usually the starting point to construct this Lorentz-breaking theory.

Different choices that included the angular part would not lead to very different physics, since the difference between \Eq{radialv} and a profile $v(r)v(\theta)\dots$ or $v(r,\theta,\dots)$ would be negligible across galactic distances. The power of $r$ in \Eqq{radialv} has been chosen in order to maintain a spatial Hausdorff dimension
\be\label{dhspace}
\dh^{\rm space}\simeq d\a
\ee
at scales where fractional effects are maximal ($r\ll r_*$ or $r\gg r_*$, depending on whether $\a<1$ or $\a>1$). In fact, in spherical coordinates the spatial volume element is $\rmd^d\bm{x}=\rmd r\,r^{d-1}\rmd\Om_{d-2}$, where $\rmd\Om_{d-2}$ is the angular volume element. Therefore, the multi-fractional measure anomalous only in the radial direction in spherical coordinates is of the form
\be
\rmd^d\bm{x}\,v(\bm{x})=\rmd r\,r^{d-1}v(r)\,\rmd\Om_{d-2}\,,
\ee
where $v(r)$ is the profile \Eq{radialv}. In the fractional regime, the radial measure scales as $\rmd r\,r^{d-1} v(r)\simeq \rmd r\,r^{d\a-1}$.

We report two useful formul\ae\ that we will need to derive the Poisson equation and Newton's potential. Using \Eq{radialv}, one can check that
\ba
\frac{\p_r v}{v} &=& \frac{d(\a-1)}{r}\cA(r)\,,\label{prvv}\\
\frac{\N^2 v}{v}&=&\frac{1}{v}\frac{1}{r^{d-1}}\p_r\left(r^{d-1}\p_r v\right)= \frac{d(\a-1)(d\a-2)}{r^2}\cA(r)\,,
\ea
where $\N^2$ is the spatial flat Laplacian and
\be\label{cAfun}
\cA(r):=\frac{|r/r_*|^{d(\a-1)}}{1+|r/r_*|^{d(\a-1)}}\,.
\ee


\subsection{Measure in cylindrical coordinates}

The same expressions of section \ref{radcor} hold in cylindrical coordinates with $r$ replaced by $R$, under some approximations we will discuss presently.

In cylindrical coordinates, we take a measure weight which is non-trivial only in the radial direction because we expect multi-fractional effects to depend on the distance from the center of the galaxy, not on the direction one is looking at. Since in cylindrical coordinates the spatial volume element is $\rmd^d\bm{x}=R\rmd R\,\rmd\vp\prod_i\rmd Z_i$, the multi-fractional measure anomalous only in the radial direction in cylindrical coordinates is
\be
\rmd^d\bm{x}\,v(\bm{x})=\rmd R\,R\,v(R)\,\rmd\vp\prod_i\rmd Z_i\,,
\ee
where
\be\label{radialvR}
v(R)=1+\left|\frac{R}{R_*}\right|^{d(\a-1)}.
\ee
Again, the power in $v$ has been chosen to keep the spatial Hausdorff dimension equal to \Eq{dhspace}, to compare it with the other models.

Notice that the weight \Eq{radialvR} is not exactly equivalent to a coordinate transformation of \Eqq{radialv} from spherical to cylindrical coordinates, since the system is not coordinate invariant except in the Einstein-theory limit $v\sim 1$. Therefore, \Eqq{radialvR} has to be taken as an \emph{Ansatz} or approximation of the effects we expect to find in a spacetime with a certain matter distribution. Since such effects usually depend on the distance from the observer \cite{Calcagni:2017ymp,CaRo2}, it is natural to confine the weight $v\neq 1$ to the radial direction both in spherical coordinates (modeling globular galaxies or the bulge of spiral galaxies) and in cylindrical coordinates (modeling the disk of spiral galaxies). 

From \Eq{radialvR},
\ba
\frac{\p_R v}{v} &=& \frac{d(\a-1)}{R}\cA(R)\,,\\
\frac{\N^2 v}{v}&=&\frac{1}{v}\frac{1}{R}\p_R\left(R\p_R v\right)= \frac{d^2(\a-1)^2}{R^2}\cA(R)\,,
\ea
where $\cA$ is given by eq.\ \Eq{cAfun} but with $r$ and $r_*$ replaced by $R$ and $R_*$.


\section{Theory \texorpdfstring{$T_1$}{T1} with ordinary derivatives}\label{secT1}

This theory is not invariant under any modified version of coordinate invariance, contrary to $T_v$ and $T_q$.


\subsection{\texorpdfstring{$T_1$}{T1}: equations of motion}

Introducing the standard Levi-Civita connection and curvature tensors \Eq{tutto}, the action of the multi-fractional theory with ordinary derivatives is \cite{fra2,frc11}
\be\label{Sg2}
S =\frac{1}{2\k^2}\int\rmd^Dx\,v\,\sqrt{|g|}\,\left[R-w(v)\p_\mu v\p^\mu v-2U(v)\right]+S_{\rm m}\,,
\ee
where $R$ is the Ricci scalar (not to be confused with the cylindrical radius $R$ in the previous section; the context will make the difference clear), we included a kinetic-like term for the measure weight where $w$ is a function of $v$, $U$ is a potential-like term for the measure weight and $S_{\rm m}$ is the baryonic matter action. There is no dark-matter component in this and the other multi-fractional theories. The prefactor 2 in front of $U$ is just a convenient notation following \cite{Calcagni:2021ipd}. While in general it could be assumed that $w=0$ because it is not required by any symmetry consideration \cite{frc11}, the Poisson equation will actually be independent of $w$.

The modified Einstein equations of the theory are \cite{frc11}
\be
\k^2\,T_{\mu\nu}^v=G_{\mu\nu}+U\,g_{\mu\nu}+g_{\mu\nu}\frac{\B v}{v}-\frac{\N_\mu\N_\nu v}{v}+w\left(\frac12 g_{\mu\nu}\p_\s v\p^\s v-\p_\mu v \p_\nu v\right),\label{ee2}
\ee
where the energy-momentum tensor of matter fields is
\be\label{tmunu}
T_{\mu\nu}^v:=-\frac{2}{\sqrt{|g|}\,v}\frac{\de S_{\rm m}}{\de g^{\mu\nu}}\,.
\ee
Notice the extra weight factor compared with \Eq{tmunugr}. 
 In general, the term $U$ cannot be set to zero before checking the consistency of the background solutions, but in this case we will do so later because we can always reabsorb it in a suitable choice of the matter energy-momentum tensor. However, this operation might be harmful for the final results and we will discuss it again in section \ref{disc}.

Taking the trace of \Eq{ee2},
\be
\k^2\,T^v=-\frac{D-2}{2}R+DU+(D-1)\frac{\B v}{v}+\frac{D-2}{2}w\p_\s v\p^\s v,\label{ee2tr}
\ee
and, replacing $R$ back into \Eq{ee2}, we get
\ba
\k^2S_{\mu\nu}^v &=& R_{\mu\nu}-\frac{1}{D-2}g_{\mu\nu}\frac{\B v}{v}-\frac{\N_\mu\N_\nu v}{v}-w\p_\mu v \p_\nu v\,,\label{ee2fin}\\
S_{\mu\nu}^v &:=& \tilde T_{\mu\nu}^v-\frac{1}{D-2}g_{\mu\nu}\tilde T^v\,,\qquad \tilde T_{\mu\nu}^v:=T_{\mu\nu}^v-\frac{U}{\k^2}\,g_{\mu\nu}\label{Smunu}\,.
\ea


\subsection{\texorpdfstring{$T_1$}{T1}: Poisson equation and Newton's potential}

We derive the modified Newton's potential of the theory as a solution of the Poisson equation.

\subsubsection{\texorpdfstring{$T_1$}{T1}: Poisson equation in spherical coordinates}

We expand the metric as Minkowski background plus a perturbation,
\be\label{getah}
g_{\mu\nu}=\eta_{\mu\nu}+h_{\mu\nu}\,.
\ee
In a weak static field with non-relativistic matter, the component
\be
h_{00}=-2\Phi
\ee
is proportional to Newton's potential $\Phi({\bm x})$. 
 Taking the 00 component of the modified Einstein equations \Eq{ee2fin} with $U=0$ for a static configuration $h_{00}=-2\Phi({\bm x})$, on the left-hand side we get $S_{00}^v \simeq [(D-3)/(D-2)]\rho$, where $\rho=T_{00}^v$ is the non-relativistic matter energy density and we ignored pressure. Expanding the right-hand side of \Eq{ee2fin} at linear order, and since the background-metric $rr$-component in spherical coordinates is $\eta^{rr}=1$, the Poisson equation is
\ba
\frac{d-2}{d-1}\k^2\rho&=&\left[\p_r^2+\frac{d-1+d(\a-1)\cA(r)}{r}\p_r+\frac{1}{r^2}\N_{S^{d-1}}\right.\nonumber\\
&&\left.+\frac{2d(\a-1)(d\a-2)}{d-1}\frac{\cA(r)}{r^2}\right]\Phi-\frac{d(\a-1)(d\a-d-1)\cA(r)}{(d-1)r^2}h^{rr}\,.
\ea
Here we assumed that the time-dependent part of the measure weight $v$ is constant at the scales where the weak-field, static, non-relativistic approximation holds. 

At this point, we recall that the condition $h_{\mu}^{\mu}=0$ is dictated by the fact that $g_{\mu}^{\mu}=D=\eta_{\mu}^{\mu}$. Choosing a gauge where $h_{\theta\theta}=0=h_{\phi\phi}$, this implies that $0=h_{00}\eta^{00}+h_{rr}\eta^{rr}=-h_{00}+h_{rr}$, so that $h^{rr}=h_{rr}=-2\Phi$. Thus,
\begin{empheq}[box=\fcolorbox{black}{white}]{align}
\quad\frac{d-2}{d-1}\k^2\rho=&\left[\p_r^2+\frac{d-1+d(\a-1)\cA(r)}{r}\p_r+\frac{1}{r^2}\N_{S^{d-1}}\right.\nonumber\\
&\left.+\frac{2d(\a-1)(2d\a-d-3)\cA(r)}{(d-1)r^2}\right]\Phi\,.\label{pois1r}\quad
\end{empheq}
In the Einstein-theory or Newtonian limit $\cA\to 0$, one gets the ordinary Poisson equation \Eq{Poiord}, while in the fractional limit $\cA\to 1$ we obtain
\be\label{fract1}
\frac{d-2}{d-1}\k^2\rho\simeq\left[\p_r^2+\frac{d\a-1}{r}\p_r+\frac{1}{r^2}\N_{S^{d-1}}+\frac{2d(\a-1)(2d\a-d-3)}{(d-1)r^2}\right]\Phi\,.
\ee




\subsubsection{\texorpdfstring{$T_1$}{T1}: Newton's potential}\label{t1newt}

The Poisson equation is more difficult to solve with respect to the other two theories we will discuss below. The reason is that, although its solution is the combination of two power laws (possibly reducible to only one), the $1/r^2$ term never vanishes in $d=3$ dimensions, unless $\a=1$. This implies a difficulty when solving the Poisson equation in the sense of distributions at the point $r=0$. Here we limit the discussion to the solution in spherical coordinates for $r\neq 0$ and $d=3$. Taking a point-wise source $\rho=m\de^3(\bm{x})/v$ of mass $m$,
\be\label{fract1bis}
\frac{\k^2m}{2}\de^3(\bm{x})\simeq v\left[\p_r^2+\frac{3\a-1}{r}\p_r+\frac{18(\a-1)^2}{r^2}\right]\Phi(r)\,.
\ee
For $r\neq 0$, the radial solution of eq.\ \Eq{fract1bis} is
\be\label{phirv10}
\Phi(r)=C_1r^{\frac{2-3\a-a}{2}}+C_2 r^{\frac{2-3\a+a}{2}}\,,\qquad a:=\sqrt{3(44-21\a)\a-68}\,.
\ee
The parameter $a$ is real when
\be
0.91\approx \frac{2(11-\sqrt{2})}{21}<\a<\frac{2(11+\sqrt{2})}{21}\approx 1.18\,.
\ee
In particular, when $\a=1$, $a=1$, $\Phi(r)=C_2+C_1/r$ and one can set $C_2=0$ by a shift. 

The alternative is to consider the profile \Eq{phirv10} with purely imaginary $a$, which happens for
\be
\a<\frac{2(11-\sqrt{2})}{21}\approx 0.91\quad {\rm or}\quad \a>\frac{2(11+\sqrt{2})}{21}\approx 1.18\,.
\ee
Then, one can make the potential real provided $C_1=(c_1-c_2)r_*^{a/2}/2$ and $C_2=(c_1+c_2)r_*^{-a/2}/2$, where $c_{1,2}$ are real. In this case, 
\be\label{phirv11}
\Phi(r)=\frac{1}{r^{\frac{3\a}{2}-1}}\left[c_1 \cos\left(\frac{|a|}{2}\ln\frac{r}{r_*}\right)+c_2 \sin\left(\frac{|a|}{2}\ln\frac{r}{r_*}\right)\right].
\ee
The coefficients $c_{1,2}$ should be determined via techniques of the theory of distributions by considering the point $r=0$, but we will not do it here. We only notice that it should be possible to set $C_2=0$ by continuity with the solution in the Einstein-theory limit ($\a=1$), so that $c_1=-c_2$.

This solution wildly oscillates near the origin when $\a>2/3$ and may have other undesirable properties when applied to physical problems, such as a vanishing circular velocity at large radii. Therefore, we allow \Eq{phirv11} only for
\be
\a\leq\frac23\,.
\ee
In this range of values, the fractional limit corresponds to short scales and the potential undergoes infinitely many oscillations as it approaches the origin. When $\a<2/3$, these oscillations are progressively damped and $\Phi(0)=0$. This suggests that the singularity of Newton's potential, and perhaps other classical singularities met in Einstein's theory, may be resolved in this theory. In contrast, when
\be\label{special}
\a= \frac23\,,
\ee
oscillations are not damped but the potential remains bounded. With $C_2=0$ as argued above and $|a|/2=\sqrt{2}$, we have
\be\label{phirv1123}
\Phi(r)=c_1\left[\cos\left(\sqrt{2}\ln\frac{r}{r_*}\right)-\sin\left(\sqrt{2}\ln\frac{r}{r_*}\right)\right],\qquad \a=\frac23\,.
\ee
For dimensional reasons, $c_1\propto G m/r_*$, where $m$ is the mass of the source. 

In this paper, we will not study in detail the predictions of this theory about the rotation curve of galaxies, since we have not solved the Poisson equation completely. However, in order to get a flat curve in the fractional limit, the potential in that range of scales should be approximately constant. This suggests that the special value \Eq{special} could make the model fit data. In fact, on one hand we have just seen that the potential \Eq{phirv1123} is bounded. On the other hand, in the solution \Eq{phirv10} neither of the two power laws becomes a constant for any $\a\neq 1$. This strongly indicates that data fitting should be carried out with the rotation curve stemming from \Eq{phirv1123}, once the coefficient $c_1$ is determined. We will carry out a full calculation with the actual gravitational potential of the galaxy for the other theories, while for $T_1$ we content ourselves with a heuristic argument. The rotation curve for a spherical matter distribution is given by 
\be\label{vcircT0}
v_{\rm circ}(r)=\sqrt{\left|\frac{\rmd\Phi}{\rmd \ln r}\right|}\,.
\ee
This could be, for instance, the circular velocity of a bulge-dominated galaxy. If this bulge has size $r_0\sim r_*$, then we can approximate \Eqq{phirv1123} to
\be\label{fiap}
\Phi(r)\simeq b\frac{Gm}{r_*}\ln\frac{r}{r_*}\,,\qquad b=-c_1\sqrt{2}\frac{r_*}{Gm}\,,\qquad r=O(r_*)\,,
\ee
up to a shift. The constant $b$ is dimensionless. This logarithmic form will reappear later in the other multi-fractional theories, just like we will meet the special value \Eq{special} again in the theory $T_v$ described in the next section. Plugging \Eqq{fiap} into \Eq{vcircT0}, we get $v_{\rm circ}^2\simeq |b|Gm/r_*={\rm const}$ at scales $r=O(r_*)$. Therefore, the theory might be able to reproduce the plateaux observed in galactic rotation curves and the key ingredient is a potential which, for a point-wise source, is logarithmic in the radial distance. 

A flat rotation curve is only one of the observations that any model alternative to dark matter should explain. Another one is the empirical baryonic Tully--Fisher relation between the asymptotic circular velocity and the baryonic mass $M$ of the galaxy \cite{Tully:1977fu,McGaugh:2000sr,McGaugh:2011ac},
\be\label{tufi}
v_{\rm circ}^4(\infty)\propto M\,,
\ee
where the proportionality constant is universal for all the galaxies. Taking again our over-simplified example with $v_{\rm circ}^2\propto M/r_*$, we see that \Eqq{tufi} holds if
\be\label{strangerel}
r_*\propto \sqrt{M}\,.
\ee
We will come across this strange relation again and comment it in due course.


\section{Theory \texorpdfstring{$T_v$}{Tv} with weighted derivatives}\label{secTv}

Unlike the previous model, the derivatives in the theory $T_v$ depend on the measure weight.


\subsection{\texorpdfstring{$T_v$}{Tv}: equations of motion}

Let ${}_\b\cD_\mu$ be a weighted derivative of order $\b$, i.e., an ordinary derivative with weight factors inserted to the left and to the right:
\be\label{bD}
{}_\b\cD_\mu:=\frac{1}{v^\b}\p_\mu(v^\b\,\cdot\,)\,.
\ee
On rank-0 and rank-1 tensors, $\b=1/2$ \cite{frc11,frc13}:
\be
\cD_\mu:={}_\frac{1}{2}\cD_\mu=\frac{1}{\sqrt{v}}\p_\mu(\sqrt{v}\,\cdot\,)\,,
\ee
while on rank-2 tensors \cite{frc11}
\be
\check{\cD}_\mu:={}_\b\cD_\mu\,,\qquad \b=\frac{2}{D-2}\,.
\ee
In $D=4$ dimensions, $\check{\cD}_\mu={}_1\cD_\mu$. The Levi-Civita connection and the curvature tensors are defined with this differential structure:
\ba
{}^v\G^\rho_{\mu\nu} &:=& \tfrac12 g^{\rho\s}\left(\check{\cD}_{\mu} g_{\nu\s}+\check{\cD}_{\nu} g_{\mu\s}-\check{\cD}_\s g_{\mu\nu}\right)\,,\\
\cR^\rho_{~\mu\s\nu} &:=& \p_\s {}^v\G^\rho_{\mu\nu}-\p_\nu {}^v\G^\rho_{\mu\s}+{}^v\G^\tau_{\mu\nu}{}^v\G^\rho_{\s\tau}-{}^v\G^\tau_{\mu\s}{}^v\G^\rho_{\nu\tau}\,,\\
\cR_{\mu\nu} &:=& \cR^\rho_{~\mu\rho\nu}\,,\qquad \cR:= g^{\mu\nu}\cR_{\mu\nu}\,.
\ea
The action in the physical frame is \cite{frc11}
\ba
S &=&\frac{1}{2\k^2}\int\rmd^Dx\,v\,\sqrt{|g|}[\cR-w(v)\cD_\mu v\cD^\mu v-2U(v)]+S_{\rm m}\nonumber\\
&=&\frac{1}{2\k^2}\int\rmd^Dx\,\,\sqrt{|g|}\rme^{\frac{D-2}{2}\vp}\left[R-\Om'\p_\mu\vp\p^\mu\vp-2U\right]+S_{\rm m}\,,\label{eha}
\ea
where $U$ is a function of the weight $v$ and
\ba
\vp &:=&\frac{2}{D-2}\ln v\,,\\
\Om'&:=& \frac{(\Om-D+1)(D-2)}{4}\,,\qquad \Om:=\frac{9(D-2)w}{4}\rme^{(D-2)\vp}=\frac{9(D-2)w}{4}v^2\,.
\ea
In the second line of \Eqq{eha}, we re-expressed the action in terms of the ordinary Ricci scalar and the scalar-looking quantity $\vp$. One can map the theory to the so-called integer frame
\be\label{bargg}
\bar g_{\mu\nu}=\rme^{\vp} g_{\mu\nu}\,,
\ee
where the curvature decouples from the measure weight. It is not difficult to show that
\ba
\B &=& \rme^{\vp}\left(\bar\B-\frac{D-2}{2}\bar\p_\mu\vp\bar\p^\mu\right),\\
\sqrt{|\bar g|} &=& \rme^{\frac{D}{2}\vp} \sqrt{|g|}\,,\\
\bar R_{\mu\nu}&=& R_{\mu\nu}-\frac12g_{\mu\nu}\B\vp+\frac{D-2}{4}\left(\p_\mu\vp\p_\nu\vp- g_{\mu\nu}\p_\s\vp\p^\s\vp-2\N_\mu\N_\nu\vp\right),\label{omn3}\\
\rme^{\vp}\bar R &=&R-(D-1)\B\vp-\frac{(D-1)(D-2)}{4}\p_\mu\vp\p^\mu\vp,\label{barRR}
\ea
so that the action in the $\bar g$ frame is\footnote{This expression corrects a typo in the published version of \cite[eq.\ (5.28)]{frc11}. The arXiv version is correct.}
\be
S = \frac{1}{2\k^2}\int\rmd^Dx\,\sqrt{|\bar g|}\,\left[\bar R-\frac{(D-2)\Om}{4}\bar\p_\mu\vp\bar\p^\mu\vp-2\rme^{-\vp}U\right],\label{sgn2}
\ee
where the bar on top of $\p_\mu$ is pleonastic. Although this looks like an ordinary conformal transformation from the Jordan to the Einstein frame, in fact it is not, since $\vp$ is not a scalar field but a fixed coordinate profile establishing that rulers and clocks measure scale-dependent events in the original frame without bars, sometimes called fractional frame. By definition, the fractional frame (corresponding to what we would call the Jordan frame in scalar-tensor models) is the physical frame and one should revert to it when comparing observables with data \cite{frc11}. The equations of motion from \Eq{sgn2} are
\be
\k^2\bar T_{\mu\nu}=\bar G_{\mu\nu}+\rme^{-\vp}U\bar g_{\mu\nu}-\frac{(D-2)\Om}{4}\left(\bar\p_\mu\vp\bar\p_\nu\vp+\frac12\bar g_{\mu\nu}\bar\p_\s\vp\bar\p^\s\vp\right),\label{ee3}
\ee
where
\be\label{tmunu3}
\bar T_{\mu\nu}:=-\frac{2}{\sqrt{|\bar g|}}\frac{\de S_{\rm m}}{\de \bar g^{\mu\nu}}=\rme^{-\frac{D-2}{2}\vp}\,v\,T_{\mu\nu}^v=T_{\mu\nu}^v\,.
\ee
Taking the trace of \Eq{ee3},
\be
\k^2\bar T=-\frac{D-2}{2}\bar R+\rme^{-\vp}DU-\frac{(D-2)(D+2)\Om}{8}\bar\p_\s\vp\bar\p^\s\vp\,,\label{ee3tr}
\ee
we rewrite \Eq{ee3} as
\ba
\k^2\bar S_{\mu\nu} &=&\bar R_{\mu\nu}+\frac{\Om}{2}\left(\bar g_{\mu\nu}\bar\p_\s\vp\bar\p^\s\vp-\frac{D-2}{2}\bar\p_\mu\vp\bar\p_\nu\vp\right),\label{ee3fin}\\
\bar S_{\mu\nu} &:=& \widetilde{\bar T}_{\mu\nu}-\frac{1}{D-2}\bar g_{\mu\nu}\widetilde{\bar T}\,,\qquad\widetilde{\bar T}_{\mu\nu}:= \bar T_{\mu\nu}-\rme^{-\vp}U\bar g_{\mu\nu}\,.
\ea
Using \Eq{omn3}, we get
\ba
\k^2 S_{\mu\nu}^v &=& R_{\mu\nu}+\frac{(D-2)(1-\Om)}{4}\p_\mu\vp\p_\nu\vp-\frac{D-2-2\Om}{4}g_{\mu\nu}\p_\s\vp\p^\s\vp\nonumber\\
&&-\frac12g_{\mu\nu}\B\vp-\frac{D-2}{2}\N_\mu\N_\nu\vp\,,\label{ee4fin}
\ea
where $S_{\mu\nu}^v$ is defined in \Eq{Smunu}. As a check, we can repeat the same calculation in the fractional frame. Since from \Eq{omn3} and \Eq{barRR}
\be
\bar G_{\mu\nu} = G_{\mu\nu}+\frac{D-2}{2} \left(\frac12\p_\mu\vp\p_\nu\vp+\frac{D-3}{4} g_{\mu\nu}\p_\s\vp\p^\s\vp+g_{\mu\nu}\B\vp-\N_\mu\N_\nu\vp\right),
\ee
in the physical frame the modified Einstein equations \Eq{ee3} become
\ba
\k^2 T_{\mu\nu}^v &=& G_{\mu\nu}+U\, g_{\mu\nu}+\frac{(D-2)(1-\Om)}{4}\p_\mu\vp\p_\nu\vp
+\frac{(D-2)(D-3-\Om)}{8}g_{\mu\nu}\p_\s\vp\p^\s\vp\nonumber\\
&&+\frac{D-2}{2}\left(g_{\mu\nu}\B\vp-\N_\mu\N_\nu\vp\right).\label{ee4}
\ea
The trace equation is
\ba
\k^2\,T^v&=&-\frac{D-2}{2}R+DU+\frac{(D-2)[2(1-\Om)+D(D-3-\Om)]}{8}\p_\s\vp\p^\s\vp\nonumber\\
&&+\frac{(D-2)(D-1)}{2}\B\vp\,,\label{ee4tr}
\ea
which, combined with \Eq{ee4}, gives \Eq{ee4fin}.


\subsection{\texorpdfstring{$T_v$}{Tv}: Poisson equation and Newton's potential}

To get the Poisson equation for the perturbation $h_{00}$, we can either calculate it directly from \Eq{ee4fin} or via the expression \Eq{ee3fin} in the integer frame. Equation \Eq{ee3} looks deceptively simple but one should bear in mind that perturbing around Minkowski spacetime in the fractional frame implies an integer-frame background which is only conformally equivalent to Minkowski. Consequently, the background operators in the perturbed equation for $\bar h_{00}$ should be further expanded to make $v$-terms explicit.
 We choose the easiest route and work directly in the fractional frame. 

\subsubsection{\texorpdfstring{$T_v$}{Tv}: Poisson equation in spherical coordinates}

Plugging \Eq{getah} into \Eq{ee4fin}, expanding and noting that, in spherical coordinates, 
\ba
\p_i\vp &=& \p_r\vp=\frac{2}{d-1}\frac{\p_r v}{v}=\frac{2d(\a-1)\cA(r)}{(d-1)r}\,,\\
\N^2\vp &=& \frac{2d(\a-1)}{d-1}\frac{[d\a-2-d(\a-1)\cA(r)]\cA(r)}{r^2}\,,
\ea
we get
\ba
\hspace{-.7cm}\frac{d-2}{d-1}\k^2\rho&=& \left\{\p_r^2+\frac{d-1+d(\a-1)\cA(r)}{r}\p_r+\frac{1}{r^2}\N_{S^{d-1}}^2\right.\nonumber\\
\hspace{-.7cm}&&\quad\left.+\frac{2d(\a-1)[(d-1)(d\a-2)-2d\Om(\a-1)\cA(r)]\cA(r)}{(d-1)^2r^2}\right\}\Phi\nonumber\\
\hspace{-.7cm}&&\quad+\frac{[2d(d-1-\Om)(\a-1)\cA(r)-(d-1)(d\a-d-1)]d(\a-1)\cA(r)}{(d-1)^2r^2}h^{rr}\nonumber\\
\hspace{-.7cm}&&\quad+\frac{2U}{d-1}\,,\label{pois3r0}
\ea
where $\cA$ is given in \Eq{cAfun}. We can eliminate the $h^{rr}$ term either by choosing
\be\nonumber
\Om=\frac{(d-1)[d+1-d\a+2d(\a-1)\cA(r)]}{2d(\a-1)\cA(r)}
\ee
or, as we did in the theory $T_1$, by fixing the gauge so that $h^{rr}=-2\Phi$. The result is exactly the same in both procedures and the Poisson equation in spherical coordinates reads
\begin{empheq}[box=\fcolorbox{black}{white}]{align}
\quad\frac{d-2}{d-1}\k^2\rho=&\left\{\p_r^2+\frac{d-1+d(\a-1)\cA(r)}{r}\p_r+\frac{1}{r^2}\N_{S^{d-1}}^2\right.\nonumber\\
&\quad\left.+\frac{2d(\a-1)[2d\a-d-3-2d(\a-1)\cA(r)]\cA(r)}{(d-1)r^2}\right\}\Phi+\frac{2U}{d-1}\,.\quad\label{pois3r}
\end{empheq}

\subsubsection{\texorpdfstring{$T_v$}{Tv}: Newton's potential}\label{newv}

From now on, we set $U(v)=0$. This term is left arbitrary in the theory and it is usually fixed by self-consistency of theoretical solutions \cite{revmu,frc11}. In the present case, there is no theoretical constraint on it and we make it vanish in the lack of a better \emph{Ansatz} for which we could solve the Poisson equation. However, in section \ref{disc} we will argue that this choice may be responsible, partly or in full, for the bad small-radius fit of the model.

\begin{figure}
\bc
\includegraphics[width=10cm]{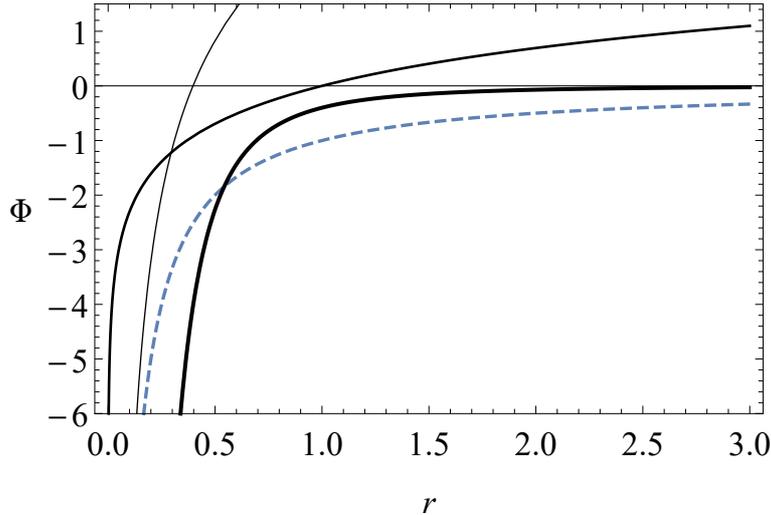}
\ec
\caption{\label{fig2} Gravitational potential in $d=3$ spatial dimensions for a point-wise source in the theory with weighted derivatives. Dashed curve: Einstein-theory limit \Eq{newt} (valid for $r\ll r_*$). Solid curves of increasing thickness: approximation \Eq{potne2} with small fractional correction with $\a=1.5$ (valid for $r\ll r_*$), logarithmic potential \Eq{Phisolvir23} (valid for $\a=2/3$ and $r\ll r_*$) and the extreme fractional limit \Eq{Phisolvir} with $\a=1.5$ (valid for $r\gg r_*$). Here $Gm=1$ and $r_*=1$.}
\end{figure}

\noindent\paragraph{Extreme fractional limit.} The asymptotic solution of the Poisson equation \Eq{pois3r} for a radial potential $\Phi(r)$ and a point-wise source $\rho=m\de^d(\bm{x})/v$ of mass $m$ can be found easily. In the Einstein-theory or Newtonian limit $\cA\to 0$, i.e., when $\a<1$ and $r\gg r_*$ or when $\a>1$ and $r\ll r_*$, one gets \Eq{eqom}, while in the fractional limit $\cA\to 1$, i.e., when $\a<1$ and $r\ll r_*$ or when $\a>1$ and $r\gg r_*$, we have
\ba
\frac{d-2}{d-1}\k^2m\frac{\de^d(\bm{x})}{v}\simeq\cC\Phi&:=&\left[\p_r^2+\frac{d\a-1}{r}\p_r+\frac{2d(\a-1)(d-3)}{(d-1)r^2}\right]\Phi\nonumber\\
&=&\left[\frac{1}{r^{d\a-1}}\p_r\left(r^{d\a-1}\p_r\,\cdot\right)+\frac{2d(\a-1)(d-3)}{(d-1)r^2}\right]\Phi\,.\label{poisrir}
\ea
Everywhere except at $r=0$, the solution is
\ba
\Phi(r) &=& C_+\,r^{-c_+}+C_-\,r^{-c_-}\,,\label{phi0}\\
c_\pm &=& \frac12\left[d\a-2\pm\sqrt{(d\a-2)^2-\frac{8d(\a-1)(d-3)}{d-1}}\right].
\ea
When $d=3$ or $\a=1$, $c_-=0$ and we can redefine the potential by a shift, thus setting $C_-=0$ from now on. The normalization constant $C_+$ can be found by looking at the solution at $r=0$. Using \Eq{dexder}, for a test function $f$
\ba
\int\! \rmd^d\bm{x}\,v\,(\cC\Phi)\,f\!\! \!&=&\! \lim_{\e\to 0}\frac{2\pi^\frac{d}{2}}{\G\left(\frac{d}{2}\right)}\!\int_\e^{+\infty}\!\!\rmd r\,r^{d-1}v\,(\cC\Phi)\, f\nonumber\\
\!\!\!&\stackrel{\textrm{\tiny\Eq{phi0}}}{=}&
\!\lim_{\e\to 0}\frac{2\pi^\frac{d}{2}}{\G\left(\frac{d}{2}\right)}\frac{1}{r_*^{d(\a-1)}}\!\!\int_\e^{+\infty}\!\!\rmd r\!\left[\p_r\left(r^{d\a-1}\p_r\Phi\right)+r^{d\a-3}\frac{2d(\a-1)(d-3)}{d-1}\Phi\right]f\nonumber\\
\!\!\!&=&\!\! \lim_{\e\to 0}\frac{2\pi^\frac{d}{2}}{\G\left(\frac{d}{2}\right)}\frac{1}{r_*^{d(\a-1)}}\!\!\int_\e^{+\infty}\!\!\rmd r\!\left[-\!\!\left(r^{d\a-1}\p_r\Phi\right)\p_r f+r^{d\a-3}\frac{2d(\a-1)(d-3)}{d-1}\Phi f\right]\nonumber\\
\!\!\!&=&\!\! \lim_{\e\to 0}\frac{2\pi^\frac{d}{2}}{\G\left(\frac{d}{2}\right)}\frac{C_+c_+}{r_*^{d(\a-1)}}\!\!\int_\e^{+\infty}\!\!\rmd r\left[r^{d\a-c_+-2}\p_r f+r^{d\a-3}\frac{2d(\a-1)(d-3)}{d-1}\Phi f\right].\nonumber\\
\ea
We can use this expression in the Einstein-theory or Newtonian limit, which is achieved by setting $\a=1$. Then, the last term vanishes, $c_+=d-2$ and one gets \Eq{Phisolv}. In the fractional limit, we are unable to proceed unless $d=3$, in which case $c_+=3\a-2$ and
\ben
\int \rmd^3\bm{x}\,v\,\cC\Phi\,f = -4\pi\frac{C_+(3\a-2)}{r_*^{3(\a-1)}}f(0)=4\pi Gm\,f(0)\,,
\een
yielding
\be\label{Phisolvir}
\Phi(r)\simeq-\frac{1}{3\a-2}\frac{Gm}{r_*}\left|\frac{r_*}{r}\right|^{3\a-2}.
\ee
The asymptotic limits \Eq{newt} and \Eq{Phisolvir} are shown in Fig.\ \ref{fig2}.

The special value \Eq{special} reappears in this theory in this regime. In the limit $\a\to 2/3$, the potential \Eq{Phisolvir} becomes (Fig.\ \ref{fig2})
\be\label{Phisolvir23}
\Phi(r)\simeq \frac{Gm}{r_*}\ln\frac{r}{r_*}\,,
\ee
up to a divergent constant $\propto (3\a-2)^{-1}$ that can be eliminated by a shift. This potential can be obtained directly from the Poisson equation \Eq{poisrir} setting $d=3$ and $\a=2/3$ therein. It coincides with the approximated potential \Eq{fiap} in the theory $T_1$ for $r\sim r_*$ and the same value of the fractional exponent $\a=2/3$.

The logarithmic behaviour implies that the rotation velocity 
\be\label{vcircT}
v_{\rm circ}(r)=\sqrt{\left|\frac{1}{\sqrt{v}}\frac{\rmd(\sqrt{v}\Phi)}{\rmd \ln r}\right|}
\ee
tends to a constant plus a logarithm, i.e., approximately a plateau:
\ba
v_{\rm circ}^2&=&\left|\frac{\rmd\Phi}{\rmd \ln r}+\frac{r\p_r v}{2v}\Phi\right| \stackrel{\textrm{\tiny \Eq{prvv}}}{=} \left|\frac{\rmd\Phi}{\rmd \ln r}+\frac{3(\a-1)}{2}\cA(r)\Phi\right|\label{vcircgen}\\
&\simeq& \left|\frac{\rmd\Phi}{\rmd \ln r}-\frac{1}{2}\Phi\right|\simeq\frac{Gm}{r_*}\left|1-\frac{1}{2}\ln\frac{r}{r_*}\right|.\label{v2circ}
\ea
Since values in the half-line $\a<1$ correspond to an extreme fractional limit valid at $r\ll r_*$, then, if $r_*$ is much larger than the galaxy scale and $\a$ is close to $2/3$, one can obtain a self-consistent model in the extreme fractional regime \Eq{Phisolvir} but not an almost flat rotation curve, since the logarithm in \Eq{v2circ} dominates. This negative conclusion, essentially due to the $\p_r v/v$ term in $v_{\rm circ}$ (absent in the theory $T_1$), will be confirmed later.

\noindent\paragraph{Small fractional corrections.} An intermediate form of the Newtonian potential when the fractional correction is small can be obtained by expanding the Poisson equation \Eq{pois3r} for small $\cA\simeq |r/r_*|^{d(\a-1)}$ with a delta source for a radial solution:
\ba
\hspace{-.5cm}\frac{d-2}{d-1}\k^2m\,\de^d(\bm{x})\left[1-\left|\frac{r}{r_*}\right|^{d(\a-1)}\right]&\simeq&\left(\p_r^2+\frac{d-1}{r}\p_r\right)\Phi(r)\nonumber\\
\hspace{-.5cm}&&+\frac{d(\a-1)}{r}\left|\frac{r}{r_*}\right|^{d(\a-1)}\!\left[\p_r+\frac{2(2d\a-d-3)}{(d-1)r}\right]\!\Phi(r).\nonumber
\ea
Splitting the potential $\Phi=\Phi_0+\Phi_1$ into a zero-order part $\Phi_0$ solving the ordinary Poisson equation \Eq{eqom} and a first-order part $\Phi_1$ and dropping higher-order terms, one has to solve
\ba
\hspace{-1cm}-\frac{d-2}{d-1}\k^2m\,\de^d(\bm{x})\left|\frac{r}{r_*}\right|^{d(\a-1)} &=& \left(\p_r^2+\frac{d-1}{r}\p_r\right)\Phi_1(r)\nonumber\\
&&+\frac{d(\a-1)}{r}\left|\frac{r}{r_*}\right|^{d(\a-1)}\!\left[\p_r+\frac{2(2d\a-d-3)}{(d-1)r}\right]\!\Phi_0(r)\nonumber\\
&=&\left(\p_r^2+\frac{d-1}{r}\p_r\right)\Phi_1(r)\nonumber\\
&&+\frac{4Gm d(\a-1)\G\left(\frac{d}{2}\right)}{\pi^{\frac{d}{2}-1}}\left|\frac{r}{r_*}\right|^{d(\a-1)}\frac{d^2-d-4\a d+8}{(d-1)^2r^d},\label{poius}
\ea
where $\Phi_0$ is given by \Eq{Phisolv}. The solution is
\be
\Phi_1(r) = \Phi_0(r)\frac{8+d(d-1-4\a)}{(d-1)(2-2d+d\a)}\left|\frac{r}{r_*}\right|^{d(\a-1)},\nonumber
\ee
so that
\be\label{potne}
\Phi(r)\simeq\Phi_0(r)\left[1-\frac{8+d(d-1-4\a)}{(d-1)(2d-2-d\a)}\left|\frac{r}{r_*}\right|^{d(\a-1)}\right].
\ee
In $d=3$ spatial dimensions (Fig.\ \ref{fig1}),
\be\label{potne2}
\Phi(r)\simeq\Phi_0(r)\left[1-\frac{7-6\a}{4-3\a}\left|\frac{r}{r_*}\right|^{3(\a-1)}\right].
\ee
When $\a\to 4/3$, the potential becomes
\be\label{filog}
\Phi(r)\simeq-\frac{Gm}{r}+\frac{Gm}{r_*}\,\ln\frac{r}{r_*}\,,
\ee
up to a shiftable constant. This potential can be obtained directly from \Eqq{poius} setting $d=3$ and $\a=4/3$. Curiously, the second term in \Eq{filog} coincides with the potential \Eq{Phisolvir23} obtained in the extreme fractional regime for a different value of the fractional exponent, $\a=2/3$. It also coincides with the approximated potential \Eq{fiap} in the theory $T_1$ for $\a=2/3$ and $r\sim r_*$.

Therefore, since for $\a>1$ the fractional correction is small when $r\ll r_*$, the rotation velocity \Eq{vcircT} tends to a constant when $\a$ is close to the special value
\be\label{special2}
\a=\frac43\,.
\ee
This could provide a self-consistent explanation of galaxy rotation curves if $r_*$ was much larger than the typical galactic size. Using \Eqq{vcircgen},
\be
v_{\rm circ}^2\simeq\left|\frac{\rmd\Phi}{\rmd \ln r}+\frac{r}{2r_*}\Phi\right|\simeq\frac{Gm}{r_*}\left|\frac{r_*}{r}+\frac{1}{2}+\frac{r}{2r_*}\ln\frac{r}{r_*}\right|.\label{v2circ2}
\ee
The first term in the circular velocity dominate at short scales. The second term dominates at large scales and may correspond to the plateau we are looking for. The third term should be negligible for $r\lesssim r_*$. Once again, formul\ae\ such as \Eq{v2circ2} are meant as a preview of the possible outcome one can get from a full analysis of galactic curves (section \ref{Tvrotcu2}).


\subsection{\texorpdfstring{$T_v$}{Tv}: thin-disk exponential-density rotation curve}\label{Tvrotcu}

Let \Eq{PhIG} be the full solution of the Poisson equation with point-wise source with mass $m$. The multi-fractional generalization of \Eqq{potetgr} is
\be\label{potet}
\Phi^{\rm galaxy}(R)=\int_\cV\rmd^3\bm{x'}\,v(\bm{x'})\,\rho(\bm{x'})\,\cG(\bm{x}-\bm{x'})\,.
\ee
The extreme asymptotic limits of $m\cG$ are given by \Eqqs{newt} and \Eq{Phisolvir} or \Eq{Phisolvir23}, while \Eq{potne2} and \Eq{filog} include a correction to \Eq{newt} in the limit of small fractional correction $|r/r_*|^{3(\a-1)}\ll 1$. The gravitational field in the theory $T_v$ in cylindrical coordinates is
\be\label{gifi}
g(R) = -\cD_R \Phi^{\rm galaxy}(R) = -\frac{1}{\sqrt{v(R)}}\frac{\rmd}{\rmd R}\left[\sqrt{v(R)}\,\Phi^{\rm galaxy}(R)\right],
\ee
while the circular velocity is (Appendix \ref{app0})
\be\label{vcircTv}
v_{\rm circ}(R) = \sqrt{\left|R\,g(R)\right|}\,.
\ee
We used the derivative $\cD_R$ in \Eqq{gifi} because the gravitational field is a rank-1 tensor \cite{frc11,frc13} (we omitted the unit vector in the expression).

\subsubsection{\texorpdfstring{$T_v$}{Tv}: extreme fractional limit}

In this approximation, when $\a\neq 2/3$ one can use the potential \Eq{Phisolvir}, so that \Eqq{potet} with $r_*=R_*$ becomes
\be\label{exfraphi}
\Phi^{\rm disk}(R) \simeq-\frac{1}{3\a-2}\frac{G}{R_*}\int_\cV\rmd^3\bm{x'}\,\left|\frac{R'}{R_*}\right|^{3(\a-1)}\rho(\bm{x'})\,\left(\frac{R_*}{|\bm{x}-\bm{x'}|}\right)^{3\a-2},
\ee
while when $\a=2/3$ one takes \Eq{Phisolvir23} and
\be\label{exfraphi2}
\Phi^{\rm disk}(R) \simeq \frac{G}{R_*}\int_\cV\rmd^3\bm{x'}\,\left|\frac{R'}{R_*}\right|^{-1}\rho(\bm{x'})\,\ln\frac{|\bm{x}-\bm{x'}|}{R_*}\,.
\ee
Both expressions are calculated in Appendix \ref{appC1} and they are given by \Eqqs{potphimulti} (approximated to \Eq{potphimulti2}) and \Eq{potphilog}. The gravitational field and circular velocity are found plugging these expressions into \Eqqs{gifi} and \Eq{vcircTv}, respectively. In particular,
\be\label{gifi2}
g(R)=-\frac{\rmd\Phi^{\rm galaxy}(R)}{\rmd R}-\frac{3(\a-1)}{2R}\Phi^{\rm galaxy}(R)\,.
\ee

\begin{figure}
\bc
\includegraphics[width=10cm]{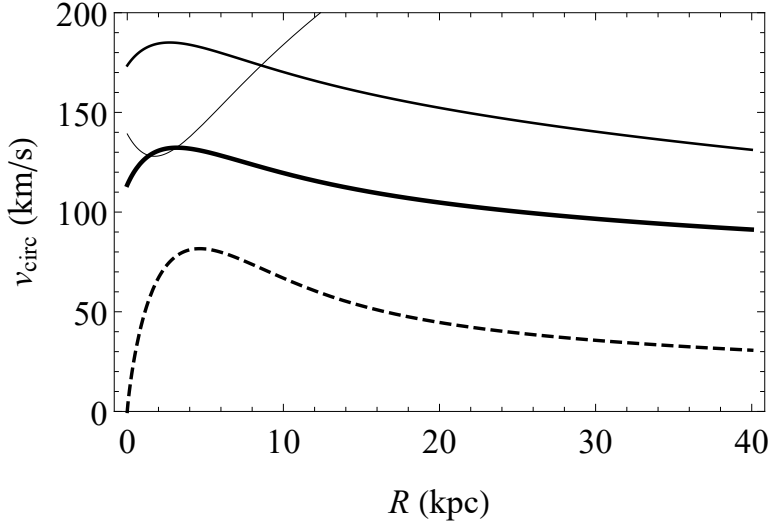}
\ec
\caption{\label{fig3} Rotation curve of baryonic matter in the thin-disk approximation and an exponential matter density profile in Einstein's theory (\Eqq{vcirc0gr}, dashed curve) and in the multi-fractional theory $T_v$ with weighted derivatives in the extreme fractional limit (\Eq{vcircTv}, \Eq{exfraphi} or \Eq{exfraphi2}, and \Eqqs{gifi2}, where $\a<1$ and $R\ll R_*$) for $\a=0.4$, $\a=2/3$ and $\a=0.8$ (solid curves of increasing thickness). Here $M=1.72\times 10^{40}\,{\rm kg}$, $R_0=2.16\,{\rm kpc}$ and $R_*=40\,{\rm kpc}$.}
\end{figure}

To explore in a preliminary way how the rotation curve is modified in this theory, we compare the thin-disk analytic approximation in the extreme fractional limit with the Einstein-theory result \Eq{vcirc0gr} (Fig.\ \ref{fig3}) and, for reference, the values of $M$ and $R_0$ of NGC6503.
These results are not conclusive because they are not based on actual matter density data but they illustrate the possibilities of the model. As one can see in Fig.\ \ref{fig3}, depending on the value of $\a$ one can enhance the tail or lift and flatten the whole curve, although without ever forming a plateau at the peak height, even when $\a=2/3$. 

\subsubsection{\texorpdfstring{$T_v$}{Tv}: small fractional corrections}

Expanding \Eq{potet}, for $\a\neq 4/3$ we have
\bs\label{diskgas2}\ba
\Phi^{\rm disk}(R) &\simeq& \Phi^{\rm disk}_0(R)+\de\Phi^{\rm disk}_1(R)+\de\Phi^{\rm disk}_2(R)\,,\label{phitotTv}\\
\Phi^{\rm disk}_0(R) &=& -G\int_\cV\rmd^3\bm{x'}\,\frac{\rho(\bm{x'})}{|\bm{x}-\bm{x'}|}\,,\label{Tvphi0}\\
\de\Phi^{\rm disk}_1(R) &=& -G\int_\cV\rmd^3\bm{x'}\,\left|\frac{R'}{R_*}\right|^{3(\a-1)}\frac{\rho(\bm{x'})}{|\bm{x}-\bm{x'}|}\,,\label{dephi1}\\
\de\Phi^{\rm disk}_2(R) &=&-G\frac{7-6\a}{(3\a-4)r_*^{3(\a-1)}}\int_\cV\rmd^3\bm{x'}\,\frac{\rho(\bm{x'})}{|\bm{x}-\bm{x'}|^{4-3\a}}\,.\label{dephi2}
\ea\es
In cylindrical coordinates, the integration volume \Eq{cylinv} is decorated with a measure $v(R')$. The measure weight $v(R')$ is assumed to have cylindrical symmetry because we want to test how the rotation curve of baryonic matter deviates from Einstein's theory in a region of space with multi-fractional measure of galactic size $\sim R_0\ll R_*$. Within this region, one considers infinitesimal masses $\rmd m$ each generating a radial potential which is in turn multi-fractional. 

In the thin-disk approximation, the lowest-order term $\Phi_0$ is the standard Newtonian potential \Eq{PhiGR} calculated in section \ref{dmrev3}. The terms \Eq{dephi1} and \Eq{dephi2} are calculated in Appendix \ref{appC2} using some formul\ae\ of Appendices \ref{appA} and \ref{appB}.

The gravitational field in the theory $T_v$ is \Eq{gifi}:
\ba
\hspace{-1cm} g^{\rm disk}(R) &\simeq& -\frac{\rmd}{\rmd R}\left[\Phi^{\rm disk}_0(R)+\de\Phi^{\rm disk}_1(R)+\de\Phi^{\rm disk}_2(R)\right]-\Phi_0(R) \frac{1}{\sqrt{v(R)}}\frac{\rmd}{\rmd R}\sqrt{v(R)}\nonumber\\
&\simeq& -\frac{\rmd}{\rmd R}\left[\Phi^{\rm disk}_0(R)+\de\Phi^{\rm disk}_1(R)+\de\Phi^{\rm disk}_2(R)\right]-\frac{3(\a-1)}{2R}\left|\frac{R}{R_*}\right|^{3(\a-1)}\Phi^{\rm disk}_0(R),\label{gthin}
\ea
while the circular velocity to leading order in the fractional limit is \Eq{vcircTv} with $g$ given by \Eq{gthin}, which can be further expanded for small corrections. This model is self-consistent at all galactic scales only if $R_0\ll R_*$ and $\a>1$.

\begin{figure}
\bc
\includegraphics[width=10cm]{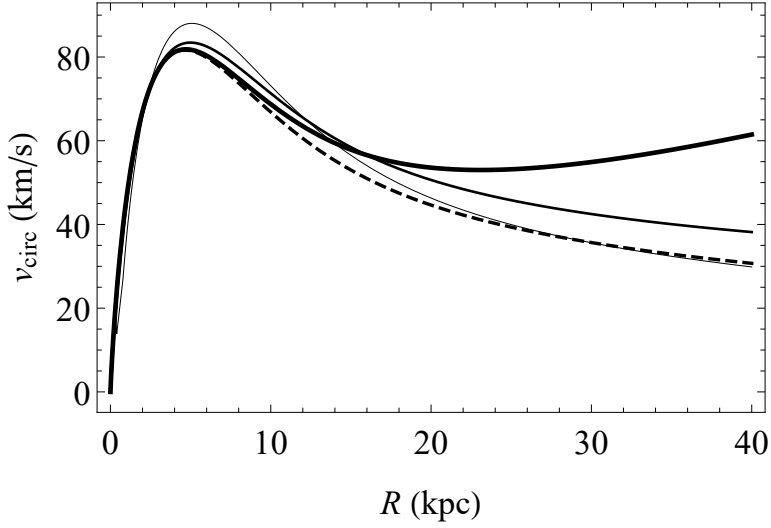}
\ec
\caption{\label{fig4} Rotation curve of baryonic matter in the thin-disk approximation and an exponential matter density profile in Einstein's theory (\Eqq{vcirc0gr}, dashed curve) and in the multi-fractional theory $T_v$ with weighted derivatives in the limit of small corrections (\Eqqs{gthin} and \Eq{vcircTv}, where $\a>1$ and $R\ll R_*$) for $\a=1.1$, $\a=4/3$ and $\a=1.9$ (solid curves of increasing thickness). Here $M=1.72\times 10^{40}\,{\rm kg}$, $R_0=2.16\,{\rm kpc}$ and $R_*=40\,{\rm kpc}$.}
\end{figure}

Figure \ref{fig4} compares the velocity \Eq{vcircTv} from the gravitational field \Eq{gthin} in the thin-disk analytic approximation with small fractional corrections with the Einstein-theory result \Eq{vcirc0gr}, using again the values of $M$ and $R_0$ of NGC6503.

Depending on the value of $\a$, one can enhance the peak or lift the damping tail, although without ever forming a plateau at the peak height, even when $\a=4/3$. 


\subsection{\texorpdfstring{$T_v$}{Tv}: data-based rotation curve}\label{Tvrotcu2}

In this sub-section, we finally compare the predictions of the theory $T_v$ with SPARC data, assuming that the matter component is purely baryonic.

\subsubsection{\texorpdfstring{$T_v$}{Tv}: extreme fractional limit}

The profiles $\Phi^{\rm disk+gas}(R)$ and $\Phi^{\rm bulge}(R)$ and the corresponding gravitational fields $g^{\rm disk+gas}$ and $g^{\rm bulge}$ for a thick disk and a generic matter density profile are calculated in Appendix \ref{appD1}: \Eqqs{diskgas1}, \Eq{phibulgeTv}, \Eq{gdiskgas1} and \Eq{gbulgeTv} for $\a\neq 2/3$, and \Eqqs{phidisk23}, \Eq{phibulge23}, \Eq{gdisk23} and \Eq{gbulge23} for $\a=2/3$. The circular velocity in cylindrical coordinates is
\be
v_{\rm circ}(R) = \sqrt{\left|R\,g(R)\right|}\,,\qquad g(R)= g^{\rm disk+gas}(R)+g^{\rm bulge}(R)\,.
\ee
We checked that the $n$-expansion of \Eq{vcircTv} in Gegenbauer polynomials converges quickly for various values of $\a$. For instance, when $\a=1$ one recovers the Einstein-theory curve already at $n=5$, with a good approximation already at $n=3$.\footnote{The series for $\a=1$ actually has a better numerical convergence than the formul\ae\ of section \ref{dmrev4} and, in fact, we used the $n_{\rm max}=5$ expression to plot Fig.\ \ref{fig1}.}

The model in the extreme fractional regime does not fit the data of any of the three galaxies (Appendix \ref{appD1c}).

\subsubsection{\texorpdfstring{$T_v$}{Tv}: small fractional corrections}

The total potential is the sum of terms as in \Eqq{phitotTv}, extended to a gas and a bulge component:
\ba
\Phi^{\rm galaxy}(R) &=& \left[\Phi^{\rm disk+gas}_0(R)+\Phi^{\rm bulge}_0(R)\right]+\left[\de\Phi^{\rm disk+gas}_1(R)+\de\Phi^{\rm bluge}_1(R)\right]\nonumber\\
&&+\left[\de\Phi^{\rm disk+gas}_2(R)+\de\Phi^{\rm bulge}_2(R)\right].\label{phitotTv2}
\ea
All its components for a thick disk and a generic matter density profile are calculated in Appendix \ref{appD2}: \Eqqs{potweak1}, \Eq{potweak2}, \Eq{potweak3}, \Eq{potweak4}, \Eq{potweak5} and \Eq{potweak6} for $\a\neq 4/3$ and \Eqqs{potweak1}, \Eq{potweak2}, \Eq{potweak9}, \Eq{potweak10}, \Eq{potweak11} and \Eq{potweak12} for $\a=4/3$.

\begin{figure}
\bc
\includegraphics[width=10cm]{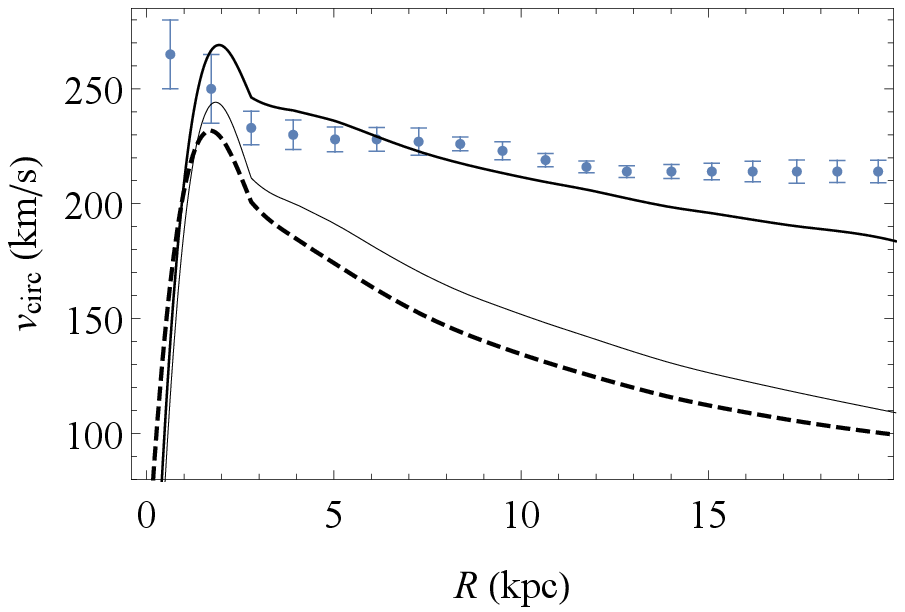}\\
\includegraphics[width=10cm]{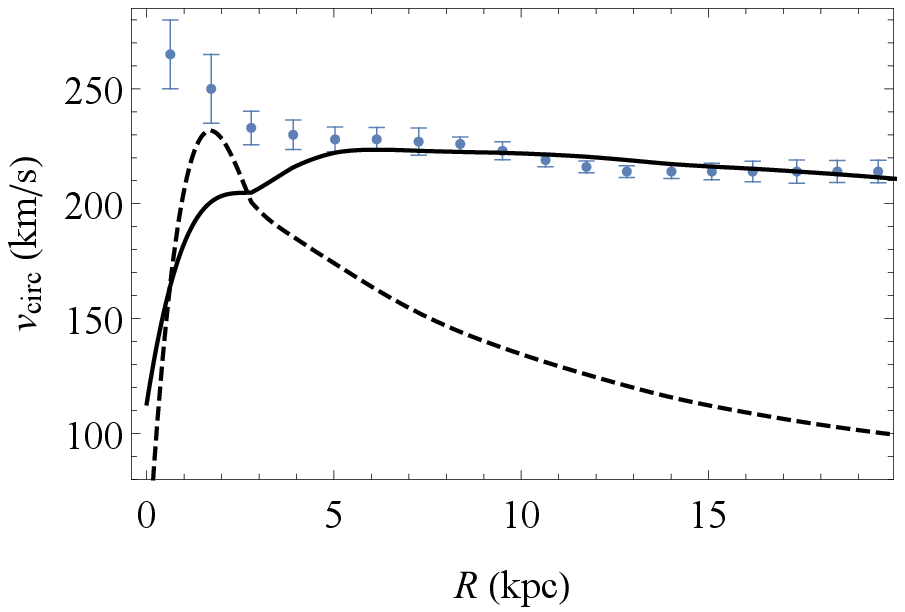}
\ec
\caption{\label{fig5} Comparison of the data points of NGC7814 with the $n=3$ rotation curve of the theory $T_v$ with weighted derivatives in the limit of weak fractional corrections with a data-based matter density profile for $\a=1$ (Einstein's theory, dashed curve), $\a=1.1,1.3$ and $R_*=4\,{\rm kpc}$ (top panel, increasing thickness) or $\a=4/3$ and $R_*=5.3\,{\rm kpc}$ (bottom panel).}
\end{figure}

\begin{figure}
\bc
\includegraphics[width=10cm]{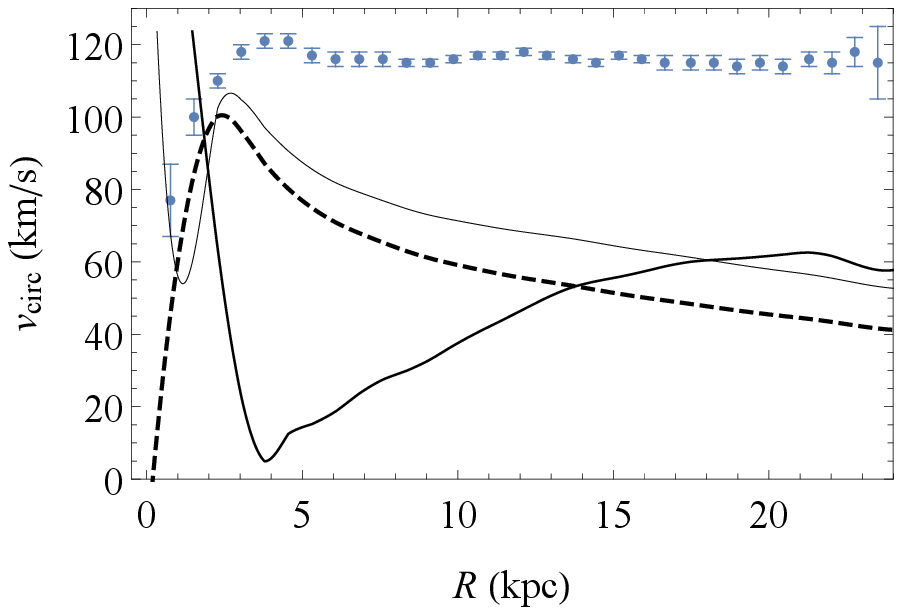}\\
\includegraphics[width=10cm]{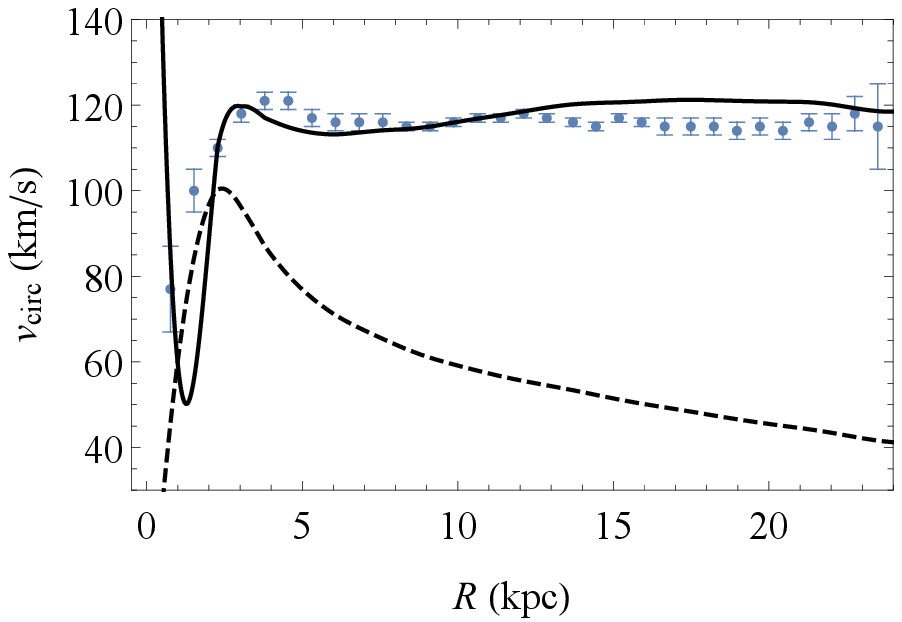}\\
\ec
\caption{\label{fig6} Comparison of the data points of NGC6503 with the $n=3$ rotation curve of the theory $T_v$ with weighted derivatives in the limit of weak fractional corrections with a data-based matter density profile for $\a=1$ (Einstein's theory, dashed curve), $\a=1.1,1.6$ and $R_*=5\,{\rm kpc}$ (top panel, increasing thickness) or $\a=4/3$ and $R_*=3.5\,{\rm kpc}$ (bottom panel).}
\end{figure}

\begin{figure}
\bc
\includegraphics[width=10cm]{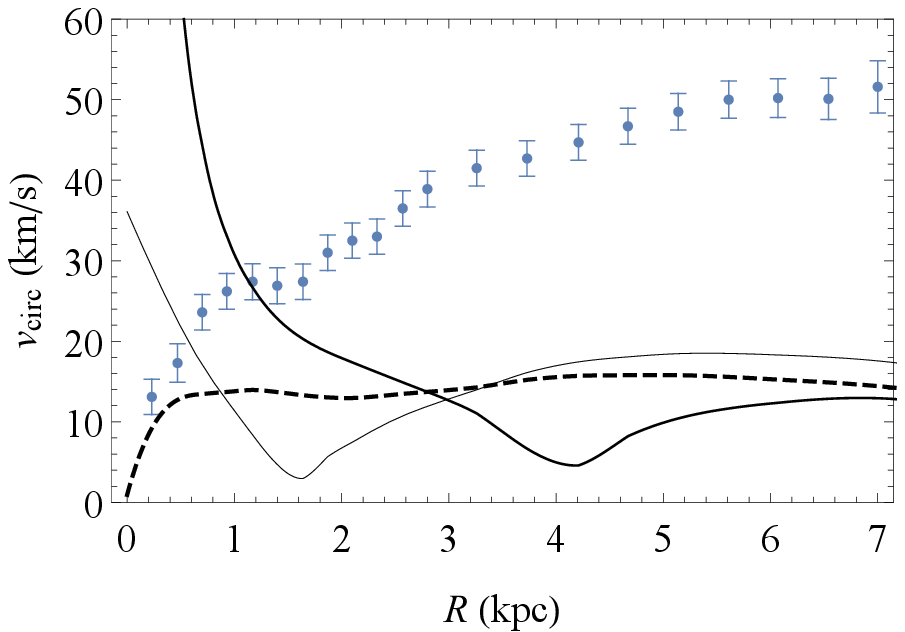}\\
\includegraphics[width=10cm]{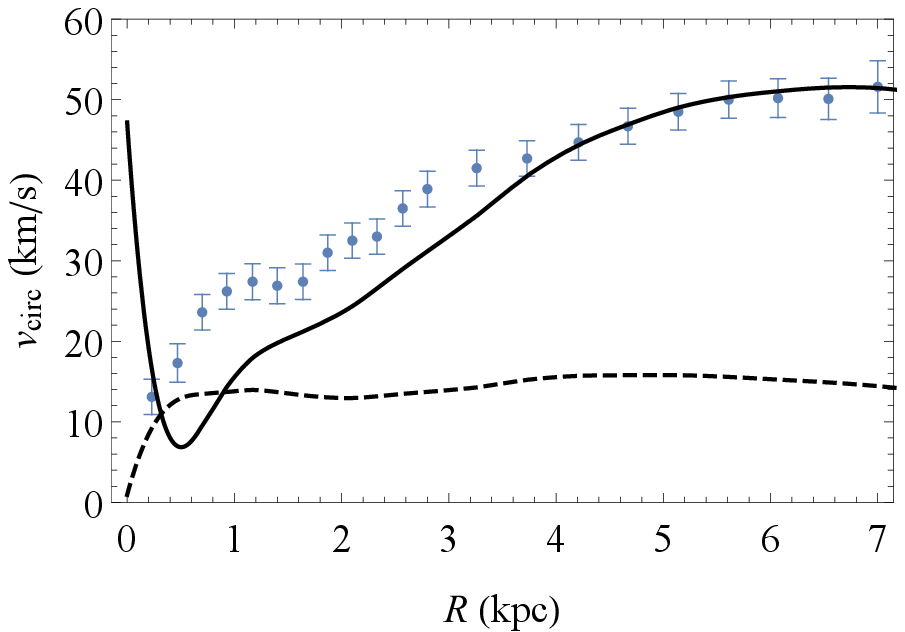}\\
\ec
\caption{\label{fig7} Comparison of the data points of NGC3741 with the $n=3$ rotation curve of the theory $T_v$ with weighted derivatives in the limit of weak fractional corrections with a data-based matter density profile for $\a=1$ (Einstein's theory, dashed curve), $\a=1.1,1.6$  and $R_*=5\,{\rm kpc}$ (top panel, increasing thickness) or $\a=4/3$ and $R_*=0.7\,{\rm kpc}$ (bottom panel).}
\end{figure}

The gravitational field is the data-based version of \Eqq{gthin}:
\bs\label{ggalweak}\ba
g^{\rm galaxy}(R) &=& \left[g^{\rm disk+gas}_0(R)+g^{\rm bulge}_0(R)\right]+\left[\de g_1^{\rm disk+gas}(R)+\de g_1^{\rm bulge}(R)\right]\nonumber\\
&&+\left[\de g_2^{\rm disk+gas}(R)+\de g_2^{\rm bulge}(R)\right],\label{ggalweak0}\\
g^{\rm disk+gas}_0(R) &=& -\frac{\rmd\Phi^{\rm disk+gas}_0(R)}{\rmd R}-\frac{3(\a-1)}{2R}\left|\frac{R}{R_*}\right|^{3(\a-1)}\Phi^{\rm disk+gas}_0(R)\,,\label{ggalweak1}\\
g^{\rm bulge}_0(R) &=& -\left.\frac{\rmd\Phi^{\rm bulge}_0(r)}{\rmd r}\right|_{r=R}-\frac{3(\a-1)}{2R}\left|\frac{R}{R_*}\right|^{3(\a-1)}\Phi^{\rm bulge}_0(R)\,,\label{ggalweak2}\\
\de g_1^{\rm disk+gas}(R)&=& -\frac{\rmd\de\Phi^{\rm disk+gas}_1(R)}{\rmd R}\,,\label{ggalweak3}\\   
\de g_1^{\rm bulge}(R)&=& -\left.\frac{\rmd\de\Phi^{\rm bulge}_1(r)}{\rmd r}\right|_{r=R}\,,\label{ggalweak4}\\  
\de g_2^{\rm disk+gas}(R)&=& -\frac{\rmd\de\Phi^{\rm disk+gas}_2(R)}{\rmd R}\,,\label{ggalweak5}\\
\de g_2^{\rm bulge}(R) &=&-\left.\frac{\rmd\de\Phi^{\rm bulge}_2(r)}{\rmd r}\right|_{r=R}\,,\label{ggalweak6}
\ea\es
and is reported in \Eqqs{gweak1}, \Eq{gweak2}, \Eq{gweak3}, \Eq{gweak4}, \Eq{gweak5} and \Eq{gweak6} for $\a\neq 4/3$, and \Eqqs{gweak7}, \Eq{gweak8}, \Eq{gweak9}, \Eq{gweak10}, \Eq{gweak11} and \Eq{gweak12} for $\a= 4/3$. The rotation curves of the three galaxies for some values of $\a$ and $R_*$ are shown in Figs.\ \ref{fig5}--\ref{fig7}. We took up to three modes in the calculation because, in this regime and for all values of $\a$ and $R_*$, the zero mode receives sizable corrections only from the $n=1,2,3$ contributions, while higher-order modes are negligible. As one can see, although the rotation curve is lifted at some points with respect to the one in Einstein's theory, the model is unable to fit the data when $\a\neq 4/3$.

When $\a= 4/3$, the model matches the large-$R$ data of the three galaxies, while it fails to fit small-$R$ data, showing that this model needs further improvement, in order to fully fit the galactic rotation data. However, it should be noted that, using the values for the $R_{*}$ parameter from the bottom panels in Figs.\ \ref{fig5}--\ref{fig7} combined with the respective galactic masses, one obtains a MOND-like acceleration parameter 
\be\label{accpam}
a_{0}\simeq \frac{GM}{R_*^2}\approx 0.72-2.1\times 10^{-10} {\rm m\,s}^{-2}\,,
\ee
comparable with the actual value of the MOND acceleration $a_{0}= 1.2\times 10^{-10} {\rm m\,s}^{-2}$ \cite{McGaugh:2016leg}. This connection is similar to the one used by Giusti \cite{Giusti:2020rul,Giusti:2020kcv} and Varieschi \cite{Varieschi:2020ioh,Varieschi:2020dnd,Varieschi:2020hvp} in their fractional models.

\begin{figure}
\bc
\includegraphics[width=9cm]{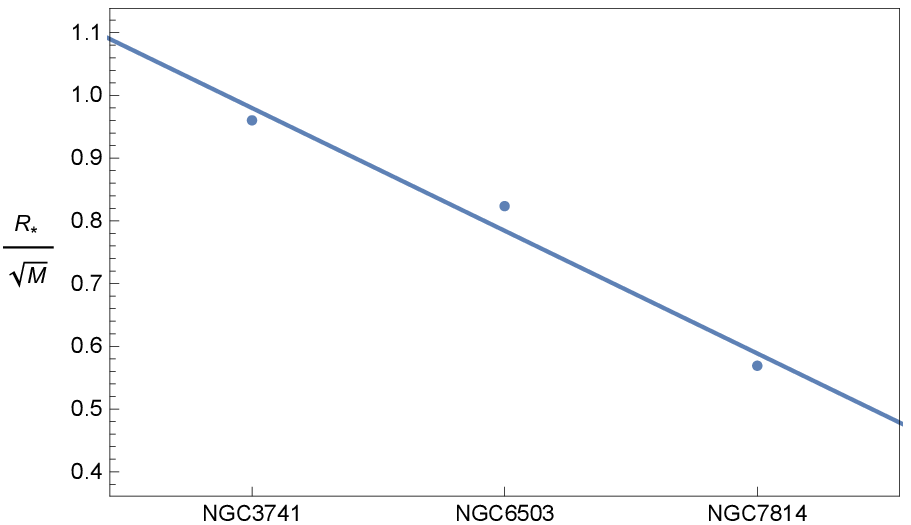}
\ec
\caption{\label{fig8} Ratio \Eq{strangerel2} (in units ${\rm m}\,{\rm kg}^{-1/2}$) for the galaxies NGC3741, NGC6503 and NGC7814 in the case $\a=4/3$.}
\end{figure}

To check whether the acceleration parameter \Eq{accpam} may indicate a universal behaviour of rotation curves, we consider once again the baryonic Tully--Fisher relation \Eq{tufi}. In section \ref{t1newt}, we found that the Tully--Fisher relation could hold in the theory $T_1$ provided the characteristic radius $r_*$ was proportional to the square root of the mass of the galaxy, \Eqq{strangerel}. For the theory $T_v$, we find the same intriguing relation. From \Eq{v2circ2}, we see that the asymptotic limit of the circular velocity obeys the Tully--Fisher relation if \Eq{strangerel} holds. In cylindrical coordinates,
\be\label{strangerel2}
\frac{R_*}{\sqrt{M}}={\rm const}\,.
\ee
Plotting this ratio for the best-fit values of $R_*$ found in the $\a=4/3$ cases, we see that we do not get a constant (Fig.~\ref{fig8}); the slope of the line is non-zero above $3\s$. However, this is not conclusive, as we will discuss in section \ref{disc}.


\section{Theory \texorpdfstring{$T_q$}{Tq} with \texorpdfstring{$q$}{q}-derivatives}\label{secTq}

Thanks to the structure of the $q$-derivatives, this theory is the simplest of the three multi-fractional scenarios with ordinary differential structure.


\subsection{\texorpdfstring{$T_q$}{Tq}: equations of motion}

All equations of Einstein's theory look the same in this theory, except that coordinates $x^\mu$ are replaced by the composite coordinates (no summation over $\mu$)
\be
q(x^\mu)=\int\rmd x^\mu\,v_\mu(x^\mu)\,.
\ee
In particular, all ordinary derivatives are replaced by the operators (no summation over $\mu$)
\be\label{qder}
\frac{\p}{\p q^\mu(x^\mu)}=\frac{1}{v_\mu(x^\mu)}\frac{\p}{\p x^\mu}\,.
\ee
Defining \cite{frc11}
\ba
{}^q\G^\rho_{\mu\nu} &:=& \tfrac12 g^{\rho\s}\left(\frac{1}{v_\mu}\p_{\mu} g_{\nu\s}+\frac{1}{v_\nu}\p_{\nu} g_{\mu\s}-\frac{1}{v_\s}\p_\s g_{\mu\nu}\right)\,,\label{leciq}\\
{}^q R^\rho_{~\mu\s\nu} &:=& \frac{1}{v_\s}\p_\s {}^q\G^\rho_{\mu\nu}-\frac{1}{v_\nu}\p_\nu {}^q\G^\rho_{\mu\s}+{}^q\G^\tau_{\mu\nu}\,{}^q\G^\rho_{\s\tau}-{}^q\G^\tau_{\mu\s}\,{}^q\G^\rho_{\nu\tau}\,,\label{riemq}\\
{}^q R_{\mu\nu}&:=& {}^q R^\rho_{~\mu\rho\nu}\,,\qquad {}^q R:= g^{\mu\nu}\, {}^q R_{\mu\nu}\,,\\
{}^q G_{\mu\nu}&:=&{}^q R_{\mu\nu}-\frac12g_{\mu\nu}{}^q R\,,
\ea
the action is
\be\label{Sgq}
S =\frac{1}{2\k^2}\int\rmd^Dx\,v\,\sqrt{|g|}\,{}^q R+S_{\rm m}\,,
\ee
while the modified Einstein equations are
\be\label{eeq}
\k^2\,T_{\mu\nu}={}^qG_{\mu\nu}\,,\qquad T_{\mu\nu}:=-\frac{2}{\sqrt{|g|}}\frac{\de S_{\rm m}}{\de g^{\mu\nu}}\,.
\ee
Taking the trace and plugging it back,
\be\label{eeqfin}
\k^2\,S_{\mu\nu}={}^qR_{\mu\nu}\,,\qquad S_{\mu\nu}:=T_{\mu\nu}-\frac{1}{D-2}g_{\mu\nu}T\,.
\ee


\subsection{\texorpdfstring{$T_q$}{Tq}: Poisson equation and Newton's potential}

The Poisson equation and gravitational potential of the theory $T_q$ are much simpler than in the other models.

\subsubsection{\texorpdfstring{$T_q$}{Tq}: Poisson equation in spherical coordinates}

Ignore log oscillations and consider the radial composite coordinate \Eq{qavg} and the measure weight \Eq{radialtildev}:
\be\label{defqr}
q(r)=r+\frac{r_*}{\a}\left|\frac{r}{r_*}\right|^\a,\qquad \tilde v(r)=\p_r q(r)=1+\left|\frac{r}{r_*}\right|^{\a-1}.
\ee
The Poisson equation is immediately obtained from its expression in Einstein's theory,
\begin{empheq}[box=\fcolorbox{black}{white}]{align}
\quad\frac{d-2}{d-1}\k^2\rho=&\left[\p_{q(r)}^2+\frac{d-1}{q(r)}\p_{q(r)}+\frac{1}{q^2(r)}\N_{S^{d-1}}\right]\Phi\label{pois2r0}\\
=&\frac{1}{\tilde v^2}\left\{\p_r^2+\left[(d-1)\frac{\tilde v}{q}-\frac{\p_r\tilde v}{\tilde v}\right]\p_r+\frac{\tilde v^2}{q^2}\N_{S^{d-1}}^2\right\}\Phi\,.\quad\label{pois2r}
\end{empheq}
When $\a<1$ and $r\ll r_*$, or when $\a>1$ and $r\gg r_*$, $q\simeq r \tilde v/\a$ and this expression is approximated to
\be
\quad\frac{d-2}{d-1}\k^2 \tilde v^2\rho\simeq\left[\p_r^2+\frac{(d-2)\a+1}{r}\p_r+\frac{\a^2}{r^2}\N_{S^{d-1}}^2\right]\Phi\,.
\ee

\subsubsection{\texorpdfstring{$T_q$}{Tq}: Newton's potential}

\begin{figure}
\bc
\includegraphics[width=10cm]{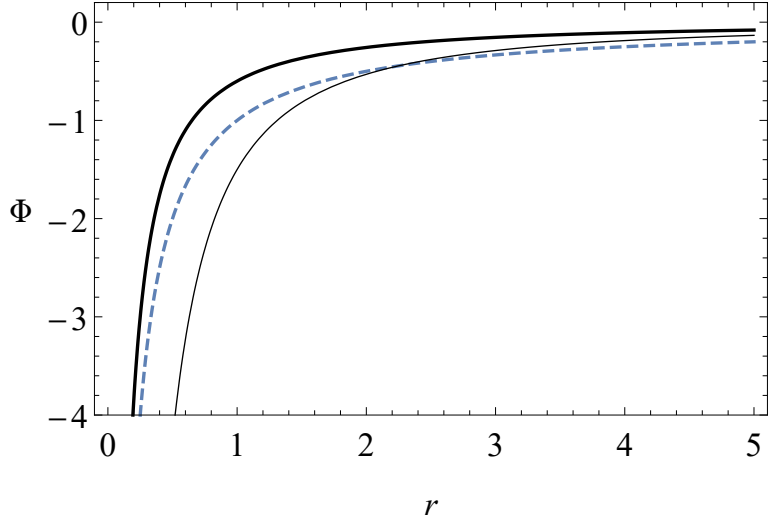}
\ec
\caption{\label{fig9} Gravitational potential in $d=3$ spatial dimensions for a point-wise source in the theory with $q$-derivatives with $\a>1$ ($\a=1.5$ in this example), in the Einstein-theory limit \Eq{newt} ($r\ll r_*$, dashed curve), in the fractional limit \Eq{phiir} ($r\gg r_*$, solid thin curve) and for the exact expression \Eq{exaq} (solid thick curve). Here $Gm=1$ and $r_*=1$.}
\end{figure}

The solution of the Poisson equation for a radial potential and a point-wise source of mass $m$ is more easily found from \Eq{pois2r0} rather than \Eq{pois2r}. Treating $q$ as a coordinate, the equation to solve is
\be
\left[\p_{q}^2+\frac{d-1}{q}\p_{q}\right]\Phi(\bm{x},\bm{x'})=\frac{d-2}{d-1}\k^2 m\,\de^d[q(\bm{x})-q(\bm{x'})]\,,
\ee
where $\de^d[q(\bm{x})-q(\bm{x'})]=\de[q(x^1)-q({x'}^1)]\cdots\de[q(x^d)-q({x'}^d)]$. The solution is found as before. Setting $\bm{x'}=0$,
\be\label{Phisolq}
\Phi(r)=-\frac{4\G\left(\frac{d}{2}\right)}{(d-1)\pi^{\frac{d}{2}-1}}\frac{Gm}{[\cQ(r)]^{d-2}}\simeq-\frac{4\G\left(\frac{d}{2}\right)}{(d-1)\pi^{\frac{d}{2}-1}}\frac{Gm}{[q(r)]^{d-2}}\,.
\ee
In $D=d+1=4$ dimensions (Fig.\ \ref{fig9}),
\be\label{exaq}
\Phi(r)=-\frac{Gm}{q(r)}\,,
\ee
which reduces to the standard Newtonian potential \Eq{newt} in the Einstein-theory limit, while in the fractional limit we obtain
\be\label{phiir}
\Phi(r)\simeq-\a\frac{Gm}{r}\left(\frac{r_*}{r}\right)^{\a-1}.
\ee


\subsection{\texorpdfstring{$T_q$}{Tq}: thin-disk exponential-density rotation curve}\label{thinq}

To get an extended matter configuration, one can consider \Eq{Phisolq} for an infinitesimal mass $\rmd m=\rho\,\rmd\cV$ and then integrate over the $q$-volume $\cV$ of the source. Restoring $\bm{x'}\neq 0$ and changing integration variable $q(\bm{x'})\to \bm{z'}$,
\ba
\Phi^{\rm disk}(\bm{x})&=&-\frac{4G\G\left(\frac{d}{2}\right)}{(d-1)\pi^{\frac{d}{2}-1}}\int_\cV\frac{\rmd^d q(\bm{x'})\,\rho[q(\bm{x'})]}{|q(\bm{x})-q(\bm{x'})|^{d-2}}\nonumber\\
&=&-\frac{4G\G\left(\frac{d}{2}\right)}{(d-1)\pi^{\frac{d}{2}-1}}\int_\cV\frac{\rmd^d \bm{z'}\,\rho(\bm{z'})}{|q(\bm{x})-\bm{z'}|^{d-2}}\,.
\ea
This is the standard expression in ordinary gravity except for the $q$-dependence in the denominator. In $d=3$ dimensions,
\be
\Phi^{\rm disk}(\bm{x}) =-G\int_\cV\frac{\rmd^3 \bm{z'}\,\rho(\bm{z'})}{|q(\bm{x})-\bm{z'}|}\,.\label{Phisolqfinapp}
\ee
This is the starting point to build a model of a galaxy (or its bulge) and its rotation curve. The calculation is identical to the standard one reported in section \ref{Tvrotcu} \cite{Mannheim:2005bfa}, the only difference being the replacement $R\to \cQ(R)$ for the radial coordinate in \Eqq{1r2},
\be
\frac{1}{|q(\bm{x})-\bm{z'}|}= \sum_{m=-\infty}^{+\infty}\int_0^{+\infty}\rmd y\,\rme^{\rmi m(\phi-\phi')-|Z-Z'|y}J_m[\cQ(R)y]\,J_m(R'y),\label{1r1bq}
\ee
and, at the last step in the derivation, the approximation $\cQ(R)\simeq q(R)$:
\be
\Phi^{\rm disk}(R) \simeq-\frac{GM}{R_0}\frac{q(R)}{2R_0}\left\{I_0\left[\frac{q(R)}{2R_0}\right]\,K_1\left[\frac{q(R)}{2R_0}\right]-I_1\left[\frac{q(R)}{2R_0}\right]\,K_0\left[\frac{q(R)}{2R_0}\right]\right\}.
\ee

The gravitational field in the $Z=0$ plane and in the radial direction is $g(R)=-\rmd\Phi(R)/\rmd q(R)$.
The circular velocity with respect to $q$-coordinates is $v_{q,{\rm circ}}=\sqrt{|q(R)\,g(R)|}$, whose expression is \Eq{vcirc0gr} with $R$ replaced by $q(R)$. Noting that the physical circular velocity is \cite{trtls}
\be
v_{\rm circ} = v_{q,{\rm circ}}\,\frac{\rmd R}{\rmd q(R)}=\frac{v_{q,{\rm circ}}}{v(R)}\,,
\ee
we get
\be\label{vcirc}
v_{\rm circ} = \frac{1}{v(R)}\frac{q(R)}{2R_0}\sqrt{\frac{2GM}{R_0}\left|I_1\left[\frac{q(R)}{2R_0}\right]\,K_1\left[\frac{q(R)}{2R_0}\right]-I_0\left[\frac{q(R)}{2R_0}\right]\,K_0\left[\frac{q(R)}{2R_0}\right]\right|}.
\ee

\begin{figure}
\bc
\includegraphics[width=10cm]{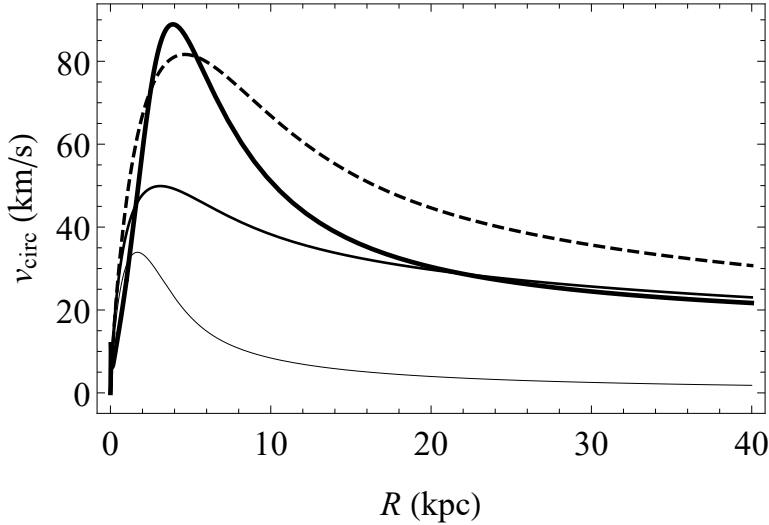}
\ec
\caption{\label{fig10} Rotation curve of baryonic matter in the thin-disk approximation and an exponential matter density profile \Eq{thinrho} in Einstein's theory (\Eqq{vcirc0gr}, dashed curve) and in the multi-fractional theory $T_q$ with $q$-derivatives (\Eqq{vcirc}) for $\a=1.5$ and $A=0=B$, $\a=0.5$ and $A=0=B$, and $\a=0.5$ and 5 harmonics all with amplitudes $A=0.2$, $B=0$ and parameter $N=10^{10}$ (solid curves of increasing thickness). Here $M=1.72\times 10^{40}\,{\rm kg}$, $R_0=2.16\,{\rm kpc}$ and $R_*=40\,{\rm kpc}$.}
\end{figure}

It is easy to check that this function never ``improves'' its counterpart \Eq{vcirc0gr} from Einstein's theory for any value and sign of the fractional exponent $\a$, since it always decays at large $R$ instead of reaching a plateau (Fig.\ \ref{fig10}). The reason is that, if $\a<0$ or $0<\a<1$, $q(R)\to R$ at large scales and the curve deviates from Einstein's theory only at small scales $R<R_*$, quickly converging to Newtonian gravity when $R>R_*$. When $\a>1$, the rotation curve deviates from the standard case at large scales but the effect is of a suppression of velocity rather than the desired enhancement.

Contrary to the other two multi-fractional models of galaxy disks, in the theory $T_q$ it is easy to consider the effect of logarithmic oscillations in the spacetime measure. We checked that the inclusion of a few harmonics of log oscillations does not improve the fit because the modulation of the log oscillations never generates a plateau, even for a minimum-frequency harmonic. An example is given in Fig.\ \ref{fig9}. 


\subsection{\texorpdfstring{$T_q$}{Tq}: data-based rotation curve}\label{Tqrotcu2}

The rotation curve of this theory for the actual matter density profile is easy to write down: 
\be
v_{{\rm circ}}(R)=\frac{v_{{\rm circ},0}[q(R)]}{v(R)}\,,
\ee
where $v_{{\rm circ},0}[q(R)]$ is given by \Eqqs{gdisk0}, \Eq{gbulge0} and \Eq{vcirc02} with $R$ replaced by $q(R)$ everywhere. 
 From this relation, we immediately foresee that the theory does not comply with the baryonic Tully--Fisher formula, since $v_{{\rm circ}}^4=v_{{\rm circ},0}^4/v^4\propto M^2/(R^2 v^4)$ and one would need $v\propto (M/R^2)^{1/4}=M^{1/4}R^{-1/2}$. However, the measure weight is assumed to be universal and independent of the mass of the object.

Indeed, we checked that the model does not fit any of the three galaxies' rotation curves. In the absence of log oscillations, varying $\a$ and $R_*$ one obtains rotation curves always below the data at their maximum and decreasing as $R$ increases. Log oscillations modulate these patterns or, if the amplitudes are large enough, they produce a strongly oscillatory pattern not matching data. In fact, while including only one harmonic does not improve the fit because the modulation of the log oscillations never generates a plateau, even for a minimum-frequency harmonic, adding several harmonics (we checked up to 10) creates a relatively well-known pattern of destructive and constructive interference \cite{Calcagni:2017via,Calcagni:2020ads} resulting in spikes or sharp features across the whole range of scales, incompatible with the observed rotation curve. We do not show these plots here.


\section{Conclusions}\label{disc}

In this paper, we derived the Poisson equation in Cartesian, spherical and cylindrical coordinates from covariant modified Einstein equations in spacetimes with varying dimension. In particular, we considered multi-fractional theories with integer-order derivatives. The gravitational potentials for a point-wise or an extended matter source are special static solutions of the Poisson equation with immediate physical applications. Using these solutions, we entertained the possibility to describe the observed rotation curve of galaxies in terms of geometric effects.

The theory $T_1$ with ordinary derivatives does not easily lead to an analytic expression of the Newtonian potential and we could not fully determine the normalization constants. However, an inspection of the approximated potential led to fixing the fractional exponent $\a$ to the value \Eq{special}. It would be interesting to make a more sophisticated analysis to see if this value eventually gives a good fit of data, as suggested by the results of \cite{Varieschi:2021rzk}.

In contrast, the theory $T_q$ with $q$-derivatives is fully solvable but, unfortunately, it is unable to fit the SPARC data for NGC7814, NGC6503 and NGC3741, either in the absence or in the presence of log oscillations in the spatial measure. We believe this negative result to be robust and we conclude that this theory does not offer an alternative to dark matter, just like it does not provide an alternative to dark energy \cite{Calcagni:2020ads}.

The theory $T_v$ with weighted derivatives fares better than the others and it can fit the SPARC data for the three galaxies at large radius, thus explaining the anomalous velocity profile of ordinary matter as an effect of geometry, more precisely, of a varying Hausdorff dimension of space. For the theory $T_v$, the spatial Hausdorff dimension of the galaxy is
\be\label{dhgalax}
\dh^{\rm galaxy}=3\a\approx 4\,,
\ee
stating that, despite the static approximation in which the galaxy is described, its dynamics still ``feels'' four dimensions. This result is tantalizing because it matches the independent findings for the same theory in the context of dark energy \cite{Calcagni:2020ads}. In some yet poorly understood way, dimensional flow in symmetry-reduced systems embedded in the theory $T_v$ tends to recover four spacetime dimensions. In the homogenous, spatial-independent cosmological setting, which is one-dimensional, the late-time universe undergoes an accelerating phase with $\dh\approx 4$. In the static, time-independent setting of spherical or axisymmetric gravitational potentials, galaxy rotation curves behave as if the spacetime around the galaxy had $\dh\approx 4$ dimensions. Just like in \cite{Calcagni:2020ads}, to our regret we do not have an explanation of this phenomenon. It underscores a sort of conservation (or, more precisely, restoration) law for the dimension of spacetime, which is preserved asymptotically when some symmetry reduction is performed. It is as if geometry is attempting to restore four dimensions in all the settings where some dimensions are lost by such reduction: one-dimensional homogeneous and isotropy cosmology in the case of \cite{Calcagni:2020ads}, a three-dimensional static background in the present case. 

While the isolated finding of \cite{Calcagni:2020ads} might have been accidental, its recovery in the independent setting of galactic physics makes us believe that this is not the case and that the restoration law is plausible. These results are very promising but provisional because they do not hold for small radii, they are not based on a full profile of the potential valid at all scales and, finally, they have not been checked on a wider sample of galaxies. The poor fit at small radii should find an explanation by relaxing one or more of the approximations. One possibility is to take a radial scale $0\ll R_*<R_0$ and solve the Poisson equation numerically, in the hope of preserving the large-$R$ fit while getting a good small-$R$ behaviour. However, we do not expect substantial changes in such extended analysis because the plateau was found only in a model where $R_*\gtrsim R_0$ and at scales where the terms we ignored are sub-dominant. 

Another possibility, which we warned the reader about in section \ref{newv}, is to set $U(v)\neq 0$. This would require a more complicated analysis of the Poisson equation and a wholly new derivation of the rotation curve we have not carried here, but that would deserve attention in the future. Just to mention one subtlety, the fact that the $R$-dependence of $U$ can only come from the measure weight is constraining. For instance, suppose we want to preserve the successful large-$R$ rotation curve of the theory $T_v$, that is, in the regime with small factional corrections with $\a=4/3$. This implies that we should choose a profile tending to a negligible constant when $R\gtrsim R_0$. On the other hand, by definition, in the regime with small factional corrections the measure weight $v$ tends to unity and $U$ tends to a constant. Therefore, at small radii the only possible non-trivial correction is a non-zero constant that can be neglected at large $R$. For instance, in terms of a simple polynomial $U(v)= a+b v^n$ we would have $U\simeq a+b(1+nR/R_*)$ and the constant entering the analysis would be $a+b$. In the Poisson equation, however, we should plug the full profile $U[v(R)]$, which would change the solution $\Phi(R)$ and all that follows.



An explanation of rotation curves alone is not sufficient to claim that an alternative to dark matter has been found. In parallel, the baryonic Tully--Fisher relation \Eq{tufi} should be verified. For the theories $T_1$ and $T_v$, this translates into the condition \Eq{strangerel2} stating that the characteristic radius $R_*$ is proportional to the square root of the mass. In other words, since $R_*$ is galaxy-dependent, the multi-fractional geometry can only work as a local model of the spacetime embedding the galaxy. But, on top of that, the radius-mass relation \Eq{strangerel2} should hold. We have seen that it does not for the theory $T_v$ with small fractional corrections and $\a=4/3$ (Fig.~\ref{fig8}). However, one cannot conclude that this model is ruled out as an alternative explanation of dark matter because, on one hand, we only checked \Eq{strangerel2} for three galaxies out of the hundreds available and, on the other hand, we used the values of $R_*$ found from the best fits of the circular velocity. In turn, the theoretical curves assumed in the fit rely on the approximations discussed above. Using the full model without approximating the fractional corrections to be small or large, and possibly also including a non-zero $U(v)$, would almost certainly lead not only to the fixing of the rotation curves at small radii, but also to different best-fit values for $R_*$. The condition \Eq{strangerel2} will be fairly tested only then. The origin of this puzzling condition, if true, should also be explained. From Fig.~\ref{fig8}, we see that the theory predicts a sort of constant surface density
\be
\frac{M}{R_*^2}\lesssim O(1)\,{\rm kg}\,{\rm m}^{-2}\,.
\ee
How this formula connects with the multi-scale geometry of the theory remains to be seen. Its character suggests a dynamical origin, since it relates the non-dynamical non-metric sector of geometry ($R_*$) with the mass of the galaxy ($M$), the latter being related to the metric part of the geometry via the modified Einstein's equations.

Another complementary aspect of the rotation curves to be checked is the so-called radial acceleration relation (RAR), a phenomenological formula extrapolated from all SPARC data which connects the observed radial acceleration with the one predicted by the observed baryonic distribution \cite{Lelli:2016zqa,Chae:2020omu}. It fits well the data of this catalogue of 175 galaxies but it may yield poor fits when applied to a very reduced sample such as the one considered here \cite{Varieschi:2020hvp}. Therefore, to test the RAR properly, both in its original form and in the presence of an external field effect as in MOND models, it would be recommendable to use the whole SPARC catalogue, or at least a sizable sub-sample of galaxies, which will be done in the future after fixing the model in $T_v$ by relaxing the approximations mentioned above. 

Moreover, any theory beyond Einstein gravity should be tested against observations of the Bullet cluster \cite{Clowe:2006eq}, galaxy clusters \cite{All11}, gravitational lensing \cite{Massey:2010hh,DES:2021gua} and the cosmic microwave background \cite{Aghanim:2018eyx}, among others. To date, the theory $T_v$ has not been applied to any of these physical systems and it would be important to do so in order to accumulate evidence, as it has been done in the case of Moffat's scalar-tensor-vector modified gravity model \cite{Moffat:2005si,Brownstein:2007sr,Moffat:2010gt,Israel:2016qsf}. As a preliminary result, we note that the theory $T_v$ can explain in part, but perhaps not fully, the observations of the Bullet cluster, without however bridging the gap between baryonic-matter-based and dark-matter-based physics. We use the qualitative argument of \cite{Moffat:2010gt} (updated with the values of \cite{Lage:2014yxa}), which is orientative but not conclusive.

According to this argument, one crudely approximates the Bullet cluster as the linear superposition of two point sources, the main cluster with dynamical mass $M_{\rm main}=2\times 10^{15}\,M_\odot$ and the ``bullet'' with dynamical mass $M_{\rm bullet}=5\times 10^{14}\,M_\odot$ \cite{Lage:2014yxa}, where both masses are found from gravitational lensing observations. These masses include both baryonic and dark matter. In standard gravity, the escape velocity of the main and bullet clusters is found by setting to zero both the total (kinetic and potential) energy of the two clusters and the total momentum \cite{Moffat:2010gt}:
\ba
&&\frac12\left(M_{\rm main} v_{\rm main}^2+M_{\rm bullet} v_{\rm bullet}^2\right)-\frac{GM_{\rm main}M_{\rm bullet}}{r}=0\,,\\
&&M_{\rm main}v_{\rm main}+M_{\rm bullet}v_{\rm bullet}=0\,.\label{momenbul}
\ea
At a separation of $r=2.5\,{\rm Mpc}$ \cite{Lage:2014yxa}, one gets an infall velocity of $|v_{\rm main}-v_{\rm bullet}|\approx 2933\,{\rm km\,s}^{-1}$. In contrast, baryonic matter amounts to about $M_{\rm baryon}/M_{\rm tot}=0.14$ of the total, which would give an infall velocity three times smaller than (37\% of) the observed one, $|v_{\rm main}-v_{\rm bullet}|\approx 1097\,{\rm km\,s}^{-1}$. Repeating the estimate for the only multi-fractional theory that could partly explain the galaxy rotation curves, the theory $T_v$ with potential \Eq{filog}, we have a system given by \Eqq{momenbul} and
\be
\frac12\left(M_{\rm main} v_{\rm main}^2+M_{\rm bullet} v_{\rm bullet}^2\right)-\frac{GM_{\rm main}M_{\rm bullet}}{r}+\frac{GM_{\rm main}M_{\rm bullet}}{r_*}\,\ln\frac{r}{r_*}=0\,.
\ee
Taking the baryonic masses not including dark matter, the maximum of the infall velocity is obtained at $r_*= 6.8\,{\rm Mpc}$ and is $|v_{\rm main}-v_{\rm bullet}|=1283\,{\rm km\,s}^{-1}$, slightly higher than the standard Newtonian value but, still, only 47\% of the actual value. This mildly encouraging estimate should be refined by a much more rigorous calculation.

As a last observable to consider, we mention a directional effect in the acceleration experienced by the matter of the Milky Way, the radial acceleration being greater than the vertical one. Recent results show that MOND-like models are disfavored relative to a simple dark-matter halo model \cite{Lisanti:2018qam}. We add this datum to the list of benchmarks to test multi-fractional theories as alternatives to dark matter. The fact that not even MOND models explain convincingly the directionality of gravitational acceleration should make one reflect that, perhaps, a viable alternative to dark matter should \emph{not} follow MOND-like dynamics too closely. In this sense, the absence of features such as the external field effect might not be necessarily detrimental to the model.
 
Other avenues to explore in the future may come across other multi-fractional theories, those with fractional operators \cite{Calcagni:2021ipd,mf2}. A phenomenological Newtonian model with fractional Laplacian seems able to fit galaxy rotation curves \cite{Giusti:2020rul,Giusti:2020kcv} but it has not been embedded in a covariant theory. The multi-fractional theories $T_\g$ offer an opportunity both to embed the model of \cite{Giusti:2020rul,Giusti:2020kcv} and to further test the idea that a dynamics with fractional operators is a viable alternative to dark matter.


\section*{Acknowledgments}

G.C.\ is supported by the I+D grant PID2020-118159GB-C41 of the Spanish Ministry of Science and Innovation. G.V.\ is supported by the Frank R.\ Seaver College of Science and Engineering, Loyola Marymount University.



\appendix


\section{Rotation curves in Einstein's theory and SPARC data}\label{dmrev}


\subsection{General relativity: equations of motion}\label{dmrev1}

We work in a $D$-dimensional spacetime with signature $(-,+,\dots,+)$ and indicate with $d=D-1$ the number of spatial directions. For the physical application of the model to galaxy rotation curves, we will eventually set $D=4$ and $d=3$.

Via the standard Levi-Civita connection and curvature tensors,
\bs\label{tutto}\ba
\hspace{-1.5cm}&&\textrm{Levi-Civita connection:}\hspace{.5cm} \Gamma^\rho_{\mu\nu} := \frac12 g^{\rho\s}\left(\p_\mu g_{\nu\s}+\p_\nu g_{\mu\s}-\p_\s g_{\mu\nu}\right),\label{lecico}\\
\hspace{-1.5cm}&&\textrm{Riemann tensor:}\hspace{1.6cm} R^\rho_{~\mu\sigma\nu}:= \p_\sigma \Gamma^\rho_{\mu\nu}-\p_\nu \Gamma^\rho_{\mu\sigma}+\Gamma^\tau_{\mu\nu}\Gamma^\rho_{\sigma\tau}-\Gamma^\tau_{\mu\sigma}\Gamma^\rho_{\nu\tau}\,,\label{rite}\\
\hspace{-1.5cm}&&\textrm{Ricci tensor:}\hspace{2.3cm} R_{\mu\nu}:= R^\rho_{~\mu\rho\nu}\,,\\
\hspace{-1.5cm}&&\textrm{Ricci scalar:}\hspace{2.35cm} R:= g^{\mu\nu}R_{\mu\nu}\,,\\
\hspace{-1.5cm}&&\textrm{Einstein tensor:}\hspace{1.8cm} G_{\mu\nu}:= R_{\mu\nu}-\tfrac12g_{\mu\nu}R\,,\label{Eiten}
\ea\es
the action of Einstein's theory is
\be\label{Sg2gr}
S =\frac{1}{2\k^2}\int\rmd^Dx\,\sqrt{|g|}\,R+S_{\rm m}\,,
\ee
where $\k^2=8\pi G$ is proportional to Newton's constant $G$, $g$ is the determinant of the metric and $S_{\rm m}$ is the baryonic matter action. The Einstein equations are
\be
\k^2\,T_{\mu\nu}=G_{\mu\nu}\,,\label{ee2gr}
\ee
where the energy-momentum tensor is
\be\label{tmunugr}
T_{\mu\nu}:=-\frac{2}{\sqrt{|g|}}\frac{\de S_{\rm m}}{\de g^{\mu\nu}}\,.
\ee
For later purposes, we recast the equations of motion \Eq{ee2gr} in a more convenient form. Taking their trace,
\be
\k^2\,T=-\frac{D-2}{2}R\,,\label{ee2trgr}
\ee
and replacing $R$ back into \Eq{ee2gr}, we get
\be\label{ee2fingr}
\k^2S_{\mu\nu} = R_{\mu\nu}\,,\qquad S_{\mu\nu} := T_{\mu\nu}-\frac{1}{D-2}g_{\mu\nu}T\,,\qquad T=T_\s^{\ \s}.
\ee


\subsection{General relativity: Poisson equation and Newton's potential}\label{dmrev2}

To find the gravitational potential in a static matter configuration, we expand the metric into Minkowski background $\eta_{\mu\nu}$ and a perturbation $h_{\mu\nu}$, as in eq.\ \Eq{getah}. Expanding the Levi-Civita connection and the curvature terms, a textbook exercise shows that
\ba
\de^{(1)}\Gamma^\s_{\mu\nu} &=& \frac12\left(\p_\mu h_\nu^\s+\p_\nu h_\mu^\s-\p^\s h_{\mu\nu}\right),\label{leciden}\\
\de^{(1)} R_{\mu\nu}	 &=& -\frac12\B h_{\mu\nu}+ \p^\s\p_{(\mu}h_{\nu)\s}-\frac12\p_{(\mu}\p_{\nu)} h\,,\label{dRmn}\\
\de^{(1)} R         &=& -\B h+\p^\s\p^\tau h_{\s\tau}\,,\label{dRg}\\
\de^{(1)} G_{\mu\nu}   &=& -\frac12\B h_{\mu\nu}+ \p^\s\p_{(\mu}h_{\nu)\s}-\frac12\eta_{\mu\nu}\p^\s\p^\tau h_{\s\tau}+\frac12\left[\eta_{\mu\nu}\B-\p_{(\mu}\p_{\nu)}\right] h\,.
\ea
Also, for any two scalars $A$ and $B$
\ba
A \de^{(1)}\B B &=& -A h^{\mu\nu} \p_\mu\p_\nu B + A\de^{(1)}\N_\mu \p^\mu B = -A h^{\mu\nu} \p_\mu\p_\nu B + A\de^{(1)}\Gamma_{\mu\s}^\mu \p^\s B\nonumber\\
&=& -A h^{\mu\nu} \p_\mu\p_\nu B\,,\label{usefu1}\\
A \de^{(1)}(\N_\mu\N_\nu) B &=& A \de^{(1)}\N_\mu\p_\nu B = -A \de^{(1)}\Gamma_{\mu\nu}^\s\p_\s B\nonumber\\
&=&  -\frac12A\left(\p_\mu h_\nu^\s+\p_\nu h_\mu^\s-\p^\s h_{\mu\nu}\right)\p_\s B\,,\label{usefu2}
\ea
where in the last expression we used the definition of covariant derivative on a covariant vector, $\N_\mu A_\nu:=\p_\mu A_\mu-\Gamma_{\mu\nu}^\s A_\s$.

In any covariant, $D$-dimensional, torsion-free theory expanded around Minkowski background, it is possible to choose the transverse-traceless gauge $\p^\mu h_{\mu\nu}=0=h_{\mu}^{\mu}$ \cite{MTW}, so that $\de^{(1)} R=0$ and $\de^{(1)} G_{\mu\nu} =\de^{(1)} R_{\mu\nu} =-\frac12\B h_{\mu\nu}$. We do not use the transverse-traceless gauge anywhere in the paper.

\subsubsection{General relativity: Poisson equation}

In the presence of matter, it would be inconsistent to equate $\de^{(1)} G_{00}$ to the matter energy density $\rho=T_{00}$ because one should also perturb the energy-momentum tensor at the linear order in the metric.\footnote{In Einstein's theory, the energy-momentum tensor for non-relativistic matter is such that $|T_{ij}|\ll |T_{00}|$, which implies $|G_{ij}|\ll |G_{00}|$, i.e., $R_{ij}\simeq g_{ij}R/2$. In turn, in the weak-field approximation \Eq{getah} this implies $R\simeq \eta^{\mu\nu}R_{\mu\nu}=-R_{00}+\sum_i R_{ii}\simeq-R_{00}+(D-1)R/2$, that is, $R\simeq 2R_{00}/(D-3)$. Therefore, $\de^{(1)} G_{\mu\nu}$ does not include all the contributions of the gravitational field in the Einstein equations.} One way to account for that contribution is to extract the trace part $T=T_\mu{}^\mu$ of the energy-momentum tensor, which is the reason why we recast \Eq{ee2gr} as \Eq{ee2fingr}.

Taking the 00 component of the Einstein equations \Eq{ee2fingr} for a static configuration $h_{00}=-2\Phi({\bm x})$, on the left-hand side we get $S_{00} \simeq [(D-3)/(D-2)]\rho$, where $\rho=T_{00}$ is the matter energy density and we ignored the pressure part in the trace, since we are in a non-relativistic regime. Expanding the right-hand side of \Eq{ee2fingr} at linear order,
\be\label{pois1gr}
\frac{d-2}{d-1}\k^2\rho = \N^2\Phi\,,
\ee
where $d=D-1$ and $\N^2$ is the spatial flat Laplacian. In spherical coordinates,
\be
\N^2=\frac{1}{r^{d-1}}\p_r\left(r^{d-1}\p_r \,\cdot\,\right)+\frac{1}{r^2}\N_{S^{d-1}}^2=\p_r^2+\frac{d-1}{r}\p_r+\frac{1}{r^2}\N_{S^{d-1}}^2\,,\label{rlapl}
\ee
where $\N_{S^{d-1}}^2$ is the spherical Laplacian. The Poisson equation \Eq{pois1gr} then reads
\be\label{Poiord}
\frac{d-2}{d-1}\k^2\rho=\left[\p_r^2+\frac{d-1}{r}\p_r+\frac{1}{r^2}\N_{S^{d-1}}\right]\Phi\,.
\ee

Using instead cylindrical coordinates, one defines a radial, an azimuth and an elevation variable $x_1=R\cos\vp$, $x_2=R\sin\vp$, $x_i=Z_i$, where $i=3,\dots,d$:
\be\label{cylapl}
\N^2=\frac{1}{R}\p_R\left(R\p_R \,\cdot\,\right)+\frac{1}{R^2}\p^2_\vp+\N_{(i)}^2=\p_R^2+\frac{1}{R}\p_R+\frac{1}{R^2}\p^2_\vp+\N_{(i)}^2\,,
\ee
where $\N_{(i)}^2$ is the Laplacian in Cartesian coordinates $x_i$, $i=3,\dots,d$. The Poisson equation is then
\be\label{Poiord2}
\frac{d-2}{d-1}\k^2\rho=\left[\p_R^2+\frac{1}{R}\p_R+\frac{1}{R^2}\p^2_\vp+\N_{(i)}^2\right]\Phi\,.
\ee

\subsubsection{General relativity: Newton's potential}

The solution of the Poisson equation \Eq{Poiord} for a radial potential $\Phi(r)$ and a point-wise source $\rho=m\de^d(\bm{x})$ of mass $m$ is found with standard techniques. We have
\ba
\frac{d-2}{d-1}\k^2m\de^d(\bm{x})=\cC\Phi&:=&\left(\p_r^2+\frac{d-1}{r}\p_r\right)\Phi(r)\label{eqom}\\
&=&\frac{1}{r^{d-1}}\p_r\left(r^{d-1}\p_r\Phi\right).\label{poisrirgr}
\ea
Everywhere except at $r=0$, the solution is
\be
\Phi(r) = C_+\,r^{-c_+}+C_-\,,\qquad c_+ = d-2\,.\label{phi0gr}
\ee
We can redefine the potential by a shift, thus setting $C_-=0$ from now on. The normalization constant $C_+$ can be found by looking at the solution at $r=0$. First, recall that the radial Dirac distribution $\de(r)$ is related to $\de^d(\bm{x})$ by
\be\label{dexder}
\int\rmd r\,\de(r) = 1 = \int \rmd^d\bm{x}\,\de^d(\bm{x})=\frac{2\pi^\frac{d}{2}}{\G\left(\frac{d}{2}\right)}\int\rmd r\,r^{d-1}\de^d(\bm{x}),
\ee
so that for a test function $f$ one has
\ba
\int \rmd^d\bm{x}\,(\cC\Phi)\,f &=& \lim_{\e\to 0}\frac{2\pi^\frac{d}{2}}{\G\left(\frac{d}{2}\right)}\int_\e^{+\infty}\rmd r\,r^{d-1}(\cC\Phi)\, f\nonumber\\
&\stackrel{\textrm{\tiny\Eq{poisrirgr}}}{=}&
\lim_{\e\to 0}\frac{2\pi^\frac{d}{2}}{\G\left(\frac{d}{2}\right)}\int_\e^{+\infty}\rmd r\left[\p_r\left(r^{d-1}\p_r\Phi\right)\right]f\nonumber\\
&=&-\lim_{\e\to 0}\frac{2\pi^\frac{d}{2}}{\G\left(\frac{d}{2}\right)}\int_\e^{+\infty}\!\rmd r\,\left(r^{d-1}\p_r\Phi\right)\p_r f\nonumber\\
&=& \lim_{\e\to 0}\frac{2\pi^\frac{d}{2}}{\G\left(\frac{d}{2}\right)}C_+(d-2)\int_\e^{+\infty}\!\rmd r\p_r f\nonumber\\
&=&-\frac{2\pi^\frac{d}{2}}{\G\left(\frac{d}{2}\right)}C_+(d-2) f(0)\nonumber\\
&\stackrel{\textrm{\tiny\Eq{eqom}}}{=}& \frac{d-2}{d-1}8\pi Gm\int \rmd^d\bm{x}\,\de^d(\bm{x}) f(\bm{x})=\frac{d-2}{d-1}8\pi Gm\,f(0)\,,\nonumber
\ea
thus fixing the constant $C_+$ and the solution \Eq{phi0gr}:
\be\label{Phisolv}
\Phi(r)=\Phi_0(r):=-\frac{4\G\left(\frac{d}{2}\right)}{(d-1)\pi^{\frac{d}{2}-1}}\frac{Gm}{r^{d-2}}\,.
\ee
In $d=3$ spatial dimensions, this reduces to the standard Newtonian potential
\be\label{newt}
\Phi(r)=\Phi_0(r)= -\frac{Gm}{r}\,.
\ee
Here and in the following, a subscript 0 denotes a quantity (potential, gravitational field, rotation velocity) in Einstein's theory.


\subsection{General relativity: thin-disk exponential-density rotation curve}\label{dmrev3}

Call
\be\label{PhIG}
\Phi(r)=m\cG(\bm{x}-\bm{x'})
\ee
the solution of the Poisson equation with point-wise source with mass $m$, where we factorized the mass dependence. The gravitational potential of a galaxy can be obtained taking an infinitesimal mass $\rmd m=\rho\,\rmd\cV$ and integrating $\cG$ over the volume $\cV$ of the galaxy. In Einstein's theory in three spatial dimensions, the potential of the matter distribution is
\be\label{potetgr}
\Phi^{\rm galaxy}_0(\bm{x})=\int_\cV\rmd^3\bm{x'}\,\rho(\bm{x'})\,\cG(\bm{x}-\bm{x'})\,,
\ee
which is calculated as follows \cite{Mannheim:2005bfa}. Assume that the galaxy has cylindrical symmetry, and that the spherical bulge component is negligible. In cylindrical coordinates, the integration volume is
\be\label{cylinv}
\int_\cV\rmd^3 \bm{x'}=\int_0^{+\infty}\rmd R'\, R'\int_0^{2\pi}\rmd\phi'\int_{-\infty}^{+\infty}\rmd Z'\,.
\ee
The Green's function is \Eqq{newt}, $\cG=-G/|\bm{x}-\bm{x'}|$. Using \Eqq{1r2},
\be
\frac{1}{|\bm{x}-\bm{x'}|}= \sum_{m=-\infty}^{+\infty}\int_0^{+\infty}\rmd y\,\rme^{\rmi m(\phi-\phi')-|Z-Z'|y}J_m(Ry)\,J_m(R'y),\label{1r1b}
\ee
where $J_m$ is the Bessel function of the first kind. Assuming that the galaxy completely lies on the $Z=0$ plane, the matter energy distribution only depends on the coordinates $R'$ and $Z'$ and, in fact, it is symmetric under a reflection $Z'\to -Z'$. Therefore, $\rho(\bm{x'})=\rho(R',|Z'|)$ and
\ba
\hspace{-1cm}\Phi^{\rm disk}_0(R) &=&-G\int_0^{+\infty}\rmd R'\,R'\int_0^{2\pi}\rmd\phi'\int_{-\infty}^{+\infty}\rmd Z'\,\rho(R',|Z'|)\nonumber\\
&&\times\sum_{m=-\infty}^{+\infty}\int_0^{+\infty}\rmd y\,\rme^{\rmi m(\phi-\phi')-|Z'|y}J_m(Ry)\,J_m(R'y)\nonumber\\
&=&-2\pi G\int_0^{+\infty}\rmd y\,J_0(Ry)\int_0^{+\infty}\rmd R'\,R'\,J_0(R'y)\int_{-\infty}^{+\infty}\rmd Z'\,\rme^{-|Z'|y}\rho(R',|Z'|)\,,\label{phidisk1}
\ea
where only the zero mode $m=0$ gives a non-zero angular integral. The gravitational field in the $Z=0$ plane and in the radial direction is 
\be
g^{\rm disk}_0(R)=-\frac{\rmd\Phi^{\rm disk}_0(R)}{\rmd R}\,, 
\ee
while the circular velocity is 
\be\label{vgengr}
v_{{\rm circ},0}^{\rm disk}=\sqrt{|R\,g^{\rm disk}_0(R)|}
\,.
\ee
                                  
At this point, we can decide whether to employ an analytic model of the galaxy or to use real data. In this sub-section, we study the analytic model, later showing how it deviates from the actual prediction for real galaxies.

In the thin-disk approximation, one assumes that the galaxy has a total mass $M$, is dominated by the disk component with characteristic scale length $R_0$ and that the disk is infinitely thin. Then, one can take the matter distribution
\be\label{thinrho}
\rho(R',Z')=\frac{M}{2\pi R_0^2}\,\rme^{-\frac{R'}{R_0}}\de(Z')\,,
\ee
so that
\ba
\Phi^{\rm disk}_0(R) &=&-\frac{GM}{R_0^2}\int_0^{+\infty}\rmd y\,J_0(Ry)\int_0^{+\infty}\rmd R'\,R'\,J_0(R'y)\,\rme^{-\frac{R'}{R_0}}\int_{-\infty}^{+\infty}\rmd Z'\,\rme^{-|Z'|y}\de(Z')\nonumber\\
&=&-\frac{GM}{R_0^2}\int_0^{+\infty}\rmd y\,J_0(Ry)\int_0^{+\infty}\rmd R'\,R'\,J_0(R'y)\,\rme^{-\frac{R'}{R_0}}\nonumber\\
&=&-GM\int_0^{+\infty}\rmd y\,\frac{J_0(Ry)}{(1+R_0^2 y^2)^\frac32}\nonumber\\
&=&-\frac{GM}{R_0}\frac{R}{2R_0}\left[I_0\left(\frac{R}{2R_0}\right)\,K_1\left(\frac{R}{2R_0}\right)-I_1\left(\frac{R}{2R_0}\right)\,K_0\left(\frac{R}{2R_0}\right)\right],\label{PhiGR}
\ea
where $I_m$ and $K_m$ are the modified Bessel functions of the first and second kind, respectively. The gravitational field $g_0(R)$ can be calculated using the relations
\be
I_0'(z)=I_1(z)\,,\quad I_1'(z)=I_0(z)-\frac{I_1(z)}{z}\,,\quad K_0'(z)=-K_1(z)\,,\quad K_1'(z)=-K_0(z)-\frac{K_1(z)}{z}\,,
\ee
so that
the circular velocity is
\be\label{vcirc0gr}
v_{{\rm circ},0}^{\rm disk} = \frac{R}{2R_0}\sqrt{\frac{2GM}{R_0}\left|I_1\left(\frac{R}{2R_0}\right)\,K_1\left(\frac{R}{2R_0}\right)-I_0\left(\frac{R}{2R_0}\right)\,K_0\left(\frac{R}{2R_0}\right)\right|}\,.
\ee


\subsection{General relativity: data-based rotation curve}\label{dmrev4}

In this sub-section, we take the route of calculating \Eqq{phidisk1} with actual surface-luminosity data. We also lift the thin-disk approximation by taking a thick-disk profile \cite{vdK88}
\be\label{thick}
\rho(R',|Z'|)= \Sigma^{\rm disk}(R')\,\frac{\rme^{-\frac{|Z'|}{h_Z}}}{2h_Z}\,,
\ee
where $\Sigma^{\rm disk}(R')$ is the surface matter distribution of the disk and $h_Z$ is the vertical scale height, related to the radial scale $R_0$ by \cite{Lelli:2016zqa,Ber10}
\be
h_Z =0.196\left(\frac{R_0}{{\rm kpc}}\right)^{0.663}{\rm kpc}\,.
\ee
Other thick-disk profiles are possible \cite{Mannheim:2005bfa,Ber10} but \Eq{thick} will be sufficient for our aim. Since
\be
\int_{-\infty}^{+\infty}\rmd Z'\,\rme^{-|Z'|y}\frac{\rme^{-\frac{|Z'|}{h_Z}}}{2h_Z}=\int_0^{+\infty}\rmd Z'\,\frac{\rme^{-\frac{1+h_Zy}{h_Z}Z'}}{h_Z}=\frac{1}{1+h_Zy}\,,
\ee
then \Eqq{phidisk1} becomes
\be
\Phi^{\rm disk}_0(R) =-2\pi G\int_0^{+\infty}\rmd y\,\frac{J_0(Ry)}{1+h_Zy}\int_0^{+\infty}\rmd R'\,R'\,J_0(R'y)\,\Sigma^{\rm disk}(R')\,.\label{phidisk2}
\ee
The thin-disk limit corresponds to $h_Z\to 0$. 

So far, we computed only the disk component but the total surface distribution of baryonic matter is made of three contributions: one from the disk, one from the gas (including helium) and one from the bulge of the galaxy. Each of these contributions is measured separately from its surface luminosity $\Sigma_L$ and it is converted to a mass distribution via an appropriate mass-to-light ratio \cite{Varieschi:2020hvp}. The gas contribution is axisymmetric just like the disk component, so that its potential $\Phi^{\rm gas}_0(R)$ and circular velocity $v_{{\rm circ},0}^{\rm gas}$ are given by \Eqqs{phidisk1} and \Eq{vgengr} with $\Sigma^{\rm disk}(R')$ replaced by $\Sigma^{\rm gas}(R')$. Additionally, we have \cite{Varieschi:2020hvp}
\bs\label{Sigma}\ba
&&\Sigma(R')=\Sigma^{\rm disk}(R')+\Sigma^{\rm gas}(R')\,,\\
&&\Sigma^{\rm disk}(R') = 0.50\,\Sigma_L^{\rm disk}(R')\,\frac{M_\odot}{L_\odot}\,,\\
&&\Sigma^{\rm gas}(R') = 1.33\,\Sigma_L^{\rm gas}(R')\,\frac{M_\odot}{L_\odot}\,,\\
&&\Sigma^{\rm bulge}(R') = 0.70\,\Sigma_L^{\rm bulge}(R')\,\frac{M_\odot}{L_\odot}\,,
\ea\es
where the surface luminosities $\Sigma_L$ are available from the SPARC database \cite{Lelli:2016zqa}. The values of the mass-to-light ratios are chosen to be the same as those used in \cite{McGaugh:2016leg} as well as in other SPARC papers. In other studies such as more recent SPARC papers \cite{Chae:2020omu,Chae:2021dzt}, the mass-to-light ratios have been treated as free parameters when fitting all the SPARC galaxies with a certain general model. However, even these more detailed studies seem to confirm our choice for the values of the three parameters as a reasonable one.

The bulge component requires some extra work because it has spherical instead of axial symmetry. Going back to \Eqq{potetgr}, instead of expression \Eq{phidisk1} in cylindrical coordinates we have, in three spatial dimensions, the volume element
\be
\int_\cV\rmd^3 \bm{x'}=\int_0^{+\infty}\rmd r'\, {r'}^2\int_0^{\pi}\rmd\theta'\,\sin\theta'\int_0^{2\pi}\rmd\vp'
\ee
and
\ba
\Phi^{\rm bulge}_0(r) &=& -G\int_0^\pi\rmd\theta'\,\sin\theta'\int_0^{2\pi}\rmd\vp'\int_0^{+\infty}\rmd r'\,{r'}^2\frac{\rho^{\rm bulge}(r')}{|\bm{x}-\bm{x'}|}\,\nonumber\\
&=& -2\pi G\sum_{n=0}^{+\infty}\int_0^\pi\rmd\theta'\,\sin\theta'\,P_n(\cos\theta')\int_0^{+\infty}\rmd r'\,{r'}^2\frac{r_<^n}{r_>^{n+1}}\,\rho^{\rm bulge}(r')\nonumber\\
&=&-4\pi G\int_0^{+\infty}\rmd r'\,\frac{{r'}^2}{r_>}\,\rho^{\rm bulge}(r')\nonumber\\
&=&-\frac{4\pi G}{r}\int_0^r\rmd r'\,{r'}^2\,\rho^{\rm bulge}(r')-4\pi G\int_r^{+\infty}\rmd r'\,r'\,\rho^{\rm bulge}(r')\,,\label{phibulge}
\ea
where we used \Eqq{1r30r} with $\cos\g=\cos\theta'$, $r_>$ (respectively, $r_<$) is the largest (smallest) between $r$ and $r'$, the $n=0$ mode is the only one giving a non-zero contribution and 
\be\label{rhobu}
\rho^{\rm bulge}(r') = 0.70\,\rho_L^{\rm bulge}(r')\frac{M_\odot}{L_\odot}.
\ee
To get the luminosity volume density $\rho_L^{\rm bulge}(r')$, one converts the observed surface luminosity $\Sigma_L^{\rm bulge}(R')$ via the formula \cite{BiTr}
\be
\rho_L^{\rm bulge}(r')=-\frac{1}{\pi}\int_{r'}^{+\infty}\frac{\rmd R'}{\sqrt{{R'}^2-{r'}^2}}\frac{\rmd\Sigma_L^{\rm bulge}}{\rmd R'}\,.
\ee
In the SPARC database, each surface luminosity distribution $\Sigma_L$ is given in $L_\odot/{\rm pc}^2$, solar luminosity per (parsec)$^2$. To convert $\Sigma_L$ into ${\rm kg}/{\rm pc}^2$ units and $\rho^{\rm bulge}$ into ${\rm kg}/{\rm pc}^3$ units, one further multiplies \Eqqs{Sigma} and \Eq{rhobu} by a conversion factor $1.989\times 10^{30} ({\rm kg}/M_\odot)$. The total baryonic mass $M$ of the galaxy can be obtained from integrating the total area energy density $\Sigma^{\rm disk}+\Sigma^{\rm gas}+\Sigma^{\rm bulge}$ over the disk or, equivalently, the area density $\Sigma^{\rm disk}+\Sigma^{\rm gas}$ over the disk and the volume density $\rho^{\rm bulge}$ over the bulge,
\ba
M&=&2\pi\int_0^{+\infty}\rmd R'\,R'\left[\Sigma(R')+\Sigma^{\rm bulge}(R')\right]\nonumber\\
&=&2\pi\int_0^{+\infty}\rmd R'\,R'\,\Sigma(R')+4\pi\int_0^{+\infty}\rmd r'\,{r'}^2\,\rho^{\rm bulge}(r')\,.
\ea
In this expression as well as in \Eqqs{phidisk2} and \Eq{phibulge}, the upper extremum of the integrals in $R'$ and $r'$ is actually taken to be a finite value $R_{\rm max}$, much larger than $R_0$ and slightly larger than the radial coordinate of the last data point.

The total potential is
\be
\Phi^{\rm galaxy}_0=\Phi^{\rm disk}_0+\Phi^{\rm gas}_0+\Phi^{\rm bulge}_0\,,
\ee
while the gravitational fields are
\ba
g_0^{\rm disk,gas}(R) &=&-\frac{\rmd\Phi^{\rm disk,gas}_0}{\rmd R}\nonumber\\
&=&-2\pi G\int_0^{+\infty}\rmd y\,y\frac{J_1(Ry)}{1+h_Zy}\int_0^{+\infty}\rmd R'\,R'\,J_0(R'y)\,\Sigma^{\rm disk,gas}(R')\,,\label{gdisk0}\\
g_0^{\rm bulge}(R) &=&-\left.\frac{\rmd\Phi^{\rm bulge}_0}{\rmd r}\right|_{r=R}\nonumber\\
&=&-\frac{4\pi G}{R^2}\int_0^R\rmd r'\,{r'}^2\,\rho^{\rm bulge}(r')\,,\label{gbulge0}
\ea
where we used $\rmd J_0(Ry)/\rmd R=-y\, J_1(Ry)$. Finally, the total rotation velocity is
\be
v_{{\rm circ},0}(R) = \sqrt{R\left|g_0^{\rm disk}(R)+g_0^{\rm gas}(R)+g_0^{\rm bulge}(R)\right|}\,.\label{vcirc02}
\ee

The three galaxies selected in \cite{McGaugh:2016leg} have different properties.
\begin{itemize}
\item NGC7814 is a bulge-dominated spiral galaxy $14.4\,{\rm Mpc}$ away with all three components, a radial scale $R_0=2.54\,{\rm kpc}$ and a total baryonic mass $M=8.26\times 10^{40}\,{\rm kg}$. In this case, we set $R_{\rm max}=20\,{\rm kpc}$.
\item NGC6503 is a disk-dominated dwarf spiral galaxy $6.3\,{\rm Mpc}$ away with two components (a thick disk and gas) and a negligible bulge ($\Sigma^L_{\rm bulge}=0$), a radial scale $R_0=2.16\,{\rm kpc}$ and a total baryonic mass $M=1.72\times 10^{40}\,{\rm kg}$. In this case, we set $R_{\rm max}=24\,{\rm kpc}$.
\item NGC3741 is a gas-dominated irregular spiral galaxy $3.2\,{\rm Mpc}$ away with two components (a thick disk and gas) and a negligible bulge ($\Sigma^L_{\rm bulge}=0$), a radial scale $R_0=0.20\,{\rm kpc}$ and a total baryonic mass $M=5.06\times 10^{38}\,{\rm kg}$. In this case, we set $R_{\rm max}=7.5\,{\rm kpc}$.
\end{itemize}

\begin{figure}
\bc
\includegraphics[width=9cm]{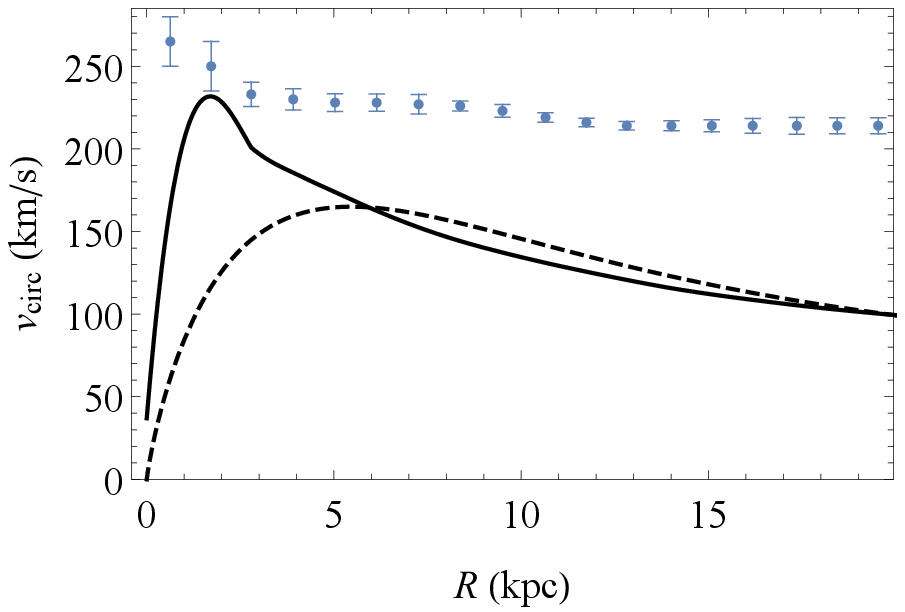}\\
\includegraphics[width=9cm]{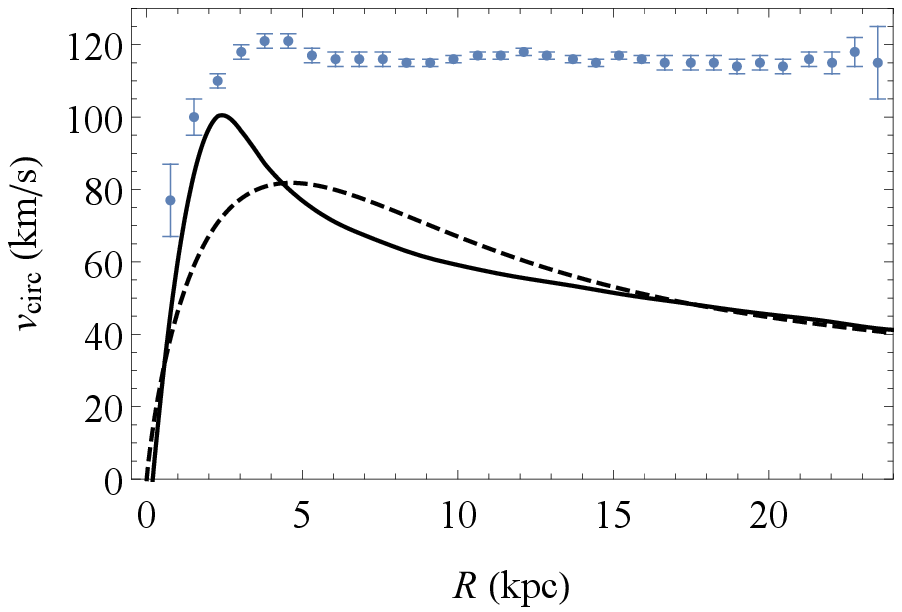}\\
\includegraphics[width=9cm]{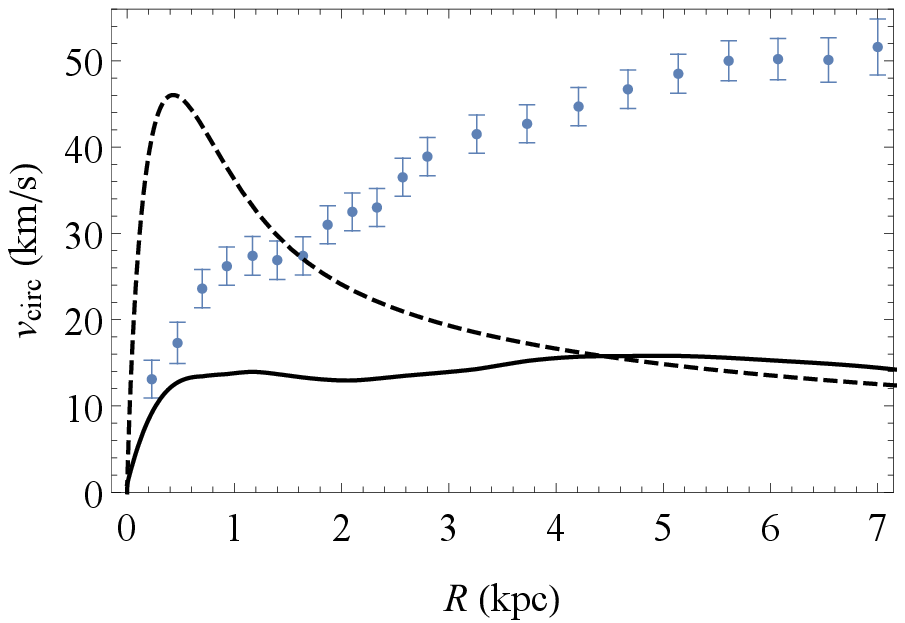}
\ec
\caption{\label{fig1} SPARC data points for the rotation curves of NGC7814 (top panel), NGC6503 (middle panel) and NGC3741 (bottom panel) compared with the prediction of Einstein's theory using the observed baryonic mass distribution (solid curve) or the exponentially decaying thin-disk profile \Eq{thinrho} (dashed curve).}
\end{figure}

The rotation curve \Eq{vcirc02} is shown in Fig.\ \ref{fig1} together with the thin-disk expression \Eq{vcirc0gr} with exponentially decaying matter distribution. As is well known, the Newtonian limit of Einstein's theory fails to reproduce the observed rotation curve of galaxies. The point here is that, when testing a theory where the matter component is purely baryonic, it is important to use the full matter distribution obtained by data rather than assuming a specific radial profile such as \Eq{thinrho}. As one can appreciate, using the actual baryonic matter density profile with a thick-disk assumption changes the rotation curve considerably, especially at small radii. The main change stems from the use of the actual profile $\Sigma(R')$, since the exponential \emph{Ansatz} \Eq{thinrho} deviates from the actual matter density at small radii. On the other hand, one can check that the effect of approximating the disk from thick to thin is comparatively smaller and it would amount to an increase in the peak of the curve \cite{Varieschi:2020hvp}.


\section{Poisson equation in Cartesian coordinates}\label{carcor}

The procedure to obtain the Poisson equation is to expand the metric as in \Eqq{getah} and expand to linear order the 00 component of the modified Einstein equations for a static configuration $h_{00}=-2\Phi({\bm x})$ and a non-relativistic pressureless matter source $S_{00}\simeq [(D-3)/(D-2)]\rho$. We assume that the measure weight $v$ is time-independent at the scales where the weak-field, static, non-relativistic approximation holds.

In the theory $T_1$ with $U=0$,
\be\label{pois1}
\frac{d-2}{d-1}\k^2\rho = \left(\N^2+\frac{\p^i v}{v}\p_i+\frac{2}{d-1}\frac{\N^2 v}{v}\right)\Phi-\frac{1}{d-1}h^{ij}\frac{\p_i\p_j v}{v}\,.
\ee

In the theory $T_v$,
\ba
\frac{d-2}{d-1}\k^2\rho &=& \left[\N^2+\frac{d-1}{2}\p^i\vp\p_i+\frac{d-1-2\Om}{2}\p_i\vp\p^i\vp+\N^2\vp\right]\Phi\nonumber\\
&&\quad+\frac12h^{ij}\left(\frac{d-1-2\Om}{2}\p_i\vp\p_j\vp-\p_i\p_j\vp\right)+\frac{2U}{d-1}\,.\label{pois5}
\ea

In the theory $T_q$, following the standard calculation in Einstein's theory, it is easy to see that the Poisson equation is
\be\label{pois2}
\frac{d-2}{d-1}\k^2\rho={}^q\N^2\Phi\,,
\ee
where the Laplacian is made of $q$-derivatives \Eq{qder}.


\section{Circular velocity in the theory \texorpdfstring{$T_v$}{Tv}}\label{app0}

Recall that velocities in the theory $T_v$ are defined through the weighted derivative $\cD_t=v^{-1/2}_0(t)\,\p_t[v^{1/2}_0(t)\,\cdot\,]$, where $v_0(t)$ is the time-dependent part of the factorizable measure. In particular, the angular velocity is $v_\om:=\cD_t\theta$, where $\theta$ is an angle. In this theory, we can always transform to an ``integer frame'' where classical mechanics or field theory formally reduce to the standard ones with ordinary derivatives. Physical observables are calculated in the original frame with $\cD$ derivatives.

Therefore, we can define an (unphysical) angular velocity in the integer frame
\be
\tilde v_\om = \sqrt{v(t)}\, v_\om = \frac{\rmd\tilde\theta}{\rmd t}\,,
\ee
where $\tilde\theta= \sqrt{v_0(t)}\,\theta$. The circular velocity associated to a distance $r \rmd\tilde\theta$ is
\be
\tilde v_{\rm circ} = r\,\frac{\rmd\tilde\theta}{\rmd t} = r \tilde v_\om\,.
\ee
In turn, the change in the circular velocity while changing the angle infinitesimally is equal to $\tilde v_{\rm circ} \rmd\tilde\theta$, so that the corresponding acceleration is
\be
\tilde g = \tilde v_{\rm circ}\, \frac{\rmd\tilde\theta}{\rmd t} = \tilde v_{\rm circ}\, \tilde v_\om = \frac{\tilde v_{\rm circ}^2}{r}\,.
\ee
At this point, we must convert velocities and accelerations in the integer frame to physical velocities and accelerations in the fractional frame. Regarding velocity, similarly to the expression linking $v_\om$ and $\tilde v_\om$ one has 
\be
\tilde v_{\rm circ} = \sqrt{v_0(t)}\, v_{\rm circ}\,,
\ee
while
\be
\tilde g = \tilde v_{\rm circ}\,\tilde v_\om = [\sqrt{v_0(t)} v_{\rm circ}] [\sqrt{v_0(t)} v_\omega] = v_0(t) v_{\rm circ} v_\omega = v_0(t)\, g\,,
\ee
where we used $g = v_{\rm circ} v_\omega = v_{\rm circ}\, \cD_t\theta$. However, we also have $\tilde g =v_0(t)\, v_{\rm circ}^2/r$, hence $g = v_{\rm circ}^2/r$ and \Eqq{vcircTv}.


\section{Multipole expansion of inverse-power potentials}\label{appA}

In this appendix, we recall some useful formul\ae\ for the multipole expansion of inverse-power potentials
\be
\frac{1}{|\bm{x}-\bm{x'}|^\la}\,,\qquad \la\in\mathbb{R}\,,
\ee
in terms of cylindrical coordinates in $d$ spatial dimensions. Cylindrical coordinates are the axial radial distance $R=x_1^2+x_2^2$, the elevation $Z_i=x_i$ with $i=3,\dots, d$ and the azimuth $\phi$, such that $x_1=R\,\cos\phi$ and $x_2=R\,\sin\phi$.

For $\la=1$, one has \cite{CoTo,CRTBCB,Coh12}
\be\label{1r1}
\frac{1}{|\bm{x}-\bm{x'}|}=\frac{1}{\pi\sqrt{RR'}}\sum_{m=-\infty}^{+\infty}\rme^{\rmi m(\phi-\phi')}Q_{m-\frac12}\left[\frac{R^2+{R'}^2+\sum_i(Z_i-Z_i')^2}{2RR'}\right],
\ee
where $Q_s$ is the Legendre function of the second kind of order $s$. In terms of the hypergeometric functions \cite[formula 8.820.2]{GR}, 
\be
Q_s(z)=\frac{\sqrt{\pi}\,\G(s+1)}{2^{s+1}\G\left(s+\frac32\right)}\frac{1}{z^{s+1}}{}_2F_1\left(\frac{s+2}{2},\frac{s+1}{2};\frac{2s+3}{2};\frac{1}{z^2}\right).
\ee
Formula \Eq{1r1} is equivalent to another one, popular from an exercise of Jackson's textbook on electrodynamics \cite[exercise 3.16]{Jac98}:
\be\label{1r2}
\frac{1}{|\bm{x}-\bm{x'}|}=\sum_{m=-\infty}^{+\infty}\int_0^{+\infty}\rmd y\,\rme^{\rmi m(\phi-\phi')-|Z-Z'|y}J_m(Ry)\,J_m(R'y)\,,
\ee
where $|Z-Z'|=\sqrt{\sum_i(Z_i-Z_i')^2}$. In fact, eqs.\ \Eq{1r1} and \Eq{1r2} can be related to each other via the identity \cite[formula 6.612.3]{GR} 
\be
\frac{1}{\pi\sqrt{RR'}}Q_{m-\frac12}\left[\frac{R^2+{R'}^2+|Z-Z'|^2}{2RR'}\right]=\int_0^{+\infty}\rmd y\,\rme^{-|Z-Z'|y}J_m(Ry)\,J_m(R'y)\,,
\ee
after continuing it analytically beyond its validity range $m>-1/2$.

When $\la$ is real, one can make use of an expansion (which we derive below) in terms of Gegenbauer polynomials, valid for $\la>-1/2$ and $\la\neq 0$. Gegenbauer polynomials are the generating functionals of the expression \cite{GR,CoKa} 
\be
\frac{1}{(1-2za+a^2)^\la}=\sum_{n=0}^{+\infty}C_n^\la(z)\,a^n\,,\qquad |a|<1\,,\qquad 0\neq\la>-\frac12\,.
\ee
In terms of the hypergeometric functions \cite[formula 8.932.1]{GR}, 
\be\label{CiFi}
C^\la_n(z)=\frac{2^n\,\G(\la+n)}{n!\G(\la)}\,z^n{}_2F_1\left(-\frac{n}{2},-\frac{n-1}{2};1-n-\la;\frac{1}{z^2}\right).
\ee
Consider the distance in spherical coordinates
\be\label{temp}
|\bm{x}-\bm{x'}|=\sqrt{r^2+{r'}^2-2rr'\cos\g}=r_>\sqrt{1+\left(\frac{r_<}{r_>}\right)^2-2\frac{r_<}{r_>}\,\cos\g}\,,
\ee
where $r=|\bm{x}|$, $r'=|\bm{x'}|$, $\g$ is the angle between the vectors $\bm{x}$ and $\bm{x'}$, $\cos\g=\bm{x}\cdot\bm{x'}/rr'$, and $r_>$ (respectively, $r_<$) is the largest (smallest) between $r$ and $r'$. In $d=3$ dimensions, $\cos\g=\cos\theta\,\cos\theta'+\sin\theta\,\sin\theta'\,\cos(\vp-\vp')$. Taking an arbitrary inverse power of \Eq{temp} and identifying $a=r_</r_>$ and $z=\cos\g$, one gets \cite{CoKa}
\be\label{1r30}
\frac{1}{|\bm{x}-\bm{x'}|^\la}=\sum_{n=0}^{+\infty}C_n^\frac{\la}{2}(\cos\g)\,\frac{r_<^n}{r_>^{n+\la}}\,,\qquad 0\neq\la>-\frac12\,.
\ee
When $\la=1$, $C_n^{1/2}=P_n$ are the Legendre polynomials of the first kind and one recovers yet another formula for the inverse-power-law potential \cite{CRTBCB}:
\be\label{1r30r}
\frac{1}{|\bm{x}-\bm{x'}|}=\sum_{n=0}^{+\infty}P_n(\cos\g)\,\frac{r_<^n}{r_>^{n+1}}\,.
\ee
Finally, \Eqq{1r30} is transformed to a cylindrical coordinate system. In galaxy models in $d=3$ dimensions, 
 the discussion is limited to the $Z=0$ plane and the $\theta=0$ direction (so that $\cos\g=\cos\theta'$ in polar coordinates), in which case
\be\label{rR'}
r=R\,,\qquad r'=\sqrt{{R'}^2+{Z'}^2}\,,\qquad \cos\g=\frac{R'\cos\phi'}{\sqrt{{R'}^2+{Z'}^2}}\,.
\ee
The thick-disk expression in cylindrical coordinates of \Eq{1r30} is obtained by replacing \Eq{rR'} therein:
\be\label{1r3thick}
\frac{1}{|\bm{x}-\bm{x'}|^\la}=\sum_{n=0}^{+\infty}C_n^\frac{\la}{2}\left(\frac{R'}{\sqrt{{R'}^2+{Z'}^2}}\cos\phi'\right)\,\frac{R_<^n}{R_>^{n+\la}}\,,\qquad 0\neq\la>-\frac12\,.
\ee
In thin-disk models, the further constraint $Z'=0$ is imposed, so that \cite{Varieschi:2020dnd}
\be\label{1r3}
\frac{1}{|\bm{x}-\bm{x'}|^\la}=\sum_{n=0}^{+\infty}C_n^\frac{\la}{2}(\cos\phi')\,\frac{R_<^n}{R_>^{n+\la}}\,,\qquad 0\neq\la>-\frac12\,,
\ee
where $R_>={\rm max}(R,\sqrt{{R'}^2+{Z'}^2})$ and $R_<={\rm min}(R,\sqrt{{R'}^2+{Z'}^2})$.

When $\la$ is complex-valued, other expansions exist \cite{Coh11,Coh12}. We do not show them here because \Eqq{1r3} already covers all the values of $\la$ appearing in multi-fractional theories, except $\la=0$. In this case, one can use the expansion in polar coordinates \cite{CoKa}
\be\label{logxr}
\ln|\bm{x}-\bm{x'}|=\ln r_>-\sum_{n=1}^{+\infty}\frac{\cos(n\g)}{n}\left(\frac{r_<}{r_>}\right)^n,
\ee
or, as it is easy to convince oneself looking at \Eqq{temp} in cylindrical coordinates and in the $Z=0$ plane, the equivalent expression in cylindrical coordinates
\be\label{logx}
\ln|\bm{x}-\bm{x'}|=\ln R_>-\sum_{n=1}^{+\infty}\frac{\cos(n\phi')}{n}\left(\frac{R_<}{R_>}\right)^n.
\ee


\section{Useful formul\ae}\label{appB}


\subsection{Angular integrals of Gegenbauer polynomials}

Consider the integral
\be
\cB_\frac{n}{2}^\la(b):=\int_0^{2\pi}\frac{\rmd\phi'}{2\pi}\,C_{n}^\frac{\la}{2}(b\cos\phi')\,,
\ee
where $C_n^{\la/2}$ are Gegenbauer polynomials given by \Eqq{CiFi} and $0<b\leq 1$ is independent of $\phi'$. With the change of variables $z= b\cos\phi'$, the integral is recast as
\ba
\cB_\frac{n}{2}^\la(b) &=& \frac{1}{\pi}\int_{-b}^b\rmd z\,\frac{C_{n}^\frac{\la}{2}(z)}{\sqrt{b^2-z^2}}\nonumber\\
&=& \frac{2^{n}\,\G\left(\frac{\la}{2}+n\right)}{\pi\,n!\G\left(\frac{\la}{2}\right)}\int_{-b}^b\rmd z\,\frac{z^n}{\sqrt{b^2-z^2}}\,{}_2F_1\left(-\frac{n}{2},-\frac{n-1}{2};1-n-\la;\frac{1}{z^2}\right).
\ea
One can check that this integral is identically zero for odd $n$:
\be
\cB_\frac{n}{2}^\la(b)=0\,,\qquad n\,{\rm odd}\,,
\ee
while for even $n\to 2n$ we find
\be\label{cBnb}
\cB_n^\la(b)=\frac{1}{n!}\frac{\G\left(1-\frac{\la}{2}\right)}{\G\left(1-\frac{\la}{2}-n\right)}\,{}_2F_1\left(-n,\frac{\la}{2}+n;1;b^2\right)\,,\qquad \cB_0^\la(b)=1\,.
\ee
In particular, for $b=1$
\be\label{cBn}
\cB_n^\la := \cB_n^\la(1)=\left[\frac{\G\left(1-\frac{\la}{2}\right)}{n!\G\left(1-\frac{\la}{2}-n\right)}\right]^2= \frac{1}{2^{2n}(n!)^2}\prod_{p=0}^{n-1}(\la+2p)^2\,,\qquad \cB_0^\la=1\,.
\ee

Another integral of interest is
\be
\cE_\frac{n}{2}^\la:=\int_0^\pi\rmd\theta'\,\sin\theta'\,C_n^\frac{\la}{2}(\cos\theta')\,.
\ee
Calling $z= \cos\theta'$,
\be
\cE_\frac{n}{2}^\la = \int_{-1}^1\rmd z\,C_n^\frac{\la}{2}(z)=\frac{1+(-1)^n}{\G(2+n)}\frac{\G(\la-1+n)}{\G(\la-1)}\nonumber\,,
\ee
which vanishes for odd $n$. Therefore,
\be\label{cEn}
\cE_n^\la=\frac{2}{(2n+1)!}\frac{\G(\la-1+2n)}{\G(\la-1)}\,.
\ee


\subsection{Radial integrals}

Two radial integrals of interest are
\ba
\cC_n^{\b,\la}(R)&:=&\frac{1}{R_0}\int_0^{+\infty}\rmd R'\,\rme^{-\frac{R'}{R_0}}{R'}^{\b}\frac{R_<^{2n}}{R_>^{\la+2n}}\nonumber\\
&=&\frac{1}{R_0}\frac{1}{R^{\la+2n}}\int_0^R\rmd R'\,\rme^{-\frac{R'}{R_0}}{R'}^{\b+2n}+R^{2n}\int_R^{+\infty}\rmd R'\,\rme^{-\frac{R'}{R_0}}{R'}^{\b-\la-2n}\nonumber\\
&=& R_0^{\b-\la}\left\{\left(\frac{R_0}{R}\right)^{\la+2n}\left[\G(\b+1+2n)-\G\left(\b+1+2n,\frac{R}{R_0}\right)\right]\right.\nonumber\\
&&\qquad\quad\left.+\left(\frac{R}{R_0}\right)^{2n}\G\left(\b-\la+1-2n,\frac{R}{R_0}\right)\right\}\!,\label{cCn}
\ea
and
\ba
\cC_{\rm log}^\b(R)&:=&\frac{1}{R_0}\int_0^{+\infty}\rmd R'\,\rme^{-\frac{R'}{R_0}} {R'}^\b\,\ln \frac{R_>}{R_*}\nonumber\\
&=&\frac{1}{R_0}\left(\ln\frac{R}{R_*}\int_0^R\rmd R'\,\rme^{-\frac{R'}{R_0}} {R'}^\b+\int_R^{+\infty}\rmd R'\,\rme^{-\frac{R'}{R_0}} {R'}^\b\,\ln \frac{R'}{R_*}\right)\nonumber\\
&=&R_0^\b\left[\G(1+\b)\,\ln\frac{R}{R_*}+G_{20}^{30}\left(\left.\frac{R}{R_0}\right|~\begin{matrix}  & 1 & 1 \\ 0 & 0 & 1+\b\end{matrix}\right)\right],\label{Clog}
\ea
where $G$ is the Meijer function.


\section{Galaxy potential in the theory \texorpdfstring{$T_v$}{Tv} with exponential matter density}\label{appC}


\subsection{Calculation of \Eqqs{exfraphi} and \Eq{exfraphi2}}\label{appC1}

The disk contribution to the potential is
\ba
\Phi^{\rm disk}(R) &\stackrel{\textrm{\tiny \Eq{1r3}}}{=}& -\frac{1}{3\a-2}\frac{MG}{R_0^2}\int_0^{+\infty}\rmd R'\, {R'}^{3\a-2}\int_0^{2\pi}\frac{\rmd\phi'}{2\pi}\int_{-\infty}^{+\infty}\rmd Z'\,\rme^{-\frac{R'}{R_0}}\de(Z')\nonumber\\
&&\times \sum_{n=0}^{+\infty}C_n^\frac{3\a-2}{2}(\cos\phi')\,\frac{R_<^n}{R_>^{n+3\a-2}}\nonumber\\
&=& -\frac{1}{3\a-2}\frac{MG}{R_0^2}\sum_{n=0}^{+\infty}\int_0^{2\pi}\frac{\rmd\phi'}{2\pi}C_n^\frac{3\a-2}{2}(\cos\phi')\int_0^{+\infty}\rmd R'\, {R'}^{3\a-2}\rme^{-\frac{R'}{R_0}}\frac{R_<^n}{R_>^{n+3\a-2}}\nonumber\\
&\stackrel[\textrm{\tiny \Eq{cCn}}]{\textrm{\tiny \Eq{cBn}}}{=}& -\frac{1}{3\a-2}\frac{MG}{R_0}\sum_{n=0}^{+\infty}\cB_n^{3\a-2}\,\cC_n^{3\a-2,3\a-2}(R)\nonumber\\
&=& -\frac{1}{3\a-2}\frac{MG}{R_0}\sum_{n=0}^{+\infty}\cB_n^{3\a-2}\left\{\left(\frac{R}{R_0}\right)^{2n}\G\left(1-2n,\frac{R}{R_0}\right)\right.\nonumber\\
&&\left.+\left(\frac{R_0}{R}\right)^{2n+3\a-2}\left[\G(2n+3\a-1)-\G\left(2n+3\a-1,\frac{R}{R_0}\right)\right]\right\}\!.\label{potphimulti}
\ea
For $\a>0$, the series converges at all scales and, in particular, it converges to the $n=0$ mode at large scales (modes with $n>8\!-\!10$ do not contribute appreciably). Since our main interest is at large scales where Einstein's theory deviates from observations, we take the zero mode as an approximation:
\ba
\hspace{-.5cm}\Phi^{\rm disk}(R) \simeq -\frac{1}{3\a-2}\frac{MG}{R_0}\!\left\{\!\G\left(1,\frac{R}{R_0}\right)+\left(\frac{R_0}{R}\right)^{3\a-2}\!\left[\G(3\a-1)-\G\left(3\a-1,\frac{R}{R_0}\right)\right]\!\right\}\!.\label{potphimulti2}
\ea

When $\a=2/3$, the potential is \Eqq{exfraphi2}:
\ba
\Phi^{\rm disk}(R) &\stackrel{\textrm{\tiny \Eq{logx}}}{=}& \frac{G}{R_*}\int_\cV\rmd^3\bm{x'}\,\left|\frac{R'}{R_*}\right|^{-1}\rho(\bm{x'})\left[\ln \frac{R_>}{R_*}-\sum_{n=1}^{+\infty}\frac{\cos(n\phi')}{n}\left(\frac{R_<}{R_>}\right)^n\right]\nonumber\\
&=&\frac{MG}{R_0^2}\int_0^{+\infty}\rmd R'\,\rme^{-\frac{R'}{R_0}}\int_0^{2\pi}\frac{\rmd\phi'}{2\pi}\int_{-\infty}^{+\infty}\rmd Z'\,\de(Z')\nonumber\\
&&\times\left[\ln \frac{R_>}{R_*}-\sum_{n=1}^{+\infty}\frac{\cos(n\phi')}{n}\left(\frac{R_<}{R_>}\right)^n\right]\nonumber\\
&=& \frac{MG}{R_0^2}\int_0^{+\infty}\rmd R'\,\rme^{-\frac{R'}{R_0}}\left[\ln \frac{R_>}{R_*}-\sum_{n=1}^{+\infty}\left(\frac{R_<}{R_>}\right)^n\int_0^{2\pi}\frac{\rmd\phi'}{2\pi}\frac{\cos(n\phi')}{n}\right]\nonumber\\
&\stackrel{\textrm{\tiny \Eq{Clog}}}{=}& \frac{MG}{R_0}\, \cC_{\rm log}^0(R)\nonumber\\
&=& -\frac{MG}{R_0}\left[\ln \frac{R}{R_*}+\Gamma\left(0,\frac{R}{R_0}\right)\right],\label{potphilog}
\ea
where the angular integral is zero for all $n$.


\subsection{Calculation of \Eqqs{dephi1} and \Eq{dephi2}}\label{appC2}

The term \Eq{dephi1} has a purely fractional measure and the ordinary Green's function \Eq{newt}. Using \Eq{1r1b},
\ba
\de\Phi^{\rm disk}_1(R) &=& -\frac{MG}{R_0^2}\int_0^{+\infty}\rmd R'\, R'\left|\frac{R'}{R_*}\right|^{3(\a-1)}\int_0^{2\pi}\frac{\rmd\phi'}{2\pi}\int_{-\infty}^{+\infty}\rmd Z'\,\rme^{-\frac{R'}{R_0}}\de(Z')\nonumber\\
&&\times\sum_{m=-\infty}^{+\infty}\int_0^{+\infty}\rmd y\,\rme^{\rmi m(\phi-\phi')-|Z-Z'|y}J_m(Ry)\,J_m(R'y)\nonumber\\
&=&-\frac{MG}{R_0^2}\int_0^{+\infty}\rmd y\,J_0(Ry)\,\int_0^{+\infty}\rmd R'\, R'\left|\frac{R'}{R_*}\right|^{3(\a-1)}\,\rme^{-\frac{R'}{R_0}}J_0(R'y)\nonumber\\
&=&-MG\left|\frac{R_0}{R_*}\right|^{3(\a-1)}\G(3\a-1)\int_0^{+\infty}\rmd y\,J_0(Ry)\,{}_2F_1\left(\frac{3\a}{2},\frac{3\a-1}{2};1;-R_0^2y^2\right)\nonumber\\
&=&-\frac{MG}{R_0}\left|\frac{R}{R_*}\right|^{3(\a-1)}\G(3\a-1)\nonumber\\
&&\times\left[\frac{1}{3\a-2}\left|\frac{R}{R_0}\right|^{3(1-\a)} {}_2F_3\left(\frac12,\frac12;1,\frac{3-3\a}{2},\frac{4-3\a}{2};\frac{R^2}{4R_0^2}\right)\right.\nonumber\\
&& -\frac{\sqrt{\pi}}{2^{3\a-1}}\frac{\G\left(\frac{1-3\a}{2}\right)}{\G\left(\frac{2-3\a}{2}\right)\G\left(\frac{3\a-1}{2}\right)\G\left(\frac{3\a+1}{2}\right)}\left(\frac{R}{R_0}\right)^2\nonumber\\
&&\quad\times{}_2F_3\left(\frac{3\a}{2},\frac{3\a}{2};\frac32,\frac{3\a+1}{2},\frac{3\a+1}{2};\frac{R^2}{4R_0^2}\right)\nonumber\\
&&\left.-\frac{3\a\sqrt{\pi}}{2^{3\a}}\frac{\G\left(-\frac{3\a}{2}\right)}{\G\left(\frac{3-3\a}{2}\right)\G^2\left(\frac{3\a}{2}\right)}\frac{R}{R_0}\,{}_2F_3\left(\frac{3\a-1}{2},\frac{3\a-1}{2};\frac12,\frac{3\a}{2},\frac{3\a}{2};\frac{R^2}{4R_0^2}\right)\right]\!\!,\label{newt2}
\ea
where ${}_2F_3$ is the generalized hypergeometric function. This expression is singular for $\a=4/3$, in which case one has
\ba
\de\Phi^{\rm disk}_1(R) &=& -\frac{MG}{R_0^2R_*}\int_0^{+\infty}\rmd y\,J_0(Ry)\,\int_0^{+\infty}\rmd R'\,{R'}^2\,\rme^{-\frac{R'}{R_0}}J_0(R'y)\nonumber\\
&=& MG\frac{R_0}{R_*}\int_0^{+\infty}\rmd y\,J_0(Ry)\,\frac{R_0^2y^2-2}{(R_0^2y^2+1)^{5/2}}\nonumber\\
&=& \frac{2}{3\sqrt{\pi}}\frac{MG}{R_*}\left[G_{13}^{21}\left(\left.\frac{R^2}{4R_0^2}\right|~\begin{matrix}  & -\frac12 & \\ 0 & 1 & 0\end{matrix}\right)-2G_{13}^{21}\left(\left.\frac{R^2}{4R_0^2}\right|~\begin{matrix}  & \frac12 & \\ 0 & 2 & 0\end{matrix}\right)\right].
\ea

The term \Eq{dephi2} has an ordinary integration measure and a fractional Green's function. We can use the results reported in Appendix \ref{appA} about the expression of inverse powers in terms of Gegenbauer polynomials $C_n^\b$. Plugging \Eq{thinrho} and \Eq{1r3} into \Eq{dephi2}, we obtain
\ba
\de\Phi^{\rm disk}_2(R)\!\!&=&\!\frac{MG}{R_0^2}\frac{7-6\a}{(4-3\a)R_*^{3(\a-1)}}\int_0^{+\infty}\rmd R'\, R'\int_0^{2\pi}\frac{\rmd\phi'}{2\pi}\int_{-\infty}^{+\infty}\rmd Z'\,\rme^{-\frac{R'}{R_0}}\de(Z')\nonumber\\
\!\!&&\!\times\sum_{n=0}^{+\infty}C_n^\frac{4-3\a}{2}(\cos\phi')\,\frac{R_<^n}{R_>^{n+4-3\a}}\nonumber\\
\!\!&=&\!\frac{MG}{R_0^2}\frac{7-6\a}{(4-3\a)R_*^{3(\a-1)}}\sum_{n=0}^{+\infty}\int_0^{2\pi}\frac{\rmd\phi'}{2\pi}\,C_n^\frac{4-3\a}{2}(\cos\phi')\int_0^{+\infty}\rmd R'\, R'\,\rme^{-\frac{R'}{R_0}}\frac{R_<^n}{R_>^{n+4-3\a}}\nonumber\\
\!\!&\!\stackrel[\textrm{\tiny \Eq{cCn}}]{\textrm{\tiny \Eq{cBn}}}{=}&\frac{MG}{R_0}\frac{7-6\a}{(4-3\a)R_*^{3(\a-1)}}\sum_{n=0}^{+\infty}\cB_n^{4-3\a}\,\cC_n^{1,4-3\a}(R)\nonumber\\
\!\!&=&\!\frac{MG}{R_0}\frac{7-6\a}{4-3\a}\sum_{n=0}^{+\infty}\cB_n^{4-3\a}\left\{\left|\frac{R}{R_*}\right|^{3(\a-1)}\!\left(\frac{R_0}{R}\right)^{2n+1}\!\left[\G(2+2n)-\G\left(2+2n,\frac{R}{R_0}\right)\right]\right.\nonumber\\
\!\!&&\!\qquad\qquad\qquad\qquad\qquad\left.+\left|\frac{R_0}{R_*}\right|^{3(\a-1)}\!\left(\frac{R}{R_0}\right)^{2n}\G\!\left(3\a-2-2n,\frac{R}{R_0}\right)\right\}\!,\label{Phisolvir2bis}
\ea
where $\a\neq 4/3$, $r_*= R_*$ in the first line and $R_>$ and $R_<$ are, respectively, the largest and the smallest between $R$ and $R'$. Although we were unable to resum the series \Eq{Phisolvir2bis}, we checked that 
 higher-order terms with $n\neq 1$ are sub-dominant at large scales for all $\a>0$ with respect to the zero mode $n=0$,
\be
\de\Phi^{\rm disk}_2(R)\simeq\frac{MG}{R_0}\frac{7-6\a}{4-3\a}\left\{\left|\frac{R}{R_*}\right|^{3(\a-1)}\frac{R_0}{R}\left[1-\G\left(2,\frac{R}{R_0}\right)\right]+\left|\frac{R_0}{R_*}\right|^{3(\a-1)}\!\G\!\left(3\a-2,\frac{R}{R_0}\right)\right\}\!.\label{Phisolvir2bis2}
\ee
At small scales the series does not converge but, as explained above in the case of the extreme fractional limit, we are mainly focus on the large-scale behaviour.

When $\a=4/3$, we have to repeat the calculation of $\de\Phi^{\rm disk}_2$ with the logarithmic potential \Eq{filog} with $r_*=R_*$. On the $Z=0$ plane, using \Eqq{logx} we get
\ba
\de\Phi^{\rm disk}_2(R)&=& \frac{G}{R_*}\int_\cV\rmd^3\bm{x'}\,\rho(\bm{x'})\,\ln\frac{|\bm{x}-\bm{x'}|}{R_*}\label{dephi2log}\\
&=&\frac{MG}{R_*R_0^2}\int_0^{+\infty}\rmd R'\, R'\int_0^{2\pi}\frac{\rmd\phi'}{2\pi}\int_{-\infty}^{+\infty}\rmd Z'\,\rme^{-\frac{R'}{R_0}}\de(Z')\nonumber\\
&&\times\left[\ln \frac{R_>}{R_*}-\sum_{n=1}^{+\infty}\frac{\cos(n\phi')}{n}\left(\frac{R_<}{R_>}\right)^n\right]\nonumber\\
&\stackrel{\textrm{\tiny \Eq{Clog}}}{=}& \frac{MG}{R_*R_0}\, \cC_{\rm log}^1(R)\nonumber\\
&=&\frac{MG}{R_*}\left[\rme^{-\frac{R}{R_0}}+\ln\frac{R}{R_*}-{\rm Ei}\left(-\frac{R}{R_0}\right)\right],\label{logR}
\ea
where ${\rm Ei}$ is the exponential integral function.


\section{Thick-disk data-based galaxy potential and gravitational field in \texorpdfstring{$T_v$}{Tv}}\label{appD}


\subsection{Extreme fractional limit}\label{appD1}

\subsubsection{\texorpdfstring{$\a\neq 2/3$}{aneq23}}\label{appD11}

Plugging the thick-disk profile \Eq{thick} into \Eqq{potet}, for $\a\neq 2/3$ we get
\ban
\Phi^{\rm disk+gas}(R) &\stackrel{\textrm{\tiny \Eq{Phisolvir}}}{=}&-\frac{G}{(3\a-2)R_*}\int_\cV\rmd^3\bm{x'}\,\left|\frac{R'}{R_*}\right|^{3(\a-1)}\rho^{\rm disk+gas}(\bm{x'})\left|\frac{R_*}{\bm{x}-\bm{x'}}\right|^{3\a-2}\\
&\stackrel[\textrm{\tiny \Eq{1r3thick}}]{\textrm{\tiny \Eq{rR'}}}{=}& -\frac{2G}{3\a-2}\sum_{n=0}^{+\infty}\int_0^{+\infty}\rmd Z'\,\frac{\rme^{-\frac{Z'}{h_Z}}}{2h_Z}\int_0^{+\infty}\rmd R'\,\Sigma(R')\, {R'}^{3\a-2}\\
&&\times\frac{R_<^n}{R_>^{3\a-2+n}}\int_0^{2\pi}\rmd\phi'\,C_n^\frac{3\a-2}{2}\left(\sqrt{\frac{{R'}^2}{{R'}^2+{Z'}^2}}\cos\phi'\right)\\
&\stackrel{\textrm{\tiny \Eq{cBnb}}}{=}& -\frac{2\pi G}{3\a-2}\sum_{n=0}^{+\infty}\int_0^{+\infty}\rmd Z'\,\frac{\rme^{-\frac{Z'}{h_Z}}}{h_Z}\int_0^{+\infty}\rmd R'\,\Sigma(R')\, {R'}^{3\a-2}\\
&&\times\frac{R_<^{2n}}{R_>^{3\a-2+2n}}\,\cB_n^{3\a-2}\left(\sqrt{\frac{{R'}^2}{{R'}^2+{Z'}^2}}\right)\\
&=&-\frac{2\pi G}{3\a-2}\sum_{n=0}^{+\infty}\frac{1}{n!}\frac{\G\left(2-\frac{3\a}{2}\right)}{\G\left(2-\frac{3\a}{2}-n\right)}\\
&&\times\int_0^{+\infty}\rmd Z'\,\frac{\rme^{-\frac{Z'}{h_Z}}}{h_Z}\int_0^{+\infty}\rmd R'\,\Sigma(R')\, {R'}^{3\a-2}\frac{R_<^{2n}}{R_>^{3\a-2+2n}}\\
&&\times{}_2F_1\left(-n,\frac{3\a}{2}-1+n;1;\frac{{R'}^2}{{R'}^2+{Z'}^2}\right),
\ean
where $R_<$ ($R_>$) is the minimum (respectively, maximum) between $R$ and $\sqrt{{R'}^2+{Z'}^2}$. At this point, we split the integral in $Z'$ on the two intervals $[0,R]$ and $[R,+\infty)$. Consider first $Z'\in[0,R]$. If $R_>=R$ and $R_<=\sqrt{{R'}^2+{Z'}^2}$, then $R'\in[0,\sqrt{R^2-{Z'}^2}]$, while if $R_>=\sqrt{{R'}^2+{Z'}^2}$ and $R_<=R$, then $R'\in[\sqrt{R^2-{Z'}^2},+\infty)$. On the hand, when $Z'\in[R,+\infty)$, it is always true that $R<\sqrt{{R'}^2+{Z'}^2}$, for any real value of $R'$. Therefore,
\ba
\Phi^{\rm disk+gas}(R)&=&-\frac{2\pi G}{3\a-2}\sum_{n=0}^{+\infty}\frac{1}{n!}\frac{\G\left(2-\frac{3\a}{2}\right)}{\G\left(2-\frac{3\a}{2}-n\right)}\left\{\int_0^R\rmd Z'\,\frac{\rme^{-\frac{Z'}{h_Z}}}{h_Z}\right.\nonumber\\
&&\times\left[\frac{1}{R^{3\a-2+2n}}\int_0^{\sqrt{R^2-{Z'}^2}}\rmd R'\,{R'}^{3\a-2}\left({R'}^2+{Z'}^2\right)^n\right.\nonumber\\
&&\qquad\left.+R^{2n}\int_{\sqrt{R^2-{Z'}^2}}^{+\infty}\rmd R'\,\frac{{R'}^{3\a-2}}{\left({R'}^2+{Z'}^2\right)^\frac{3\a-2+2n}{2}}\right] \nonumber\\
&&\left.+\int_R^{+\infty}\rmd Z'\,\frac{\rme^{-\frac{Z'}{h_Z}}}{h_Z}R^{2n}\int_0^{+\infty}\rmd R'\,\frac{{R'}^{3\a-2}}{\left({R'}^2+{Z'}^2\right)^\frac{3\a-2+2n}{2}} \right\}\nonumber\\
&&\times\Sigma(R')\,{}_2F_1\left(-n,\frac{3\a}{2}-1+n;1;\frac{{R'}^2}{{R'}^2+{Z'}^2}\right),\label{diskgas1}
\ea
where
\be
{}_2F_1\left(-n,\frac{3\a}{2}-1+n;1;z\right) = \sum_{k=0}^n\binom{n}{k}\frac{\G\left(\frac{3\a}{2}-1+n+k\right)}{\G\left(\frac{3\a}{2}-1+n\right)}\frac{(-z)^k}{k!}\,.
\ee
Taking the first derivative in $R$ of \Eqq{diskgas1}, we get the disk-gas part of the gravitational field \Eq{gifi}
\ba
g^{\rm disk+gas}(R) &=& -\frac{\rmd\Phi^{\rm disk+gas}(R)}{\rmd R}-\frac{3(\a-1)}{2R}\Phi^{\rm disk+gas}(R)\nonumber\\
&=&\frac{2\pi G}{3\a-2}\sum_{n=0}^{+\infty}\frac{1}{n!}\frac{\G\left(2-\frac{3\a}{2}\right)}{\G\left(2-\frac{3\a}{2}-n\right)}\left\{
\int_0^R\rmd Z'\,\frac{\rme^{-\frac{Z'}{h_Z}}}{h_Z}\right.\nonumber\\
&&\times\left[-\frac{3\a-2+2n}{R^{3\a-1+2n}}\int_0^{\sqrt{R^2-{Z'}^2}}\rmd R'\,{R'}^{3\a-2}\left({R'}^2+{Z'}^2\right)^n\right.\nonumber\\
&&\qquad\left.+2n\,R^{2n-1}\int_{\sqrt{R^2-{Z'}^2}}^{+\infty}\rmd R'\,\frac{{R'}^{3\a-2}}{\left({R'}^2+{Z'}^2\right)^\frac{3\a-2+2n}{2}}\right] \nonumber\\
&&\left.+\int_R^{+\infty}\rmd Z'\,\frac{\rme^{-\frac{Z'}{h_Z}}}{h_Z}\,2n\,R^{2n-1}\int_0^{+\infty}\rmd R'\,\frac{{R'}^{3\a-2}}{\left({R'}^2+{Z'}^2\right)^\frac{3\a-2+2n}{2}} \right\}\nonumber\\
&&\times\Sigma(R')\,{}_2F_1\left(-n,\frac{3\a}{2}-1+n;1;\frac{{R'}^2}{{R'}^2+{Z'}^2}\right)-\frac{3(\a-1)}{2R}\Phi^{\rm disk+gas}(R)\nonumber\\
&=&\frac{\pi G}{3\a-2}\sum_{n=0}^{+\infty}\frac{1}{n!}\frac{\G\left(2-\frac{3\a}{2}\right)}{\G\left(2-\frac{3\a}{2}-n\right)}\left\{
\int_0^R\rmd Z'\,\frac{\rme^{-\frac{Z'}{h_Z}}}{h_Z}\right.\nonumber\\
&&\times\left[\frac{1-3\a-4n}{R^{3\a-1+2n}}\int_0^{\sqrt{R^2-{Z'}^2}}\rmd R'\,{R'}^{3\a-2}\left({R'}^2+{Z'}^2\right)^n\right.\nonumber\\
&&\qquad\left.+(4n+3\a-3)\,R^{2n-1}\int_{\sqrt{R^2-{Z'}^2}}^{+\infty}\rmd R'\,\frac{{R'}^{3\a-2}}{\left({R'}^2+{Z'}^2\right)^\frac{3\a-2+2n}{2}}\right] \nonumber\\
&&\left.+\int_R^{+\infty}\rmd Z'\,\frac{\rme^{-\frac{Z'}{h_Z}}}{h_Z}\,(4n+3\a-3)\,R^{2n-1}\int_0^{+\infty}\rmd R'\,\frac{{R'}^{3\a-2}}{\left({R'}^2+{Z'}^2\right)^\frac{3\a-2+2n}{2}} \right\}\nonumber\\
&&\times\Sigma(R')\,{}_2F_1\left(-n,\frac{3\a}{2}-1+n;1;\frac{{R'}^2}{{R'}^2+{Z'}^2}\right)\,.\label{gdiskgas1}
\ea

The bulge component of the potential is
\ba
\Phi^{\rm bulge}(r) &=& -\frac{G}{(3\a-2)r_*}\int_\cV\rmd^3\bm{x'}\,\left|\frac{r'}{r_*}\right|^{3(\a-1)}\rho^{\rm bulge}(r')\left|\frac{r_*}{\bm{x}-\bm{x'}}\right|^{3\a-2}\nonumber\\
&\stackrel{\textrm{\tiny \Eq{1r30}}}{=}& -\frac{2\pi G}{3\a-2}\sum_{n=0}^{+\infty}\int_0^\pi\rmd\theta'\,\sin\theta'\,C_n^\frac{3\a-2}{2}(\cos\theta')\int_0^{+\infty}\rmd r'\,\rho^{\rm bulge}(r')\,{r'}^{3\a-1}\,\frac{r_<^n}{r_>^{n+3\a-2}}\nonumber\\
&\stackrel{\textrm{\tiny \Eq{cEn}}}{=}& -\frac{4\pi G}{3\a-2}\sum_{n=0}^{+\infty}\frac{1}{(2n+1)!}\frac{\G(3\a-3+2n)}{\G(3\a-3)}\int_0^{+\infty}\rmd r'\,\rho^{\rm bulge}(r')\,{r'}^{3\a-1}\,\frac{r_<^{2n}}{r_>^{2n+3\a-2}}\nonumber\\
&=& -\frac{2\pi G}{(3\a-2)(3\a-4)r}\int_0^{+\infty}\!\rmd r'\,\rho^{\rm bulge}(r')\,{r'}^{3\a-2}\!\left[(r_>-r_<)^{4-3\a}-(r_>+r_<)^{4-3\a}\right]\nonumber\\
&=& -\frac{2\pi G}{(3\a-2)(3\a-4)r}\int_0^{r}\rmd r'\,\rho^{\rm bulge}(r')\,{r'}^{3\a-2}\left[(r-r')^{4-3\a}-(r+r')^{4-3\a}\right]\nonumber\\
&& -\frac{2\pi G}{(3\a-2)(3\a-4)r}\int_r^{+\infty}\rmd r'\,\rho^{\rm bulge}(r')\,{r'}^{3\a-2}\left[(r'-r)^{4-3\a}-(r'+r)^{4-3\a}\right].\nonumber\\\label{phibulgeTv}
\ea
The gravitational field is then
\ba
g^{\rm bulge}(R) &=& -\left.\frac{\rmd\Phi^{\rm bulge}(r)}{\rmd r}\right|_{r=R}-\frac{3(\a-1)}{2R}\Phi^{\rm bulge}(R)\nonumber\\
&=&\frac{2\pi G}{(3\a-2)(3\a-4)R^2}\int_0^{R}\rmd r'\,\rho^{\rm bulge}(r')\,{r'}^{3\a-2}\left\{[3(1-\a)R+r'](R-r')^{3(1-\a)}\right.\nonumber\\
&&\qquad\qquad\left.-[3(1-\a)R-r'](R+r')^{3(1-\a)}\right\}\nonumber\\
&& -\frac{2\pi G}{(3\a-2)(3\a-4)R^2}\int_R^{+\infty}\rmd r'\,\rho^{\rm bulge}(r')\,{r'}^{3\a-2}\nonumber\\
&&\qquad\qquad\times\left\{[3(1-\a)R+r'](r'-R)^{3(1-\a)}\right.\nonumber\\
&&\qquad\qquad\qquad\left.+[3(1-\a)R-r'](r'+R)^{3(1-\a)}\right\}+\frac{3(1-\a)}{2R}\Phi^{\rm bulge}(R)\nonumber\\
&=&\frac{\pi G}{(3\a-2)(3\a-4)R^2}\int_0^{R}\rmd r'\,\rho^{\rm bulge}(r')\,{r'}^{3\a-2}\nonumber\\
&&\qquad\qquad\times\left\{[3(1-\a)(R+r')+2r'](R-r')^{3(1-\a)}\right.\nonumber\\
&&\qquad\qquad\qquad\left.-[3(1-\a)(R-r')-2r'](R+r')^{3(1-\a)}\right\}\nonumber\\
&& -\frac{\pi G}{(3\a-2)(3\a-4)R^2}\int_R^{+\infty}\rmd r'\,\rho^{\rm bulge}(r')\,{r'}^{3\a-2}\nonumber\\
&&\qquad\qquad\times\left\{[3(1-\a)(R+r')+2r'](r'-R)^{3(1-\a)}\right.\nonumber\\
&&\qquad\qquad\qquad\left.+[3(1-\a)(R-r')-2r'](r'+R)^{3(1-\a)}\right\}\nonumber\\
&=&\frac{\pi G}{(3\a-2)(3\a-4)R^2}\nonumber\\
&&\times\left[\int_0^{R}\rmd r'\,\rho^{\rm bulge}(r')\,{r'}^{3\a-2}[3(1-\a)(R+r')+2r'](R-r')^{3(1-\a)}\right.\nonumber\\
&& \qquad-\int_R^{+\infty}\rmd r'\,\rho^{\rm bulge}(r')\,{r'}^{3\a-2}[3(1-\a)(R+r')+2r'](r'-R)^{3(1-\a)}\nonumber\\
&&\left.\qquad-\int_0^{+\infty}\rmd r'\,\rho^{\rm bulge}(r')\,{r'}^{3\a-2}[3(1-\a)(R-r')-2r'](R+r')^{3(1-\a)}\right]\!.\label{gbulgeTv}
\ea

\subsubsection{\texorpdfstring{$\a=2/3$}{a=23}}

Plugging the thick-disk profile \Eq{thick}, the potential \Eq{Phisolvir23} with $r_*=R_*$ and the series representation \Eq{logx} into \Eqq{potet}, for $\a=2/3$ we get
\ba
\Phi^{\rm disk+gas}(R) &=& \frac{G}{R_*}\int_\cV\rmd^3\bm{x'}\,\left|\frac{R'}{R_*}\right|^{-1}\rho^{\rm disk+gas}(\bm{x'})\,\ln\frac{|\bm{x}-\bm{x'}|}{R_*}\nonumber\\
&=& 2G\int_0^{+\infty}\rmd Z'\,\frac{\rme^{-\frac{Z'}{h_Z}}}{2h_Z}\int_0^{+\infty}\rmd R'\,\Sigma(R')\nonumber\\
&&\times\left[2\pi\ln \frac{R_>}{R_*}-\sum_{n=1}^{+\infty}\int_0^{2\pi}\rmd\phi'\frac{\cos(n\phi')}{n}\left(\frac{R_<}{R_>}\right)^n\right]\nonumber\\
&=& 2\pi G\int_0^{+\infty}\rmd Z'\,\frac{\rme^{-\frac{Z'}{h_Z}}}{h_Z}\int_0^{+\infty}\rmd R'\,\Sigma(R')\ln \frac{R_>}{R_*}\nonumber\\
&=& \pi G\int_0^{R}\rmd Z'\,\frac{\rme^{-\frac{Z'}{h_Z}}}{h_Z}\left[\ln \frac{R^2}{R_*^2}\int_0^{\sqrt{R^2-{Z'}^2}}\rmd R'\,\Sigma(R')\right.\nonumber\\
&&\quad\qquad\qquad\qquad\qquad\left.+\int_{\sqrt{R^2-{Z'}^2}}^{+\infty}\rmd R'\,\Sigma(R')\ln \frac{{R'}^2+{Z'}^2}{R_*^2}\right]\nonumber\\
&& +\pi G\int_R^{+\infty}\rmd Z'\,\frac{\rme^{-\frac{Z'}{h_Z}}}{h_Z}\left[\int_0^{+\infty}\rmd R'\,\Sigma(R')\,\ln \frac{{R'}^2+{Z'}^2}{R_*^2}\right].\label{phidisk23}
\ea
The associated gravitational field is
\ba
g^{\rm disk+gas}(R) &=& -\frac{\rmd\Phi^{\rm disk+gas}(R)}{\rmd R}+\frac{1}{2R}\Phi^{\rm disk+gas}(R)\nonumber\\
&=& -\frac{2\pi G}{R}\int_0^{R}\rmd Z'\,\frac{\rme^{-\frac{Z'}{h_Z}}}{h_Z}\int_0^{\sqrt{R^2-{Z'}^2}}\rmd R'\,\Sigma(R')+\frac{1}{2R}\Phi^{\rm disk+gas}(R)\,.\label{gdisk23}
\ea

The bulge component of the galactic potential is
\ba
\Phi^{\rm bulge}(r) &=& \frac{G}{r_*}\int_\cV\rmd^3\bm{x'}\,\left|\frac{r'}{r_*}\right|^{-1}\rho^{\rm bulge}(r')\ln\frac{|\bm{x}-\bm{x'}|}{r_*}\nonumber\\
&\stackrel{\textrm{\tiny \Eq{logxr}}}{=}& 2\pi G\int_0^\pi\rmd\theta'\,\sin\theta'\int_0^{+\infty}\rmd r'\,\rho^{\rm bulge}(r')\,r'
\left[\ln \frac{r_>}{r_*}-\sum_{n=1}^{+\infty}\frac{\cos(n\theta')}{n}\left(\frac{r_<}{r_>}\right)^n\right]\nonumber\\
&=&4\pi G\int_0^{+\infty}\rmd r'\,\rho^{\rm bulge}(r')\,r'
\left[\ln \frac{r_>}{r_*}-\sum_{n=1}^{+\infty}\frac{1}{2n(1-4 n^2)}\left(\frac{r_<}{r_>}\right)^{2n}\right]\nonumber\\
&=&2\pi G\int_0^{+\infty}\rmd r'\,\rho^{\rm bulge}(r')\,r'\nonumber\\
&&\times\left[-1+2\ln \frac{r_>}{r_*}+\frac{r^2+{r'}^2}{rr'}{\rm arctanh}\frac{r_<}{r_>}+\ln\left(1-\frac{r_<^2}{r_>^2}\right)\right]\nonumber\\
&=&2\pi G\int_0^{r}\rmd r'\,\rho^{\rm bulge}(r')\,r'\nonumber\\
&&\qquad\times\left[-1+2\ln \frac{r}{r_*}+\frac{r^2+{r'}^2}{rr'}{\rm arctanh}\frac{r'}{r}+\ln\left(1-\frac{{r'}^2}{r^2}\right)\right]\nonumber\\
&&+2\pi G\int_r^{+\infty}\rmd r'\,\rho^{\rm bulge}(r')\,r'\nonumber\\
&&\qquad\times\left[-1+2\ln \frac{r'}{r_*}+\frac{r^2+{r'}^2}{rr'}{\rm arctanh}\frac{r}{r'}+\ln\left(1-\frac{r^2}{{r'}^2}\right)\right].\label{phibulge23}
\ea
The gravitational potential reads
\ba
g^{\rm bulge}(R) \!&=&\! -\left.\frac{\rmd\Phi^{\rm bulge}(r)}{\rmd r}\right|_{r=R}+\frac{1}{2R}\Phi^{\rm bulge}(R)\nonumber\\
\!&=&\!-\frac{2\pi G}{R}\int_0^{R}\!\rmd r'\,\rho^{\rm bulge}(r')\,r'\!\left[1+\frac{R^2-{r'}^2}{Rr'}{\rm arctanh}\frac{r'}{R}\right]\nonumber\\
\!&&\!-\frac{2\pi G}{R}\int_R^{+\infty}\!\rmd r'\,\rho^{\rm bulge}(r')\,r'\!\left[1+\frac{R^2-{r'}^2}{Rr'}{\rm arctanh}\frac{R}{r'}\right]+\frac{1}{2R}\Phi^{\rm bulge}(R)\nonumber\\
\!&=&\!-\frac{\pi G}{R}\int_0^{R}\!\rmd r'\,\rho^{\rm bulge}(r')\,r'\!
\left[3-2\ln \frac{R}{R_*}-\ln\left(1-\frac{{r'}^2}{R^2}\right)+\frac{R^2-3{r'}^2}{Rr'}{\rm arctanh}\frac{r'}{R}\right]\nonumber\\
\!&&\!-\frac{\pi G}{R}\int_R^{+\infty}\!\rmd r'\,\rho^{\rm bulge}(r')\,r'\!
\left[3-2\ln \frac{r'}{R_*}-\ln\left(1-\frac{R^2}{{r'}^2}\right)+\frac{R^2-3{r'}^2}{Rr'}{\rm arctanh}\frac{R}{r'}\right]\!.\nonumber\\\label{gbulge23}
\ea

\subsubsection{Comparison with data}\label{appD1c}

\begin{figure}
\bc
\includegraphics[width=10cm]{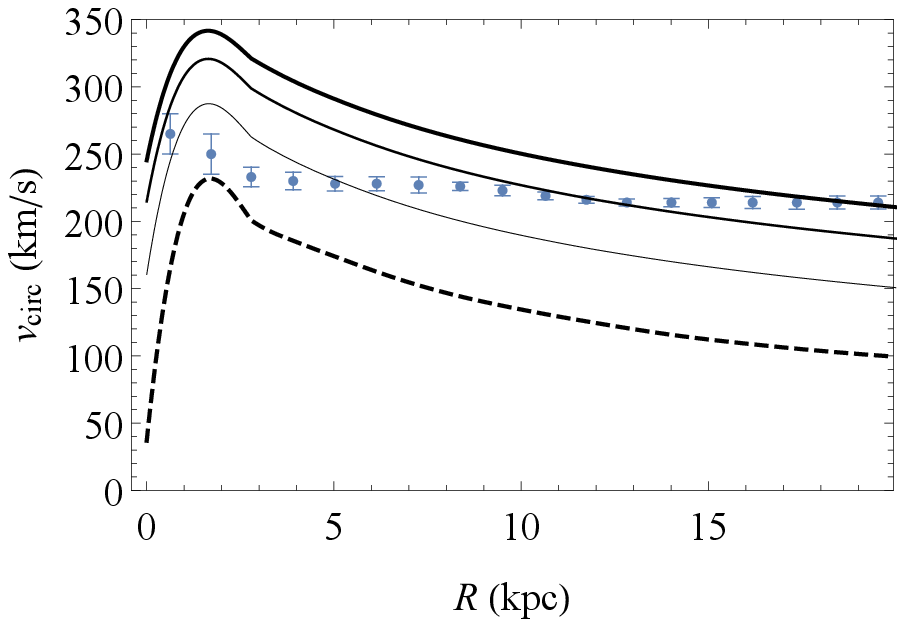}\\
\includegraphics[width=10cm]{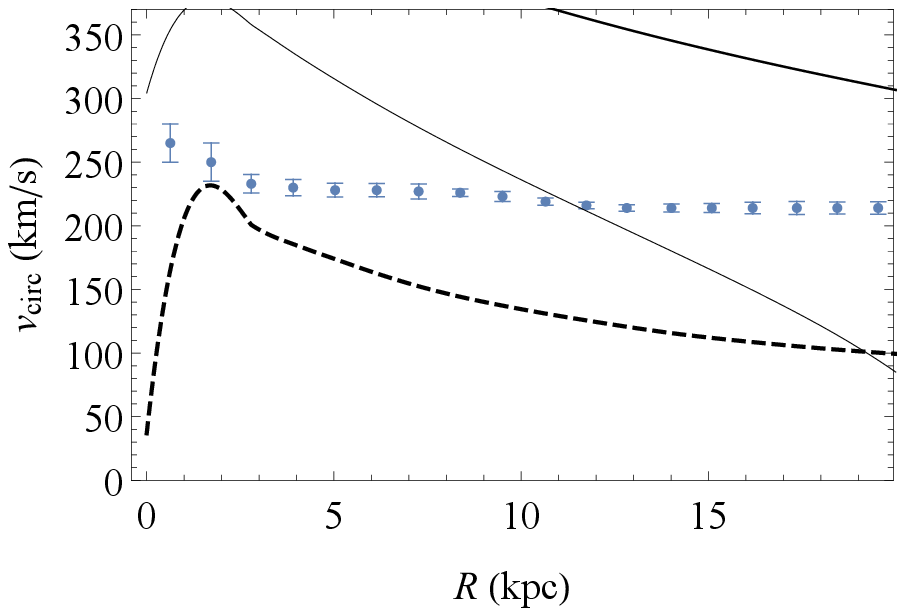}
\ec
\caption{\label{figA1} Comparison of the data points of NGC7814 with the $n=0$ rotation curve of the theory $T_v$ with weighted derivatives in the extreme fractional regime with a data-based matter density profile for $\a=1$ (Einstein's theory, dashed curve), $\a=0.90,0.86,0.84$ (top panel, increasing thickness) or $\a=2/3$ and $R_*=3,10\,{\rm kpc}$ (bottom panel, increasing thickness).}
\end{figure}

\begin{figure}
\bc
\includegraphics[width=10cm]{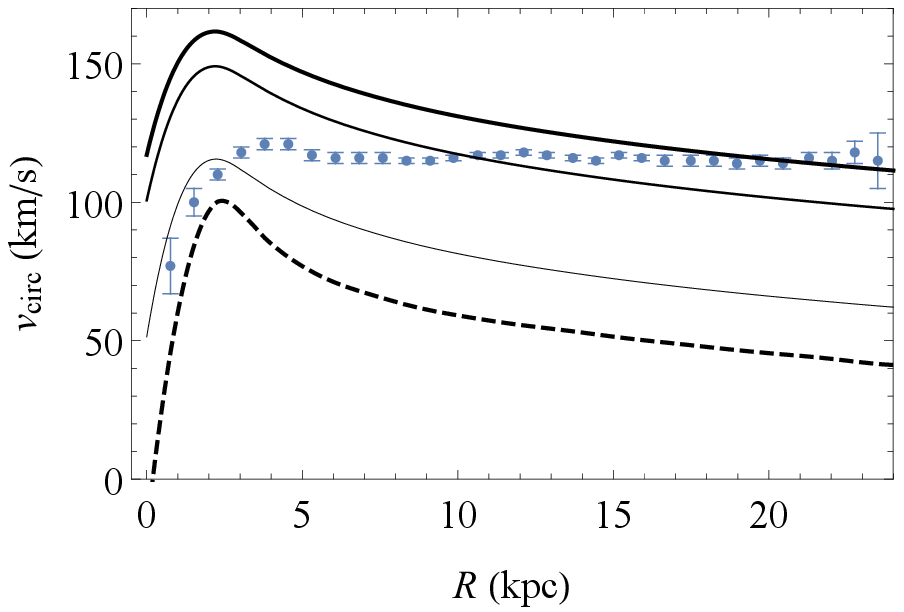}\\
\includegraphics[width=10cm]{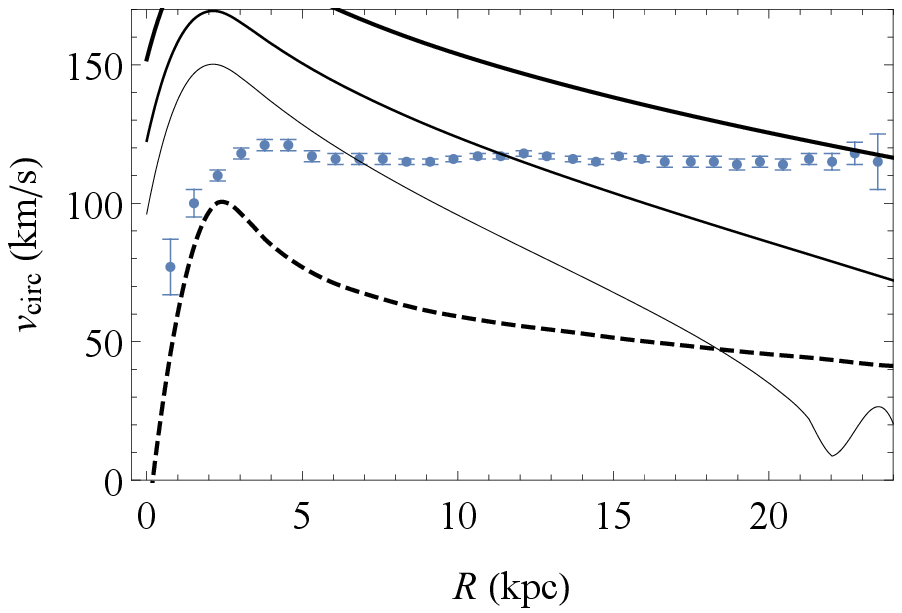}\\
\ec
\caption{\label{figA2} Comparison of the data points of NGC6503 with the $n=0$ rotation curve of the theory $T_v$ with weighted derivatives in the extreme fractional regime with a data-based matter density profile for $\a=1$ (Einstein's theory, dashed curve), $\a=0.90,0.82,0.80$ (top panel, increasing thickness) or $\a=2/3$ and $R_*=3,5,10\,{\rm kpc}$ (bottom panel, increasing thickness).}
\end{figure}

\begin{figure}
\bc
\includegraphics[width=10cm]{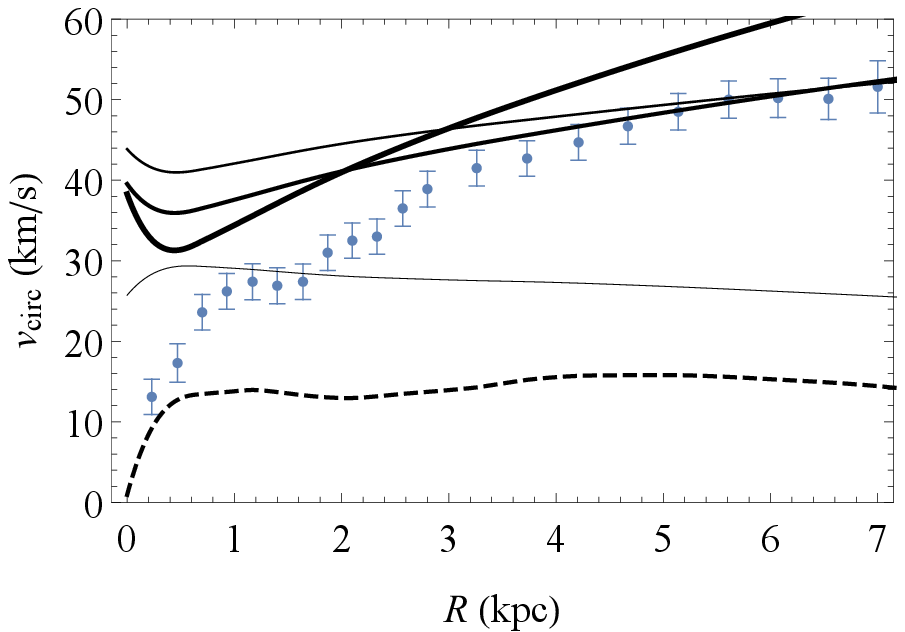}\\
\includegraphics[width=10cm]{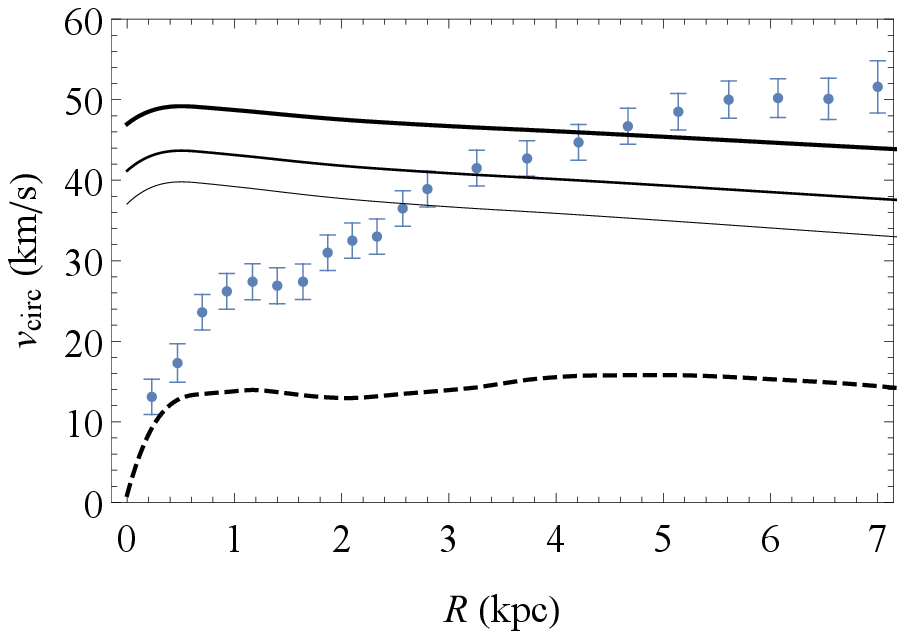}\\
\ec
\caption{\label{figA3} Comparison of the data points of NGC3741 with the $n=0$ rotation curve of the theory $T_v$ with weighted derivatives in the extreme fractional regime with a data-based matter density profile for $\a=1$ (Einstein's theory, dashed curve), $\a=0.80,0.55,0.50,0.40$ (top panel, increasing thickness) or $\a=2/3$ and $R_*=10,20,60\,{\rm kpc}$ (bottom panel, increasing thickness).}
\end{figure}

Regarding NGC7814, the model has some overlap with the data curve at $\a=0.84\!-\!0.86$, a range similar to the one found for NGC6503. As one can see in Fig.\ \ref{figA1}, as $\a$ decreases from 1 the curve is lifted and flattened, without, however, matching the data enough to warrant a more rigorous best-fit analysis. For NGC6503, as one varies $\a$ to values smaller than 1, the rotation curve is lifted up similarly to what depicted in Fig.\ \ref{fig3}, crossing the data at around $\a=0.80\!-\!0.82$ (Fig.\ \ref{figA2}), but it does not change much its shape for $\a>2/3$. For $\a<2/3$, the theoretical rotation curve changes drastically and away from the data points, as in Fig.~\ref{fig3}. Therefore, the plateau evidenced by the data is never reproduced. For NGC3741, the rotation curve is flat and low for $0.80<\a<1$ and reproduces the large-$R$ behaviour of the data for $\a=0.50\!-\!0.55$ (Fig.\ \ref{figA3}). However, the data points at small radii are never fitted, which may suggest one to look at a more refined model away from the extreme fractional regime. Negative results for all the three galaxies are obtained also when $\a=2/3$ and $R_*\gg R_0$ (i.e., for an $R_*$ consistent with the extreme fractional limit).


\subsection{Small fractional corrections}\label{appD2}

\subsubsection{\texorpdfstring{$\a\neq 4/3$}{aneq43}}

The contributions $\Phi^{\rm disk+gas}_0(R)+\Phi^{\rm bulge}_0(R)$ of the total potential \Eq{phitotTv} have already been calculated in section \ref{dmrev4} for Einstein's theory or, alternatively, in Appendix \ref{appD11} for $\a=1$:
\ba
\Phi^{\rm disk+gas}_0(R) &=& -G\int_\cV\rmd^3\bm{x'}\,\frac{\rho^{\rm disk+gas}(\bm{x'})}{|\bm{x}-\bm{x'}|}\nonumber\\
&=& -2\pi G\sum_{n=0}^{+\infty}\frac{1}{n!}\frac{\sqrt{\pi}}{\G\left(\frac{1}{2}-n\right)}\left\{\int_0^R\rmd Z'\,\frac{\rme^{-\frac{Z'}{h_Z}}}{h_Z}\right.\nonumber\\
&&\times\left[\frac{1}{R^{1+2n}}\int_0^{\sqrt{R^2-{Z'}^2}}\rmd R'\,R'\left({R'}^2+{Z'}^2\right)^n\right.\nonumber\\
&&\qquad\left.+R^{2n}\int_{\sqrt{R^2-{Z'}^2}}^{+\infty}\rmd R'\,\frac{R'}{\left({R'}^2+{Z'}^2\right)^\frac{1+2n}{2}}\right] \nonumber\\
&&\left.+\int_R^{+\infty}\rmd Z'\,\frac{\rme^{-\frac{Z'}{h_Z}}}{h_Z}R^{2n}\int_0^{+\infty}\rmd R'\,\frac{R'}{\left({R'}^2+{Z'}^2\right)^\frac{1+2n}{2}} \right\}\nonumber\\
&&\times\Sigma(R')\,{}_2F_1\left(-n,\frac{1}{2}+n;1;\frac{{R'}^2}{{R'}^2+{Z'}^2}\right),\label{potweak1}
\ea
and
\be
\Phi^{\rm bulge}_0(R) = -4\pi G\left[\frac{1}{R}\int_0^{R}\rmd r'\,\rho^{\rm bulge}(r') \,{r'}^2+\int_R^{+\infty}\rmd r'\,\rho^{\rm bulge}(r')\,r'\right].\label{potweak2}
\ee
The gravitational fields associated to these potential components are
\ba
\hspace{-.8cm}g^{\rm disk+gas}_0(R) &=& -\frac{\rmd\Phi^{\rm disk+gas}_0(R)}{\rmd R}-\frac{3(\a-1)}{2R}\left|\frac{R}{R_*}\right|^{3(\a-1)}\Phi^{\rm disk+gas}_0(R)\nonumber\\
\hspace{-.8cm}&=&2\pi G\sum_{n=0}^{+\infty}\frac{1}{n!}\frac{\sqrt{\pi}}{\G\left(\frac{1}{2}-n\right)}\left\{
\int_0^R\rmd Z'\,\frac{\rme^{-\frac{Z'}{h_Z}}}{h_Z}\right.\nonumber\\
\hspace{-.8cm}&&\times\left[-\frac{1+2n}{R^{2(1+n)}}\int_0^{\sqrt{R^2-{Z'}^2}}\rmd R'\,R'\left({R'}^2+{Z'}^2\right)^n\right.\nonumber\\
\hspace{-.8cm}&&\qquad\left.+2n\,R^{2n-1}\int_{\sqrt{R^2-{Z'}^2}}^{+\infty}\rmd R'\,\frac{R'}{\left({R'}^2+{Z'}^2\right)^\frac{1+2n}{2}}\right] \nonumber\\
\hspace{-.8cm}&&\left.+\int_R^{+\infty}\rmd Z'\,\frac{\rme^{-\frac{Z'}{h_Z}}}{h_Z}\,2n\,R^{2n-1}\int_0^{+\infty}\rmd R'\,\frac{R'}{\left({R'}^2+{Z'}^2\right)^\frac{1+2n}{2}} \right\}\nonumber\\
\hspace{-.8cm}&&\times\Sigma(R')\,{}_2F_1\left(-n,\frac{1}{2}+n;1;\frac{{R'}^2}{{R'}^2+{Z'}^2}\right)-\frac{3(\a-1)}{2R}\Phi^{\rm disk+gas}_0(R)\,,\label{gweak1}
\ea
and
\ba
g^{\rm bulge}_0(R) &=& -\frac{\rmd\Phi^{\rm bulge}_0(R)}{\rmd R}-\frac{3(\a-1)}{2R}\Phi^{\rm bulge}_0(R)\nonumber\\
&=&-\frac{4\pi G}{R^2}\int_0^{R}\rmd r'\,\rho^{\rm bulge}(r')\,{r'}^2-\frac{3(\a-1)}{2R}\,\Phi^{\rm bulge}(R)\,.\label{gweak2}
\ea

The potential term \Eq{dephi1} is simply given by \Eqqs{potweak1} and \Eq{potweak2} with a measure factor inserted in the radial integral,
\ba
\de\Phi^{\rm disk+gas}_1(R) &=& -G\int_\cV\rmd^3\bm{x'}\,\left|\frac{R'}{R_*}\right|^{3(\a-1)}\frac{\rho^{\rm disk+gas}(\bm{x'})}{|\bm{x}-\bm{x'}|}\nonumber\\
&=& -2\pi GR_*\sum_{n=0}^{+\infty}\frac{1}{n!}\frac{\sqrt{\pi}}{\G\left(\frac{1}{2}-n\right)}\left\{\int_0^R\rmd Z'\,\frac{\rme^{-\frac{Z'}{h_Z}}}{h_Z}\right.\nonumber\\
&&\times\left[\frac{1}{R^{1+2n}}\int_0^{\sqrt{R^2-{Z'}^2}}\rmd R'\left|\frac{R'}{R_*}\right|^{3\a-2}\left({R'}^2+{Z'}^2\right)^n\right.\nonumber\\
&&\qquad\left.+R^{2n}\int_{\sqrt{R^2-{Z'}^2}}^{+\infty}\rmd R'\left|\frac{R'}{R_*}\right|^{3\a-2}\frac{1}{\left({R'}^2+{Z'}^2\right)^\frac{1+2n}{2}}\right] \nonumber\\
&&\left.+\int_R^{+\infty}\rmd Z'\,\frac{\rme^{-\frac{Z'}{h_Z}}}{h_Z}R^{2n}\int_0^{+\infty}\rmd R'\left|\frac{R'}{R_*}\right|^{3\a-2}\frac{1}{\left({R'}^2+{Z'}^2\right)^\frac{1+2n}{2}} \right\}\nonumber\\
&&\times\Sigma(R')\,{}_2F_1\left(-n,\frac{1}{2}+n;1;\frac{{R'}^2}{{R'}^2+{Z'}^2}\right),\label{potweak3}\\
\de\Phi^{\rm bulge}_1(R) &=&-4\pi GR_*\left[\frac{R_*}{R}\int_0^{R}\rmd r'\left|\frac{r'}{R_*}\right|^{3\a-1}\rho^{\rm bulge}(r') +\int_R^{+\infty}\rmd r'\,\left|\frac{r'}{R_*}\right|^{3\a-2}\rho^{\rm bulge}(r')\right]\!,\nonumber\\
\label{potweak4}
\ea
where $r_*=R_*$, while the gravitational fields are
\ba
g^{\rm disk+gas}_1(R) &=& -\frac{\rmd\de\Phi^{\rm disk+gas}_1(R)}{\rmd R}\nonumber\\
&=&2\pi GR_*\sum_{n=0}^{+\infty}\frac{1}{n!}\frac{\sqrt{\pi}}{\G\left(\frac{1}{2}-n\right)}\left\{
\int_0^R\rmd Z'\,\frac{\rme^{-\frac{Z'}{h_Z}}}{h_Z}\right.\nonumber\\
&&\times\left[-\frac{1+2n}{R^{2(1+n)}}\int_0^{\sqrt{R^2-{Z'}^2}}\rmd R'\left|\frac{R'}{R_*}\right|^{3\a-2}\left({R'}^2+{Z'}^2\right)^n\right.\nonumber\\
&&\qquad\left.+2n\,R^{2n-1}\int_{\sqrt{R^2-{Z'}^2}}^{+\infty}\rmd R'\left|\frac{R'}{R_*}\right|^{3\a-2}\frac{1}{\left({R'}^2+{Z'}^2\right)^\frac{1+2n}{2}}\right] \nonumber\\
&&\left.+\int_R^{+\infty}\rmd Z'\,\frac{\rme^{-\frac{Z'}{h_Z}}}{h_Z}\,2n\,R^{2n-1}\int_0^{+\infty}\rmd R'\left|\frac{R'}{R_*}\right|^{3\a-2}\frac{1}{\left({R'}^2+{Z'}^2\right)^\frac{1+2n}{2}} \right\}\nonumber\\
&&\times\Sigma(R')\,{}_2F_1\left(-n,\frac{1}{2}+n;1;\frac{{R'}^2}{{R'}^2+{Z'}^2}\right)\,,\label{gweak3}\\
g^{\rm bulge}_1(R) &=& -\frac{\rmd\de\Phi^{\rm bulge}_1(R)}{\rmd R}=-4\pi G\frac{R_*^2}{R^2}\int_0^{R}\rmd r'\left|\frac{r'}{R_*}\right|^{3\a-1}\rho^{\rm bulge}(r')\,.\label{gweak4}
\ea

Finally, the potential term \Eq{dephi2} is
\ba
\de\Phi^{\rm disk+gas}_2(R) &=&-G\frac{7-6\a}{(3\a-4)R_*^{3(\a-1)}}\int_\cV\rmd^3\bm{x'}\,\frac{\rho^{\rm disk+gas}(\bm{x'})}{|\bm{x}-\bm{x'}|^{4-3\a}}\nonumber\\
&\stackrel{\textrm{\tiny \Eq{1r3thick}}}{=}&-G\frac{7-6\a}{(3\a-4)R_*^{3(\a-1)}}\sum_{n=0}^{+\infty}\int_0^{+\infty}\rmd Z'\,\frac{\rme^{-\frac{Z'}{h_Z}}}{h_Z}\int_0^{+\infty}\rmd R'\,\Sigma(R')\nonumber\\
&&\times\frac{R_<^{2n}}{R_>^{3\a-2+2n}}\int_0^{2\pi}\rmd\phi'\,C_n^\frac{3\a-2}{2}\left(\sqrt{\frac{{R'}^2}{{R'}^2+{Z'}^2}}\cos\phi'\right)\nonumber\\
&\stackrel{\textrm{\tiny \Eq{cBnb}}}{=}& -2\pi G\frac{7-6\a}{(3\a-4)R_*^{3(\a-1)}}\sum_{n=0}^{+\infty}\frac{1}{n!}\frac{\G\left(2-\frac{3\a}{2}\right)}{\G\left(2-\frac{3\a}{2}-n\right)}\nonumber\\
&&\times\int_0^{+\infty}\rmd Z'\,\frac{\rme^{-\frac{Z'}{h_Z}}}{h_Z}\int_0^{+\infty}\rmd R'\,\Sigma(R')\frac{R_<^{2n}}{R_>^{3\a-2+2n}}\nonumber\\
&&\times{}_2F_1\left(-n,\frac{3\a}{2}-1+n;1;\frac{{R'}^2}{{R'}^2+{Z'}^2}\right)\nonumber\\
&=& -2\pi G\frac{7-6\a}{(3\a-4)R_*^{3(\a-1)}}\sum_{n=0}^{+\infty}\frac{1}{n!}\frac{\G\left(2-\frac{3\a}{2}\right)}{\G\left(2-\frac{3\a}{2}-n\right)}\left\{\int_0^R\rmd Z'\,\frac{\rme^{-\frac{Z'}{h_Z}}}{h_Z}\right.\nonumber\\
&&\left[\frac{1}{R^{3\a-2+2n}}\int_0^{\sqrt{R^2-{Z'}^2}}\rmd R'\,\left({R'}^2+{Z'}^2\right)^n\right.\nonumber\\
&&\left.\qquad+R^{2n}\int_{\sqrt{R^2-{Z'}^2}}^{+\infty}\frac{\rmd R'}{\left({R'}^2+{Z'}^2\right)^{\frac{3\a-2+2n}{2}}}\right]\nonumber\\
&&\left.+\int_R^{+\infty}\rmd Z'\,\frac{\rme^{-\frac{Z'}{h_Z}}}{h_Z}R^{2n}\int_0^{+\infty}\frac{\rmd R'}{\left({R'}^2+{Z'}^2\right)^{\frac{3\a-2+2n}{2}}}\right\}\nonumber\\
&&\times\Sigma(R')\,{}_2F_1\left(-n,\frac{3\a}{2}-1+n;1;\frac{{R'}^2}{{R'}^2+{Z'}^2}\right),\label{potweak5}
\ea
with bulge component
\ba
\de\Phi^{\rm bulge}_2(R) &\stackrel{\textrm{\tiny \Eq{1r30}}}{=}& -2\pi G\frac{7-6\a}{(3\a-4)R_*^{3(\a-1)}}\sum_{n=0}^{+\infty}\int_0^{\pi}\rmd\theta'\,\sin\theta'\,C_n^\frac{4-3\a}{2}(\cos\theta')\nonumber\\
&&\times\int_0^{+\infty}\rmd r'\,\rho^{\rm bulge}(r')\, {r'}^2\frac{r_<^n}{r_>^{n+4-3\a}}\nonumber\\
&\stackrel{\textrm{\tiny \Eq{cEn}}}{=}&-2\pi G\frac{7-6\a}{(3\a-4)R_*^{3(\a-1)}}\int_0^{+\infty}\rmd r'\,\rho^{\rm bulge}(r')\,\frac{{r'}^2}{r_>^{4-3\a}}\nonumber\\
&&\times\sum_{n=0}^{+\infty}\frac{2}{(2n+1)!}\frac{\G(3-3\a+2n)}{\G(3-3\a)}\left(\frac{r_<}{r_>}\right)^{2n}\nonumber\\
&=&-2\pi G\frac{7-6\a}{(3\a-4)(3\a-2)R_*^{3(\a-1)}}\int_0^{+\infty}\rmd r'\,\rho^{\rm bulge}(r')\,\frac{{r'}^2}{r_>^{4-3\a}}\nonumber\\
&&\times\frac{r_>^{3-3\a}}{r_<}\left[(r_>+r_<)^{3\a-2}-(r_>-r_<)^{3\a-2}\right]\nonumber\\
&=&-2\pi G\frac{7-6\a}{(3\a-4)(3\a-2)R_*^{3(\a-1)}}\frac{1}{R}\int_0^{+\infty}\rmd r'\,\rho^{\rm bulge}(r')\,r'\nonumber\\
&&\times\left[(R+r')^{3\a-2}-(r_>-r_<)^{3\a-2}\right]\nonumber\\
&=&-2\pi G\frac{7-6\a}{(3\a-4)(3\a-2)R_*^{3(\a-1)}}\frac{1}{R}\left[\int_0^{+\infty}\rmd r'\,\rho^{\rm bulge}(r')\,r'\,(R+r')^{3\a-2}\right.\nonumber\\
&&\left.-\int_0^{R}\rmd r'\,\rho^{\rm bulge}(r')\,r'\,(R-r')^{3\a-2}-\int_R^{+\infty}\rmd r'\,\rho^{\rm bulge}(r')\,r'\,(r'-R)^{3\a-2}\right]\!.\nonumber\\\label{potweak6}
\ea
The gravitational fields are
\ba
g^{\rm disk+gas}_2(R) &=& -\frac{\rmd\de\Phi^{\rm disk+gas}_2(R)}{\rmd R}\nonumber\\
&=& 2\pi G\frac{7-6\a}{(3\a-4)R_*^{3(\a-1)}}\sum_{n=0}^{+\infty}\frac{1}{n!}\frac{\G\left(2-\frac{3\a}{2}\right)}{\G\left(2-\frac{3\a}{2}-n\right)}\left\{\int_0^R\rmd Z'\,\frac{\rme^{-\frac{Z'}{h_Z}}}{h_Z}\right.\nonumber\\
&&\left[\frac{2-3\a-2n}{R^{3\a-1+2n}}\int_0^{\sqrt{R^2-{Z'}^2}}\rmd R'\,\left({R'}^2+{Z'}^2\right)^n\right.\nonumber\\
&&\left.\qquad+2n\,R^{2n-1}\int_{\sqrt{R^2-{Z'}^2}}^{+\infty}\frac{\rmd R'}{\left({R'}^2+{Z'}^2\right)^{\frac{3\a-2+2n}{2}}}\right]\nonumber\\
&&\left.+\int_R^{+\infty}\rmd Z'\,\frac{\rme^{-\frac{Z'}{h_Z}}}{h_Z}2n\,R^{2n-1}\int_0^{+\infty}\frac{\rmd R'}{\left({R'}^2+{Z'}^2\right)^{\frac{3\a-2+2n}{2}}}\right\}\nonumber\\
&&\times\Sigma(R')\,{}_2F_1\left(-n,\frac{3\a}{2}-1+n;1;\frac{{R'}^2}{{R'}^2+{Z'}^2}\right),\label{gweak5}\\
g^{\rm bulge}_2(R) &=& -\frac{\rmd\de\Phi^{\rm bulge}_2(R)}{\rmd R}\nonumber\\
&=&2\pi G\frac{7-6\a}{(3\a-4)(3\a-2)R_*^{3(\a-1)}}\frac{1}{R^2}
\nonumber\\
&&\times\left\{\int_0^{+\infty}\rmd r'\,\rho^{\rm bulge}(r')\,r'[3(\a-1)R-r']\,(R+r')^{3(\a-1)}\right.\nonumber\\
&&\qquad-\int_0^{R}\rmd r'\,\rho^{\rm bulge}(r')\,r'[3(\a-1)R+r']\,(R-r')^{3(\a-1)}\nonumber\\
&&\qquad\left.+\int_R^{+\infty}\rmd r'\,\rho^{\rm bulge}(r')\,r'[3(\a-1)R+r']\,(r'-R)^{3(\a-1)}\right\}.\label{gweak6}
\ea

\subsubsection{\texorpdfstring{$\a=4/3$}{a=43}}

At lowest order, the potential is given by \Eqqs{potweak1} and \Eq{potweak2}, while the gravitational field is \Eqqs{gweak1} and \Eq{gweak2} with $\a=4/3$:
\ba
g^{\rm disk+gas}_0(R) &=&2\pi G\sum_{n=0}^{+\infty}\frac{1}{n!}\frac{\sqrt{\pi}}{\G\left(\frac{1}{2}-n\right)}\left\{
\int_0^R\rmd Z'\,\frac{\rme^{-\frac{Z'}{h_Z}}}{h_Z}\right.\nonumber\\
&&\times\left[-\frac{1+2n}{R^{2(1+n)}}\int_0^{\sqrt{R^2-{Z'}^2}}\rmd R'\,R'\left({R'}^2+{Z'}^2\right)^n\right.\nonumber\\
&&\qquad\left.+2n\,R^{2n-1}\int_{\sqrt{R^2-{Z'}^2}}^{+\infty}\rmd R'\,\frac{R'}{\left({R'}^2+{Z'}^2\right)^\frac{1+2n}{2}}\right] \nonumber\\
&&\left.+\int_R^{+\infty}\rmd Z'\,\frac{\rme^{-\frac{Z'}{h_Z}}}{h_Z}\,2n\,R^{2n-1}\int_0^{+\infty}\rmd R'\,\frac{R'}{\left({R'}^2+{Z'}^2\right)^\frac{1+2n}{2}} \right\}\nonumber\\
&&\times\Sigma(R')\,{}_2F_1\left(-n,\frac{1}{2}+n;1;\frac{{R'}^2}{{R'}^2+{Z'}^2}\right)-\frac{1}{2R}\Phi^{\rm disk+gas}_0(R)\,,\label{gweak7}\\
g^{\rm bulge}_0(R) &=&-\frac{4\pi G}{R^2}\int_0^{R}\rmd r'\,\rho^{\rm bulge}(r')\,{r'}^2-\frac{1}{2R}\,\Phi^{\rm bulge}(R)\,.\label{gweak8}
\ea
The next terms in the expansion are \Eq{potweak3} and \Eq{potweak4} with $\a=4/3$:
\ba
\de\Phi^{\rm disk+gas}_1(R) &=& -G\int_\cV\rmd^3\bm{x'}\,\frac{R'}{R_*}\frac{\rho^{\rm disk+gas}(\bm{x'})}{|\bm{x}-\bm{x'}|}\nonumber\\
&=& -2\pi GR_*\sum_{n=0}^{+\infty}\frac{1}{n!}\frac{\sqrt{\pi}}{\G\left(\frac{1}{2}-n\right)}\left\{\int_0^R\rmd Z'\,\frac{\rme^{-\frac{Z'}{h_Z}}}{h_Z}\right.\nonumber\\
&&\times\left[\frac{1}{R^{1+2n}}\int_0^{\sqrt{R^2-{Z'}^2}}\rmd R'\frac{{R'}^2}{R_*^2}\left({R'}^2+{Z'}^2\right)^n\right.\nonumber\\
&&\qquad\left.+R^{2n}\int_{\sqrt{R^2-{Z'}^2}}^{+\infty}\rmd R'\frac{{R'}^2}{R_*^2}\frac{1}{\left({R'}^2+{Z'}^2\right)^\frac{1+2n}{2}}\right] \nonumber\\
&&\left.+\int_R^{+\infty}\rmd Z'\,\frac{\rme^{-\frac{Z'}{h_Z}}}{h_Z}R^{2n}\int_0^{+\infty}\rmd R'\frac{{R'}^2}{R_*^2}\frac{1}{\left({R'}^2+{Z'}^2\right)^\frac{1+2n}{2}} \right\}\nonumber\\
&&\times\Sigma(R')\,{}_2F_1\left(-n,\frac{1}{2}+n;1;\frac{{R'}^2}{{R'}^2+{Z'}^2}\right)\!,\label{potweak9}\\
\de\Phi^{\rm bulge}_1(R) &=&-4\pi GR_*\left[\frac{R_*}{R}\int_0^{R}\rmd r'\frac{{r'}^3}{R_*^3}\rho^{\rm bulge}(r') +\int_R^{+\infty}\rmd r'\,\frac{{r'}^2}{R_*^2}\rho^{\rm bulge}(r')\right]\!,\label{potweak10}
\ea
leading to
\ba
g^{\rm disk+gas}_1(R) &=&2\pi GR_*\sum_{n=0}^{+\infty}\frac{1}{n!}\frac{\sqrt{\pi}}{\G\left(\frac{1}{2}-n\right)}\left\{
\int_0^R\rmd Z'\,\frac{\rme^{-\frac{Z'}{h_Z}}}{h_Z}\right.\nonumber\\
&&\times\left[-\frac{1+2n}{R^{2(1+n)}}\int_0^{\sqrt{R^2-{Z'}^2}}\rmd R'\frac{{R'}^2}{R_*^2}\left({R'}^2+{Z'}^2\right)^n\right.\nonumber\\
&&\qquad\left.+2n\,R^{2n-1}\int_{\sqrt{R^2-{Z'}^2}}^{+\infty}\rmd R'\frac{{R'}^2}{R_*^2}\frac{1}{\left({R'}^2+{Z'}^2\right)^\frac{1+2n}{2}}\right] \nonumber\\
&&\left.+\int_R^{+\infty}\rmd Z'\,\frac{\rme^{-\frac{Z'}{h_Z}}}{h_Z}\,2n\,R^{2n-1}\int_0^{+\infty}\rmd R'\frac{{R'}^2}{R_*^2}\frac{1}{\left({R'}^2+{Z'}^2\right)^\frac{1+2n}{2}} \right\}\nonumber\\
&&\times\Sigma(R')\,{}_2F_1\left(-n,\frac{1}{2}+n;1;\frac{{R'}^2}{{R'}^2+{Z'}^2}\right)\,,\label{gweak9}\\
g^{\rm bulge}_1(R) &=& -4\pi G\frac{R_*^2}{R^2}\int_0^{R}\rmd r'\frac{{r'}^3}{R_*^3}\rho^{\rm bulge}(r')\,.\label{gweak10}
\ea
Next, \Eqq{dephi2log} is
\ba
\de\Phi^{\rm disk+gas}_2(R) &=& \frac{G}{R_*}\int_\cV\rmd^3\bm{x'}\,\rho(\bm{x'})\,\ln\frac{|\bm{x}-\bm{x'}|}{R_*}\nonumber\\
&\stackrel{\textrm{\tiny \Eq{logx}}}{=}&\frac{G}{R_*}\int_0^{+\infty}\rmd Z'\,\frac{\rme^{-\frac{Z'}{h_Z}}}{h_Z}\int_0^{+\infty}\rmd R'\,\Sigma(R')\,R'\nonumber\\
&& \times\left[2\pi\ln \frac{R_>}{R_*}-\sum_{n=1}^{+\infty}\int_0^{2\pi}\rmd\phi'\,\frac{\cos(n\phi')}{n}\left(\frac{R_<}{R_>}\right)^n\right]\nonumber\\
&=&\frac{2\pi G}{R_*}\int_0^{+\infty}\rmd Z'\,\frac{\rme^{-\frac{Z'}{h_Z}}}{h_Z}\int_0^{+\infty}\rmd R'\,\Sigma(R')\,R'\,\ln \frac{R_>}{R_*}\nonumber\\
&=&\frac{2\pi G}{R_*}\int_0^{R}\rmd Z'\,\frac{\rme^{-\frac{Z'}{h_Z}}}{h_Z}\left[\ln \frac{R}{R_*}\int_0^{\sqrt{R^2-{Z'}^2}}\rmd R'\,\Sigma(R')\,R'\right.\nonumber\\
&&\qquad\qquad\left.+\int_{\sqrt{R^2-{Z'}^2}}^{+\infty}\rmd R'\,\Sigma(R')\,\ln \frac{\sqrt{{R'}^2+{Z'}^2}}{R_*}\right]\nonumber\\
&&+\frac{2\pi G}{R_*}\int_R^{+\infty}\rmd Z'\,\frac{\rme^{-\frac{Z'}{h_Z}}}{h_Z}\int_0^{+\infty}\rmd R'\,\Sigma(R')\,R'\,\ln \frac{\sqrt{{R'}^2+{Z'}^2}}{R_*}\,,\label{potweak11}\\
\de\Phi^{\rm bulge}_2(R) &\stackrel{\textrm{\tiny \Eq{logxr}}}{=}& \frac{2\pi G}{R_*}\int_0^{+\infty}\rmd r'\,\rho^{\rm bulge}(r')\,{r'}^2\left[2\ln \frac{r_>}{R_*}-\sum_{n=1}^{+\infty}\int_0^{\pi}\rmd\theta'\,\sin\theta'\frac{\cos(n\theta')}{n}\left(\frac{r_<}{r_>}\right)^n\right]\nonumber\\
&=& \frac{2\pi G}{R_*}\int_0^{+\infty}\rmd r'\,\rho^{\rm bulge}(r')\,{r'}^2\left[2\ln \frac{r_>}{R_*}-\sum_{n=1}^{+\infty}\frac{1}{n(1-4n^2)}\left(\frac{r_<}{r_>}\right)^{2n}\right]\nonumber\\
&=& \frac{2\pi G}{R_*}\int_0^{+\infty}\rmd r'\,\rho^{\rm bulge}(r')\,{r'}^2\left[2\ln \frac{r_>}{R_*}-1+\ln\left(1-\frac{r_<^2}{r_>^2}\right)\right.\nonumber\\
&&\qquad\qquad\left.+\left(\frac{r_<}{r_>}+\frac{r_>}{r_<}\right)\,{\rm arctanh}\frac{r_<}{r_>}\right]\nonumber\\
&=& \frac{2\pi G}{R_*}\int_0^{R}\rmd r'\,\rho^{\rm bulge}(r')\,{r'}^2\left[2\ln \frac{R}{R_*}-1+\ln\left(1-\frac{{r'}^2}{R^2}\right)\right.\nonumber\\
&&\qquad\qquad\left.+\frac{R^2+{r'}^2}{r'R}\,{\rm arctanh}\frac{r'}{R}\right]\nonumber\\
&&+\frac{2\pi G}{R_*}\int_R^{+\infty}\rmd r'\,\rho^{\rm bulge}(r')\,{r'}^2\left[2\ln \frac{r'}{R_*}-1+\ln\left(1-\frac{R^2}{{r'}^2}\right)\right.\nonumber\\
&&\qquad\qquad\left.+\frac{R^2+{r'}^2}{r'R}\,{\rm arctanh}\frac{R}{r'}\right],\label{potweak12}
\ea
with corresponding gravitational fields
\ba
g^{\rm disk+gas}_2(R) &=&-\frac{2\pi G}{R_*R}\int_0^{R}\rmd Z'\,\frac{\rme^{-\frac{Z'}{h_Z}}}{h_Z}\int_0^{\sqrt{R^2-{Z'}^2}}\rmd R'\,\Sigma(R')\,R'\,,\label{gweak11}\\
g^{\rm bulge}_2(R) &=& -\frac{2\pi G}{R_*R}\left[\int_0^{+\infty}\rmd r'\,\rho^{\rm bulge}(r')\,{r'}^2
+\frac{1}{R}\int_0^{R}\rmd r'\,\rho^{\rm bulge}(r')\,r'(R^2-{r'}^2)\,{\rm arctanh}\frac{r'}{R}\right.\nonumber\\
&&\qquad\qquad\left.+\frac{1}{R}\int_R^{+\infty}\rmd r'\,\rho^{\rm bulge}(r')\,r'(R^2-{r'}^2)\,{\rm arctanh}\frac{R}{r'}\right].\label{gweak12}
\ea


\end{document}